\documentclass[nohyper,12pt,letterpaper]{JHEP3}

\usepackage{amsmath,epsfig}
\usepackage[latin1]{inputenc}
\usepackage{amssymb,amsfonts}
\usepackage{array,multirow,cite}

\title{Precision Counting of Small Black Holes}

\preprint{\hepth{0507014}\\LPTHE-05-14\\LPTENS-05-21\\TIFR/TH/05-27}

\author{Atish ~Dabholkar$^{\rm a}$, Frederik ~Denef$^{\rm b}$, 
Gregory ~W.~Moore$^{\rm b}$, and
Boris ~Pioline$^{\rm c, d}$\\

\centerline{$^{\rm a}$ Department of Theoretical Physics, Tata
Institute of Fundamental Research,}

\centerline{Homi Bhabha Road, Mumbai 400005, India}

\medskip

\centerline{$^{\rm b}$Department of Physics, Rutgers University,}

\centerline{Piscataway, NJ, 08854-8019 USA}

\medskip

\centerline{$^{\rm c}$LPTHE, Universit\'es Paris 6 et 7, 4 place
Jussieu,}

\centerline{75252 Paris cedex 05, France}

\medskip

\centerline{$^{\rm d}$LPTENS, D\'epartement de Physique de l'ENS,
24 rue Lhomond,}

\centerline{75231 Paris cedex 05, France}

\bigskip
{\tt E-mail:atish@tifr.res.in, denef@physics.rutgers.edu,
gmoore@physics.rutgers.edu,pioline@lpthe.jussieu.fr} }

\abstract{
It has recently been proposed that a class of supersymmetric
higher-derivative interactions in $\CN=2$ supergravity may
encapsulate an infinite number of finite size corrections to the
microscopic entropy of certain supersymmetric black holes. If
this proposal is correct, it allows one to probe the
string theory description of
black-hole micro-states to far greater accuracy than has been
possible before. We test this
proposal for ``small'' black holes whose
microscopic degeneracies can be computed exactly by counting the
corresponding perturbative BPS states. We also study   the
``black hole partition sum'' using general properties of
of BPS degeneracies. This complements and
extends our earlier work in \cite{Dabholkar:2005by}.}

\newcommand{\BesselI}[3]{\hat I_{#1}\left( #2 \pi \sqrt{ #3 }\right)}

\renewcommand{\th}{\theta}
\renewcommand{\Im}{\mbox{Im}}
\renewcommand{\Re}{\mbox{Re}}

\newcommand{\ar}[2]{\left[ {}^{#1}_{#2} \right]}
\newcommand{\art}[2]{\left[ {}^{\frac{#1}{2}}_{\frac{#2}{2}} \right]}

\newcommand{\pa}{\partial}
\newcommand{\nn}{\nonumber}
\newcommand{\eps}{\epsilon}
\newcommand{\IR}{\mathbb{R}}
\newcommand{\IZ}{\mathbb{Z}}

\newcommand{\Zint}{\mathbb{Z}}

\newcommand{\Tr}{\mbox{Tr}}

\newcommand{\apm}{\alpha'}

\def\g{\gamma}
\def\t{\tau}

\def\b{\beta}
\def\d{\delta}

\def\la{{\lambda}}

\def\L{I}

\def\wp{{\cal P}}
\def\vt#1#2#3{ {\vartheta[{#1 \atop  #2}](#3\vert \tau)} }
\def\CH{{\cal H}}

\def\CM{{\cal M}}
\def\CF{{\cal F}}

\def\CZ{{\cal Z}}
\def\CN{{\cal N}}

\def\CP{{\cal P}}
\def\CQ{{\cal Q}}

\def\half{{\frac12}}

\def\IC{\relax\hbox{$\inbar\kern-.3em{\rm C}$}}

\def\IC{{\bf C}}
\def\IP{{\bf P}}
\def\IQ{{\bf Q}}
\def\CN{{\cal N}}
\def\CQ{{\cal Q}}
\def\CX{{\cal X}}
\def\CZ{{\cal Z}}

\def\bea{\begin{eqnarray}}
\def\eea{\end{eqnarray}}
\def\be{\begin{equation}}
\def\ee{\end{equation}}
\def\ba{\begin{align}}
\def\ea{\end{align}}
\def\bse{\begin{subequations}}
\def\ese{\end{subequations}}
\def\1F1{{}_1\!F_1}
\def\2F0{{}_2\!F_0}

\def\qeq{\, {? \atop ~} \hskip-4mm =}

\begin{document}
%\maketitle \setcounter{tocdepth}{2}
%\tableofcontents

\section{Introduction}

One of the distinct successes of string theory is that it explains
the statistical origin of the thermodynamic Bekenstein-Hawking
entropy \cite{Bekenstein:1973ur,Bekenstein:1974ax,Hawking:1974sw}
of certain supersymmetric black holes in terms of counting of
underlying micro-states \cite{Strominger:1996sh}. This has been
particularly successful in the case of dyonic black holes in
string theories with $\CN=2$ supersymmetry in four dimensions. In
the regime of large electric and magnetic charges, these black
holes possess a non-singular horizon with area much larger than
the Planck or string scale. For such ``large'' black holes, the
Bekenstein-Hawking entropy, one quarter of the horizon area in
Planck units, matches the logarithm of the number of micro-states
of specific supersymmetric brane-configurations with the same
quantum numbers \cite{Maldacena:1996gb,Johnson:1996ga}.

For black holes with large but finite area, there are subleading
corrections to the Bekenstein-Hawking formula, due to
higher-derivative
interactions in the quantum effective action. The latter alter both the
black hole geometry near the horizon, as well as the very relation
between macroscopic entropy and geometry \cite{Wald:1993nt,
Iyer:1994ys}. On the microscopic side, there are also
finite size corrections to the
entropy\footnote{Following standard practice, we define the
entropy as the Legendre transform of the logarithm of the
partition function in a given statistical ensemble.}, which
however depend on the choice of a statistical ensemble.
It is natural to ask whether the successful matching between the
Bekenstein-Hawking entropy and the string theoretical counting of
black hole micro-states continues to hold beyond leading order.

Several advances in recent years have made it possible to address this
question. By generalizing the attractor mechanism for $\CN=2$ black holes,
Cardoso, de Wit, and Mohaupt (CdWM) computed the Bekenstein-Hawking-Wald (BHW)
entropy incorporating an infinite number of
higher derivative F-type interactions
\cite{LopesCardoso:1998wt,LopesCardoso:1999cv,LopesCardoso:1999xn,
LopesCardoso:2000qm}. Revisiting this result,
Ooguri, Strominger and Vafa (OSV) conjectured
that the statistical ensemble implicit in the CdWM entropy is
a specific mixed ensemble \cite{Ooguri:2004zv}, and furthermore
that non F-type interactions can be consistently neglected provided
one restricts to a suitable supersymmetric index on the microscopic side.
If correct, this proposal opens the way to a
more detailed comparison of macroscopic and microscopic
degeneracies than has been possible thus far.

For a generic dyonic black hole, such a comparison is hampered by
our insufficient understanding of the dynamics of the D-brane micro-states.
The aim of this work is to identify and analyze a large class of examples
where microscopic degeneracies are known exactly and where a very
explicit comparison is possible exactly and to all orders in an
asymptotic expansion.  This complements and extends our
earlier work \cite{Dabholkar:2005by} where some of the main
results were announced. It should be noted that
alternative approaches have been put forward \cite{Sen:2004dp,Sen:2005pu,
Sen:2005ch,Cardoso:2004xf}, relying on different
statistical ensembles. It goes beyond the scope of this paper to relate
these two approaches.

\subsection{The Black Hole Attractor and the OSV Conjecture \label{osvintro}}

In general, the quantum effective admits an infinite series of
unknown higher-derivative corrections, which make it difficult to
determine higher-order corrections to the macroscopic entropy.
In Type IIA string theory compactified on a Calabi-Yau three-fold
$\CX$ however, there exist an  infinite series of computable
higher-derivative F-term corrections of the form $F_h(X) (^-
C^-)^2 (T^-)^{2h-2}$, where $^- C^-$ and $T^-$ denote the
anti-self-dual part of the Weyl tensor and graviphoton field strength,
and $X^I$ the K\"ahler moduli of $\CX$. The peculiarity of these
interactions is that they can be written as the integral of a chiral
density in superspace, and satisfy certain non-renormalisation properties.
In particular, they arise only at genus $h$ in type II string theory,
and the coefficient
$F_h(X^I)$ reduces to the genus $h$ vacuum amplitude in the
A-model topological string on $\CX$
\cite{Bershadsky:1993ta,Antoniadis:1993ze}. The
Bekenstein-Hawking-Wald  macroscopic entropy of BPS black holes
incorporating these interactions was computed by CdWM in
\cite{LopesCardoso:1998wt,LopesCardoso:1999cv,LopesCardoso:1999xn,
LopesCardoso:2000qm}, generalizing the standard tree-level
attractor mechanism. As noticed by the authors of
\cite{Ooguri:2004zv}, this expression takes a particularly simple
form after Legendre transform with respect to the electric
charges\footnote{The generality of this fact has been recently
clarified in \cite{Sen:2005wa}.}, 
\be \label{sbhw} S_{\rm CdWM}(p^I, q_I) = {\cal
F}_{top}(p^I, \phi^I  ) + \pi\ \phi^\L q_\L \ ,\qquad \pi q_\L =
-\frac{\partial}{\partial \phi^\L} {\cal F}_{top}(p^I,\phi^I) \ee
where \be
\label{fbhw}
 {\cal F}_{top}(p^I , \phi^I ) := - \pi \ {\rm Im}
\left[ F_{top}\left( p^I + i \phi^I, 2^8 \right) \right].
\ee
is proportional to the imaginary part of the
all-order topological string vacuum amplitude
$F_{top}(X,W^2)=\sum_{h=0}^{\infty} W^{2h-2} F_h(X^I)$
evaluated at $X^I= p^I + i \phi^I$ and $W^2=2^8$. The
attractor equations \cite{Ferrara:1995ih,Ferrara:1996dd,Ferrara:1996um}
\be
\label{att}
p^I = \Re(X^I)\ ,\qquad q_I=\Re( \partial F_{top} / \partial X^I)
\ee
controlling the fixed point behavior of the K\"ahler moduli
at the horizon follow naturally from this procedure \cite{Ooguri:2004zv}.

Based on this observation, Ooguri, Strominger and Vafa (OSV) have
proposed that the statistical ensemble implicit in the above
Bekenstein-Hawking-Wald entropy has fixed magnetic charges $p^I$,
but fluctuating electric charges $q_I$ at a fixed electric
potential $\phi^I$ \cite{Ooguri:2004zv}: 
\be 
\label{zosv} Z_{OSV}
(p^I, \phi^I) :=  \exp {{\cal F}_{OSV}(p^I , \phi^I)} :=
\sum_{q^I\in\Lambda_e} \Omega(p^I, q_I) \ e^{\pi \phi^I q_I} 
\ee 
where $\Omega(p^I,q_I)$ denotes the  number or possibly a suitable index
of micro-states with electric and magnetic charges $q_I$ and
$p^I$, and $\Lambda_e$ is the lattice of electric charges in the large
volume polarization. Put otherwise, the essence of the proposal
\cite{Ooguri:2004zv} is an equality between the {\it microscopic}
free energy ${\cal F}_{OSV}$ in the mixed statistical ensemble
\eqref{zosv} and the {\it macroscopic} free energy ${\cal
F}_{top}$ computed from the higher-derivative F-term interactions,
\be \label{osvf} {\cal F}_{OSV} (p^I, \phi^I) \equiv {\cal
F}_{top}  (p^I, \phi^I) \ee Using the relation \eqref{fbhw}
between the topological free energy ${\cal F}_{top}$ and the
topological string amplitude $F_{top}$, this equation may be
rephrased as a relation between the BPS black hole degeneracies in
type II on $\CX$ and the topological string amplitude, \be
\label{osv1} Z_{OSV} (p^I, \phi^I)  \equiv \left|
\exp\left[\frac{i\pi}{2} F_{top}( p^I + i \phi^I , 2^8 )
  \right]
\right|^2 \ .
\ee
Evaluating the sum over charges in the partition function \eqref{zosv}
by steepest descent, one indeed finds that the
Legendre transform of the entropy is equal to the topological
free energy \eqref{fbhw}, in the limit of large charges.

The proposal \eqref{osvf} goes far beyond the large charge regime
in which it was motivated, since it allows in principle to extract the
microscopic degeneracies of BPS black holes from the topological
string amplitude by means of an inverse Laplace transform, \be
\label{osvii} \Omega(p^I,q_I) \equiv \int\ [d\phi^I]\  \exp\left[
{\cal F}_{top} (p^I , \phi^I) + \pi q_I \phi^I \right]\ . \ee A
strong form of the conjecture asserts that this equation holds at
finite electric and magnetic charges, provided some yet unknown
non-perturbative contributions to the topological string amplitude
are included \cite{Ooguri:2004zv}. A weaker form states that this
equality hold to all orders in an asymptotic expansion in the
inverse of the charges \cite{Ooguri:2004zv}. One
aim of our work is make Eqs. \eqref{osvii},\eqref{osv1} more precise
and use them to study the degeneracies of finite charge black holes.
Certain proposed nonperturbative corrections to \eqref{osvii} have been
explored in \cite{Vafa:2004qa,Aganagic:2004js}, but in a rather different
context from the examples studied here.

\subsection{Small Black Holes \label{dhintro}}

For this purpose, it is useful to consider cases for which the
exact degeneracies of the micro-states are computable. Using
heterotic / type II duality, this is indeed possible for type II
black holes which are dual to the heterotic Dabholkar-Harvey (DH)
states \cite{Dabholkar:1989jt,Dabholkar:1990yf}. Recall that these
are BPS states in the perturbative heterotic spectrum, which exist
provided the conformal field theory contains a compact free boson.
The simplest example is provided by a state carrying quantized
momentum $n$ and winding number $w$ around an internal circle. The
left- and right-moving momenta are given by
\begin{equation}\label{lrmomenta}
q_{R, L} \equiv \sqrt{\frac{\apm}{2}}(\frac{n}{R} \pm
\frac{wR}{\apm}),
\end{equation}
and the vector $(q_R, q_L)$ belongs to the Narain lattice
$\Gamma^{1,1}$.
 Such a state is
half-BPS as long as it is in the right-moving superconformal
ground state but it can carry arbitrary left-moving excitations
that  satisfy the level-matching
\begin{equation}\label{virasoro}
    N- 1 = \half (q_R^2 - q_L^2) = nw,
\end{equation}
where $N$ is the left-moving excitation level. For given charges
$(n,w)$, there is a Hagedorn density $\Omega(n,w)\sim \exp( 4 \pi
\sqrt{|nw|} )$ of such states, as a result of the large degeneracy
of the left-moving excitations.

The integers $(n,w)$ can be viewed as the
quantized electric charges under the Kaluza-Klein
and Neveu-Schwarz gauge fields $g_{\mu i}$ and $B_{\mu i}$
arising by dimensional reduction along the circle.
The mass $M$ of the
state  $(n, w)$ saturates the BPS bound
\begin{equation}\label{mass}
    M^2 = q_R^2 = \left[\frac{n}{ R} +
\frac{w R}{\apm} \right]^2 = \frac{n^2}{R^2} + \frac{w^2
      R^2}{\apm^2} + \frac{2(N -1)}{\apm}
\end{equation}
where $R$ is the radius of the circle. Provided it does not become
  degenerate with another half-BPS state with which it may pair
up, the  $(n, w)$ state is therefore stable. As the string
coupling $g_H$ is increased, the de Broglie - Compton wavelength
$1/M$ of the particle becomes smaller than its Schwarzschild
radius $M l_P^2$, leading to the formation of an
extremal black hole with electric charges $(n,w)$. It is thus
tempting to compare the Bekenstein-Hawking entropy of this black
hole with the logarithm of the number of fundamental strings with
the same charges
\cite{Susskind:1993ws,Susskind:1994sm,Russo:1994ev,Sen:1995in,Horowitz:1996nw},
\be
S_{DH} = \log \Omega(n,w) \sim 4 \pi \sqrt{|nw|}
\ee
More generally, the black hole charges are characterized by an
arbitrary  charge vector $Q$ in the Narain lattice $\Gamma^{6,
22}$ and the leading entropy of the DH states in that case goes as
$S_{DH} \sim 4 \pi \sqrt{Q^2/2}$.

In contrast with the ``large'' black holes discussed above, these
 black holes are singular solutions of the tree-level
supergravity Lagrangian \cite{Sen:1994eb,Dabholkar:1995nc}, where
the horizon and the inner singularity coalesce. Their classical
entropy therefore vanishes, as a result of their carrying only
electric charge (in the natural heterotic polarization). While the
heterotic string coupling goes to zero at the singularity,
higher-derivative $\apm$ corrections are however expected to be
quite important, and, assuming the singularity is resolved,  have
been argued to lead to an entropy of the required order
\cite{Sen:1995in}. By including the tree-level $R^2$ correction to
the heterotic effective Lagrangian (or, from the type II point of
view, the large volume limit of the one-loop topological amplitude
$F_1$), it was shown recently that the black hole develops a
smooth horizon, with a similar geometry $AdS_2 \times S^2 \times
\CX$ as found in the large black hole
case \cite{Dabholkar:2004dq,Dabholkar:2004yr} (see \cite{Behrndt:1996jn,
Behrndt:1998eq,LopesCardoso:2000qm} for
earlier work on this subject). Moreover, the
Bekenstein-Hawking-Wald entropy, taking into account this $R^2$
correction, matches the microscopic entropy in leading order,
including the precise numerical coefficient
\cite{Dabholkar:2004dq,Dabholkar:2004yr}. The geometry
interpolating between the horizon and infinity has been recently
studied in \cite{Sen:2004dp,Hubeny:2004ji}. For this type of black
hole, the four-dimensional heterotic string coupling is  of order
$g_H^2 \sim 1/\sqrt{|nw|}$ at the horizon, so that the area is of
the same order as the inverse tension of the heterotic string
$l_H^2 = l_P^2 / g_H^2$ at the horizon. We shall thus refer to
these states as  ``small'' black holes, keeping in mind that, for
large charges, they are nevertheless much larger than the Planck
scale.

Recently, the OSV conjecture has been tested
for small black holes in type
IIA string theory compactified on $K3 \times T^2$, or
equivalently, in heterotic string theory on $T^6$
\cite{Dabholkar:2004yr}. Although the original proposal was
formulated for $\CN=2$ backgrounds, an extension to  the $\CN=4$
case is simpler to analyze since all gravitational F-terms vanish
except $F_1$ \cite{Harvey:1996ir}. Using the generalized attractor
formalism in
\cite{LopesCardoso:1998wt,LopesCardoso:1999cv,LopesCardoso:1999xn,
LopesCardoso:2000qm}, adapted to the $\CN=4$ setting, it was found
that the macroscopic entropy of these small black holes precisely
matches the Ramanujan-Hardy estimate for the number of heterotic
BPS states preserving 1/2 supersymmetry \cite{Dabholkar:2004yr}.
It was also shown that  even the sub-leading corrections to the
entropy computed using the OSV proposal match to all orders in an
asymptotic expansion. The super-gravity solutions for these small
black holes have been further analyzed in
\cite{Dabholkar:2004dq,Sen:2004dp,Hubeny:2004ji,Bak:2005mt}.

In this paper, we greatly extend the range of validity of the
analysis in \cite{Dabholkar:2004yr}, by studying the exact degeneracy
of small black holes in a variety of backgrounds with $\CN=4$
supersymmetry (but a different low-energy spectrum from the
``benchmark'' $K3 \times T^2$ case), or with $\CN=2$ supersymmetry
(for Calabi-Yau compactifications with a $K3$ fibration).

\subsection{Summary of Main Results}

For the reader's convenience, we summarize our main results below:

\begin{enumerate}

\item On the heterotic side, by standard orbifold techniques, the
microscopic degeneracy of the DH states can be enumerated using
{\it modular forms}. The leading microscopic entropy at large
charge can be extracted using the Hardy-Ramanujan formula. The
{\it Rademacher formula} provides a convenient way to extract
subleading corrections: it expresses the Fourier coefficients of
the modular form as a series of {\it modified Bessel functions},
where each term is exponentially suppressed (but nevertheless
exponentially growing) with respect to the previous one (see
\cite{Dijkgraaf:2000fq} for a review). In particular, all power
corrections to the leading entropy are captured by the first
Bessel function in the Rademacher expansion.

\item Retaining only the perturbative part of the topological amplitude
(i.e. discarding the Gromov-Witten instanton series), and assuming a proper
choice of contour, we find that the integral \eqref{osvii} can be computed
exactly, both in the large and small black hole case,
and expressed as a {\it modified Bessel function} of the first kind.
Using the standard asymptotic expansion of the latter, the leading term
is the Bekenstein-Hawking-Wald entropy $S_{BHW}= 4\pi \sqrt{Q^2/2}$
predicted by the generalized attractor
mechanism. In particular, due to the topological coupling $F_1$, the
entropy of small black holes is computable and
finite, as observed in \cite{Dabholkar:2004yr}. In addition, the
Bessel function captures {\it an infinite number of computable corrections in
inverse powers of the charges}.

\item In a variety of $\CN=4$ and $\CN=2$ models, we find that
the integral \eqref{osvii}, {\it neglecting the Gromov-Witten
instanton series}, reproduces precisely the {\it leading Bessel function} in
the Rademacher expansion of the degeneracies of heterotic DH states.
In other words, the OSV proposal predicts the correct degeneracies
of BPS states, {\it to all orders in an asymptotic expansion in inverse
powers of the charges}. In particular, the leading entropy
is correctly reproduced, including the corrections computed
in \cite{Maldacena:1997de}. Importantly, this success relies only on the large
volume limit of $F_1$ only (equivalently, on the heterotic tree-level $R^2$
amplitude).

\item For this all order perturbative agreement to hold, it is
important to use the {\it holomorphic} topological amplitude,
which controls the Wilsonian supergravity action, rather than the
non-holomorphic BCOV generating function, which describes the 1PI
couplings in the low-energy effective action. This is consistent
with the discussion in \cite{Verlinde:2004ck}, but in stark
contrast with the alternative approaches in \cite{Sen:2005pu,Cardoso:2004xf}
(note however that \cite{Verlinde:2004ck} has proposed a formally
equivalent formula,
using the holomorphic rather than real polarization, where
non-holomorphic anomalies are likely to play a role).
It is also important to count states with {\it arbitrary
angular momentum} $J$, as the restriction to $J=0$ states leads to
different subleading terms in the microscopic amplitude, which
would spoil agreement with the OSV prediction. In other words, the
proper statistical ensemble implicit in the
Bekenstein-Hawking-Wald entropy appears to be an ensemble with
zero angular velocity at the horizon, rather than zero angular
momentum. Finally, it is necessary to consider ratios of
degeneracies at fixed magnetic charge only, in order to cancel a
magnetic-charge dependent pre-factor ${\cal N}(p)$, which would
spoil duality invariance. For $p^0\neq 0$, a more drastic
modification is necessary, since, as shown in Section 
\ref{labhd6}, the pre-factor in general
involves both electric and magnetic charges.

\item The neglect of Gromov-Witten instantons can be rigorously
justified in $\CN=4$ cases, as all instanton corrections are
exponentially suppressed. The situation is more subtle in $\CN=2$
theories: when $\chi(\CX)\neq 0$, the series of point-like
instantons contribution becomes strongly coupled in the regime
of validity of the Rademacher formula,
$\hat q_0 \gg \hat C(p)$. The strong coupling
behavior is controlled, up to a logarithmic term, by the Mac-Mahon
function, which is exponentially suppressed in this regime. Upon
absorbing the logarithmic term into a redefinition of the
topological string amplitude $\Psi_{top}\to \lambda^{\chi/24}
\Psi_{top}$, one recovers the naive result. As for non-degenerate
instantons, they are exponentially suppressed provided all
magnetic charges are non zero. This is unfortunately not the case
for the small black holes dual to the heterotic DH states, whose
K\"ahler classes are attracted to the boundary of the K\"ahler cone at the
horizon. In this case, we cannot rigorously justify the neglect of
Gromov-Witten contributions.

\item Even in the cases where an all-order agreement is obtained,
the OSV formula appears to fail in reproducing the
{\it subleading Bessel functions}
in the Rademacher expansion of the microscopic degeneracies,
as those cannot be associated to subleading saddle points in the
contour integral \eqref{osvii}  in any obvious way\footnote{It was
recently proposed that exponentially suppressed contributions should
reflect multi-centered black hole configurations \cite{Dijkgraaf:2005bp}.}.
As a matter of fact, we encounter serious difficulties in trying
to make sense of the formula \eqref{osvii} non-perturbatively.
Due to the non-convexity of the free energy ${\cal F}$ (or,
equivalently, the instability of the mixed thermodynamical ensemble),
the convergence
of the integral can only be achieved when the potentials $\phi^I$
take imaginary values. However, at least for Calabi-Yau threefolds
admitting a $K3$ fibration, the topological string amplitude
$\Psi_{top}$ is an
{\it automorphic} form, and is very badly behaved at the boundary of moduli
space where the moduli $X^I$ become real.

\item On general grounds (e.g. if it is to satisfy the second
law of thermodynamics), the Bekenstein-Hawking-Wald
entropy, including all higher-derivative corrections, is expected
to be equal to the logarithm of the {\it total} number of micro-states.
The truncation to only F-term type higher-derivative corrections
is not expected to have a thermodynamical interpretation, unless
non-F-terms do not contribute by some non-renormalization 
property\footnote{
See \cite{Kraus:2005vz} for some recent 
interesting results in this direction.}.
On the other hand, the counting of heterotic DH states at
zero string coupling may differ from the
actual number of states in the regime where a black hole is formed,
due to the possibility of BPS states pairing up into longer multiplets.
Useful diagnostic tools to determine whether this happens are
{\it helicity supertraces}
$\Omega_n = \Tr (-1)^F J_3^n$ (where
$F$ is the space-time fermion number and $J_3$ one of the
generators of the little-group of a massive particle in
4 dimensions), namely $\Omega_2$ for 1/2 BPS states in theories
with $\CN=2$ supersymmetry, $\Omega_4$ for 1/2 BPS states in theories
with $\CN=4$ and $\Omega_6$ for 1/4 BPS states in theories with $\CN=4$.
In contrast to absolute degeneracies, helicity supertraces are
invariant under generic variations of the moduli (except for lines
of marginal stability). If cases where the degeneracies at zero coupling can
be identified with an helicity supertrace, one can reasonably assume
that they will be equal to the actual number of states in the black hole
regime (barring the unlikely possibility that long multiplets unpair
as the coupling is increased). We can then   reliably compare them with the
macroscopic BHW entropy\footnote{This differs from the interpretation
advocated in \cite{Ooguri:2004zv}, who propose to identify directly the topological
amplitude with a supersymmetric index. This is a mathematically appealing
and logically acceptable conjecture, but it has no direct bearing on the
relation between the BHW entropy and the counting of black hole micro-states.}.
In some cases however, the helicity supertraces
can be {\it exponentially}
smaller than the zero-coupling degeneracies\footnote{This
occurs e.g. in the case
of 5D black holes \cite{Vafa:1997gr}, but we shall find numerous other
examples in this work.}, and it is difficult to determine
the actual number of states at strong coupling.

\item We find that in cases where the absolute degeneracies
are equal to the helicity supertraces, the instanton-deprived
OSV proposal appears to work
successfully. This includes 1/2 BPS states in all $\CN=4$ models, as well
as BPS states in twisted sectors of $\CN=2$ orbifolds.
This suggests that, for this class of BPS black holes, non F-type
higher derivative interactions   have no effect, if not
on the geometry, at least on the Bekenstein-Hawking-Wald
entropy. It would be very
interesting to check such a non-renormalization explicitly.

\item In cases where it appears to fail, the helicity
supertraces are in general exponentially
smaller than their absolute degeneracy, due to cancellations of pairs
of DH states. This occurs in general for (i) untwisted DH states of
$\CN=2$  heterotic orbifolds, and (ii) DH states in
$\CN=4$ type II orbifolds. In case (i), the OSV prediction
appears to agree with the {\it absolute degeneracies} of untwisted DH
states to leading order ( which have the same exponential growth as
twisted DH states), but not at subleading order (as
the subleading corrections in the untwisted sector are moduli-dependent,
and uniformly smaller than in the twisted sectors).
In models where twisted and untwisted states cannot be distinguished
by their charges, the helicity supertrace $\Omega_2$ is dominated by the
contribution of the twisted sectors, and it may be consistent to
identify it with the l.h.s. of \eqref{osvii}. The situation in case (ii)
is rather different, since the helicity supertraces grow only as a power
rather than exponentially. On the macroscopic side,
$R^2$ interactions are not sufficient to resolve the singular horizon,
and higher derivative interactions are bound to become important.

\item Conversely, one may try to compute the black hole
partition function \eqref{zosv} from our knowledge of the
microscopic degeneracies, and compare to the proposed answer
\eqref{osv1}. For some choices of Calabi-Yau manifolds and of
magnetic charges, in the infinite radius limit, the degeneracies are
known exactly for arbitrary electric charges, and this program can
be carried out explicitly. Examples of this are D4-D2-D0 bound
states wrapped on a rigid divisor, or D-branes dual to heterotic DH
states.

An immediate problem which arises when attempting to compute the
partition sum \eqref{zosv} is that it is badly divergent. We solve
this by introducing a convenient and physically natural regulator,
namely an additional Boltzmann weight $e^{-\alpha H(p,q)}$, with
$H(p,q)$ the BPS energy of the given charge. This renders the
partition function finite and rigorously justifies various formal
manipulations, after which one can send $\alpha$ back to zero.

Our result is that in these cases, the \emph{polynomial} part of the
resulting free energy indeed equals the corresponding terms at the
right-hand side of \eqref{osv1}, but with an additional sum over
integral imaginary shifts of the $\phi^a$ on the right hand side.
This ensures periodicity under $\phi^a \to \phi^a + 2 i n^a$, as is
required by the definition \eqref{zosv}. In fact this summed version
of \eqref{osv1} is trivially equivalent to the integral form
\eqref{osvii}, with the $\phi^a$ integration contours running over
the entire imaginary axis.

More importantly, we find that at least for these choices of
charges, the non-perturbative part of the topological string free
energy is not reproduced; the corrections to the polynomial terms of
both sides do not match. This is true even in the limit of large
charges.

\item Despite the fact that the exact degeneracies are not known in
more general cases, one can extract some information about the
general partition function by exploiting large radius monodromy
invariance. These are integral shifts of the NS B-field, acting on
the $\phi^a$ as $\phi^a \to \phi^a + n^a \phi^0$. This induces a
spectral flow on the electric and magnetic charges, which leaves the
degeneracies unchanged, at least in the large volume limit.
Exploiting this symmetry, we argue that the BPS partition sum does
not generate the full data of the topological amplitude at any
finite magnetic charge $P$. In particular we show that the
$\phi^a$-dependence of the integrand in \eqref{osvii} predicted from
monodromy invariant BPS degeneracies is simply given by a finite sum
of Gaussians, which is to be compared to the intricate
$\phi^a$-dependence generated by the Gromov-Witten series in the
topological string free energy. The conjecture might   still
hold in a suitable asymptotic sense when  $P \to \infty$, because in
this case number of independent Gaussian terms will in general
 go to infinity.

\end{enumerate}

\subsection{Outline of the Paper}

This paper is organized as follows.

In Section 2, we illustrate our methods in the simplest example with
$\CN=4$ supersymmetry: type IIA string theory compactified
on $K3\times T^2$, or equivalently, in heterotic string theory compactified
on $T^6$, extending the analysis in \cite{Dabholkar:2004yr}.

In Section 3, we generalize this analysis to a class of $\CN=4$ models
with reduced rank, obtained as freely acting orbifolds of the IIA/$K3\times
T^2$ or Het/$T^6$ models.

In Section 4, we come to the $\CN=2$ supersymmetric case, for which the
OSV conjecture was originally formulated. After recalling the main features
of the topological string amplitude, we compute the asymptotic degeneracies
predicted by \eqref{osvii} for a particular scaling of the charges.

In Section 5, we compare this prediction to the microscopic counting
in the perturbative heterotic description. After discussing several
illustrative $\CN=2$ models, we find the asymptotic degeneracies of
DH states for arbitrary asymmetric orbifold of the heterotic string
compactified on $T^6$.

In Section 6, we reverse the approach, construct the
partition function in the mixed thermodynamical ensemble \eqref{zosv}
from our partial knowledge of the micro-canonical degeneracies,
and compare the result to the topological string amplitude.

Section 7 contains our conclusions and further comments.

In the Appendices, the reader will find a summary of the Rademacher
expansion for the Fourier coefficients of modular forms with negative
weight (Appendix A), a collection of useful modular identities (B),  an analysis of the degeneracies
of the DH states at fixed angular momentum in the Het/$T^6$ model (C)
and a detailed computation of the degeneracies of DH states in $\CN=4$
and $\CN=2$ orbifolds of the $SO(32)$ heterotic string (D),
a detailed
analysis of the asymptotic expansion of the Mac-Mahon as well as an
observation on its (non-)modularity (E).

\section{A Benchmark $\CN=4$ Example: Type $IIA/K3\times T^2$ \label{424}}

In this section, we revisit the ``benchmark'' case of
small black holes in type IIA string theory compactified
on $K3 \times T^2$, or equivalently
heterotic string compactified on $T^6$, first discussed in
\cite{Dabholkar:2004yr}. Despite the fact that this model has
$\CN=4$ supersymmetry, we shall be able to apply the $\CN=2$ attractor
formalism, provided 4 out of the 28 charges, corresponding to
gauge fields in gravitino multiplets of $\CN=2$ supersymmetry, vanish.
For this reason we shall denote this model as the Het/IIA$(4,24)$,
where the first number refers to the number of supersymmetries
in 4 dimensions, and the second to the effective
number of $\CN=2$ vector multiplets, including the graviphoton.
More general Het/IIA$(4,n_V)$ compactifications with $\CN=4$ supersymmetry
and $n_V<24$ vector multiplets will be discussed in Section 3
and Appendix \ref{oth4}.

\subsection{Review of Heterotic/Type II Duality in 4 Dimensions}

Let us consider the type IIA string compactified on $K3 \times
T^2$. The massless spectrum consists of the $\CN=4$ supergravity
multiplet together with $22$ vector multiplets. The moduli space
takes a factorized form \be \label{modspace}
\frac{SL(2,\IR)}{U(1)} \times \frac{SO(6,n_V-2,\IR)}{SO(6)\times
SO(n_V-2)} \ee with $n_V=24$, where the first factor corresponds
to the K\"ahler modulus $T$ of $T^2$, while the axio-dilaton $S$,
the complex structure modulus $U$ of $T^2$ and the geometric
moduli of $K3$ sit in the second factor. Points in
\eqref{modspace} related by an action of the duality group
$Sl(2,\IZ)\times O(\Gamma_{6,22})$ are non-perturbatively
equivalent. The gauge fields in the 22 vector multiplets originate
from the 3-form gauge field in the ten-dimensional type IIA
string, after reduction on a basis $\gamma_a, a=2\dots 24$ of
2-cycles in $H^2(K3,\IR)$. Accordingly, the electrically charged
states are D2-branes wrapped on 2-cycles $\gamma_a$, and their
magnetic counterparts are D4-branes wrapped on $T^2 \times
\gamma_a$, with charges $(q_a, p^a)$, respectively. On the other
hand, the 6 gauge fields in the $\CN=4$ supergravity multiplet
correspond to the ten-dimensional Ramond-Ramond (RR) 1-form, the
3-form reduced on $T^2$, the Kalb-Ramond 2-form reduced on either
circle of $T^2$ and the Kaluza-Klein gauge fields on $T^2$. The
corresponding electric charges are therefore carried by the
D0-brane (denoted by $q_0$), D2-brane wrapped on $T^2$ ($q_1$),
the fundamental string wrapped on $S^1 \subset T^2$ ($w^5, w^6$)
and the momentum states on $T^2$ ($n_5,n_6$), respectively; the
magnetic charges are carried by the D6-brane wrapped on $K3\times
T^2$ ($p^0$), the D4-brane wrapped on $K3$ ($p^1$), the NS5-branes
wrapped on $K3\times S^1$ ($m^5,m^6$) and the Kaluza-Klein
monopoles on $S^1\subset T^2$ ($k^5,k^6$).

One of the earliest string duality conjectures identifies this
model with the heterotic string compactified on $T^6$. The
massless spectrum is identical, but the $Sl(2)/U(1)$ complex
scalar in the supergravity multiplet is now the heterotic
axio-dilaton. The second factor in \eqref{modspace} is identified
as the Narain moduli space of the even self-dual compactification
lattice $\Gamma_{6,22}$. The 28 charges now correspond to the
Cartan subalgebra of the rank 16 ten-dimensional gauge group, the
reduction of the Kalb-Ramond two-form on $T^6$ and the
Kaluza-Klein gauge fields on $T^6$. Accordingly, the electric
charges in the natural heterotic polarization are carried by the
10-dimensional charged states, the fundamental string wound around
$S^1 \subset T^6$ and the momentum states along $S^1\subset T^6$;
the corresponding magnetic charges are carried by H-monopoles,
NS5-branes and KK5-monopoles wrapped on $T^5 \subset T^6$. The
precise map can be obtained by applying triality on an $SO(4,4)$
subgroup of the $SO(4,20)$ duality group in 6 dimensions
\cite{Kiritsis:2000zi}, and is displayed in Table \ref{tab4}
below. In particular, the $SO(6,22)$ vectors $Q,P$ of electric and
magnetic charges in the natural heterotic polarization are related
to the type II charges by \bea
Q &=& (q_0, p^1, q_a, n_5, n_6, m^5, m^6) \\
P &=& (-q_1, p^0, C_{ab} p^b, -w^6, w^5, k^5, k^6)
\eea
with $SO(6,22)$ invariant inner products
\begin{subequations}
\label{ppqqpq}
\bea
Q^2 &=& 2q_0 p^1 + q_a C^{ab} q_b + 2m^i n_i \\
P^2 &=& -2q_1 p^0 + p^a C_{ab} p^b + 2\eps_{ij} w^i k^j\\
Q\cdot P &=& p^0 q_0 - p^1 q_1 + p^a q_a + n_i k^j + \eps_{ij} m^i
w^j
\eea
\end{subequations}
The heterotic polarization is therefore obtained from the type II
large volume polarization by applying
electric-magnetic duality to the $(D4/K3, D2/T2)$ and $(F1, NS5/K3\times S^1)$
pairs.

\begin{table}
\label{tab4}
$$
\begin{array}{|c|c||l|l||l|l|}
\hline
\multicolumn{2}{|c||}{Het/T^6}      &
\multicolumn{2}{c||}{IIA/K3\times T^2} &
\multicolumn{2}{c|}{\rm Charges}
\\
\hline
KK/1       &      NS5/\hat 1     &     D0  & D2/T^2 & q_0 & q_1
\\
KK/2,3,4   &     NS5/\hat 2,\hat 3,\hat 4   &
                       D2/\gamma_{a }   &  D4/T^2\times \gamma_a & q_{a=2,3,4}  &p^{a=2,3,4}
\\
KK/5,6     &    NS5/ 6,5   &  KK/5,6  &  F1/6,5 & n_5, n_6 & -w^6, w^5
\\
\hline
F1/1       &      KKM/\hat 1     &     D4/K3   &
D6/K3\times T^2 & p^1 &  p^0
\\
F1/2,3,4   &   KKM/\hat 2,\hat 3,\hat 4   &     D2/\gamma_{a}   &
  D4/T^2\times \gamma_a  & q_{a=5,6,7} & p^{a=5,6,7}
\\
F1/5,6  &      KKM/\hat 5,\hat 6
&     NS5/\hat 6,\hat 5  &   KKM/\hat 5, \hat 6 & m^5, m^6 & k^5, k^6
\\
\hline
Q_{1,\dots,16} & HM_{1,\dots,16} & D2/\gamma_a & D4/T^2 \times \gamma_a & q_{a=8,\dots, 23} & p^{a=8,\dots, 23}
\\
\hline
\end{array}
$$
\caption{ Charge assignment in the Het/IIA$(4,24)$ model. The
vertical columns denote $O(6,22)$ vectors. Even and odd columns
are related by the Weyl reflection in $Sl(2,\IZ)$, i.e. S-duality
on the heterotic side or double T-duality on $T^2$ followed by an
exchange of the two circles on the type II side.
Abbreviations: KK/1= momentum state
along $S^1$, $NS5/\hat 1$= NS5-brane wrapped on all directions
except 1, KKM$/\hat 5$=Kaluza-Klein monopole localized in direction 5,
HM= H-monopole.}
\end{table}

\subsection{Small Black Holes and DH States in the $Het(4,24)$ Model}

The tree-level  Bekenstein-Hawking entropy for generic
BPS black holes in models with $\CN=4$ supersymmetry is given by
\be
\label{sqrpp}
S_{BH} = \pi \sqrt{(P\cdot P)(Q\cdot Q) - (P\cdot Q)^2}
\ee
in the natural heterotic polarization, such that
$P,Q$ transform as a doublet of
$SO(6,n_V-2)$ vectors under $Sl(2)$ \cite{Cvetic:1995yq}.
We shall be interested in black holes
which are dual to perturbative heterotic states, with vanishing magnetic
charge $P=0$, hence zero tree-level entropy.
In particular, let us consider a type IIA state with $q^0$ D0-brane charge
and $p_1$ D4-brane charge. This is dual to a fundamental heterotic
string with momentum $n=q_0$ and winding $w=p^1$ along one circle
in $T^6$. As we reviewed in Section 2.1,
DH heterotic states with these charges can be obtained
by tensoring the ground state of the right-moving superconformal
theory with a level $N$ excitation of the 24 left-moving bosons,
provided the level matching condition $N-1= n w$ is satisfied.
The number of distinct DH states with fixed charges $(n,w)$ is thus
$\Omega(n,w)= p_{24}(N)$, where $p_{24}(N)$ is the number of partitions
on $N$ into the sum of 24 integers (up to an overall factor of 16
corresponding
to the size of short $\CN=4$ multiplets, which we will always drop).
Accordingly, the generating function of the degeneracies of DH states is
\begin{equation}\label{partition}
\sum_{N=0}^{\infty} p_{24}(N) q^{N-1}  =
\frac{1}{ \Delta(q)},
\end{equation}
where $\Delta(q)$ is Jacobi's discriminant function
\be
\Delta(q)=\eta^{24}(q)=q\prod_{n=1}^{\infty}(1-q^n)
\ee
It should be noted that the partition function
for the degeneracies of the $D0-D4$ system can be obtained without
resorting to the dual heterotic formulation, either by computing
the Euler number of the Hilbert scheme of $K3$, or by enumerating
genus $g$ curves in $K3$ \cite{Bershadsky:1995qy,Yau:1995mv}.
Nevertheless, the heterotic
description will prove very useful in more complicated examples.
Notice that the type IIA model on $K3\times T^2$ also has
DH states with zero tree-level entropy, but those are
in general 1/4-BPS. We shall return to them in \ref{dhii}.

\subsection{Asymptotic Degeneracies and the Rademacher Formula}

In order to determine the asymptotic density of states at large $N-1=nw$,
it is convenient to extract $d(N)$ from the partition function
\eqref{partition} by an inverse Laplace transform,
\begin{equation}\label{density}
    p_{24}(N) = \frac{1}{2\pi i} \int_{\eps-i\pi}^{\eps+i\pi}
    d\b\ e^{\b (N-1)}
    \frac{16}{\Delta(e^{-\b})}.
\end{equation}
where the contour $C$ runs from $\eps-i\pi$ to $\eps+i\pi$,
parallel to the imaginary axis. One may now take the high temperature
limit $\eps\rightarrow 0$, and use the modular property of the discriminant
function (see Appendix \ref{cornu})
\begin{equation}\label{modular}
    \Delta(e^{-\b}) = \left(\frac{\b}{2\pi}\right)^{-12}
    \Delta(e^{-4 \pi^2 /\b}).
\end{equation}
As $e^{-4 \pi^2 /\b} \rightarrow 0$, we can approximate
$\Delta(q) \sim q$ and write the integral as
\begin{equation}\label{asymptotic}
   p_{24}(N) = \frac{16}{2\pi i} \int_C d\b\ \left(\frac{\b}
   { 2\pi} \right)^{12} e^{\b (N-1) + 4 \frac{\pi^2}{\b}}
\end{equation}
This integral may be evaluated by steepest descent:
the saddle point occurs at $\beta = 2\pi / \sqrt{N-1}$, leading to
the characteristic exponential growth $ p_{24}(N)  \sim
\exp{( 4 \pi \sqrt{nw})}$ for the degeneracies.

To calculate the sub-leading terms systematically in an asymptotic
expansion at large $N$,
one may recognize that \eqref{asymptotic} is proportional
to the integral representation of a modified Bessel function,
\begin{equation}\label{Besselint}
    I_\nu(z) = \left(\frac{z}{2} \right)^\nu \frac{1}{2\pi i} \int_{\epsilon
    -i\infty}^{\epsilon +i\infty} \frac{dt}{t^{\nu +1}} e^{(t
    +z^2/4t)} := \frac1{2\pi} \left(\frac{z}{4\pi} \right)^{\nu}
\hat I_\nu(z)
\end{equation}
In order to reach \eqref{Besselint} from \eqref{asymptotic}, notice
however that one should extend the contour $C$ to the whole line
$\eps+i \IR$. While this would have lead to an infinite
multiplicative factor in \eqref{density} (a Dirac delta at integer
$N$ rather than a Kronecker delta),
this is no longer a problem in \eqref{asymptotic},
where periodicity under $\beta\to\beta+2\pi i$ has been broken.
We thus obtain
\begin{equation}\label{asymptotic2}
    p_{24}(N) \sim 2^4\ \BesselI{13}{4}{N-1} \ .
\end{equation}
Using the asymptotic expansion of $\hat I_\nu(z)$ at large $z$
(see e.g. \cite{Arfken})
\begin{equation}\label{besselasymp}
    \hat I_\nu(z) \sim \frac{ e^z}{\sqrt{2}}
\left(\frac{z}{4\pi}\right)^{-\nu-\frac12} \left[ 1- \frac{(\mu
    -1)}{8z} + \frac{(\mu
    -1)(\mu -3^2)}{2!(8z)^2} - \frac{(\mu
    -1)(\mu -3^2)(\mu -5^2)}{3!(8z)^3}+ \ldots \right],
\end{equation}
where $\mu = 4\nu^2$, we can thus compute the subleading corrections
to the microscopic entropy of DH states,
\begin{equation}\label{boltzentropy}
\log \Omega(n,w) \sim
 4\pi \sqrt{|nw|} -\frac{27}{4} \log|nw|
    +\frac{15}{2}\log 2 -\frac{675}{32\pi\sqrt{|nw|}} -\frac{675}{2^8 \pi^2
 |nw|}-\ldots
\end{equation}
This is however {\it not} the complete asymptotic expansion of $\Omega(n,w)$
at large charge: indeed, there are exponentially suppressed
corrections to \eqref{asymptotic2} which can be computed by
using the general Rademacher expansion formula for the Fourier
coefficients of modular forms with weight $w<0$
(see Appendix \ref{rademacher}). For the case at hand, we have
\be
\label{rade24}
\Omega(n,w) = 2^4 \sum_{c=1}^{\infty} c^{-14}\ \mbox{Kl}(nw+1,-1;c)\
\BesselI{13}{\frac{4}{c}}{|nw|}
\ee
where ${\rm Kl}(N,-1;c)$ are the Kloosterman sums defined in \eqref{kloos},
which are uniformly bounded by $|c|$. Although each term
is exponentially suppressed with respect to the previous one in the
sum, they all become large at large charge.
%Moreover, the
%terms with $c>1$ in this expansion
%cannot be associated in any obvious fashion to subleading saddle
%points in the integral \eqref{density}. Instead they arise upon
%representing the partition function as the Poincaré series of its
%polar part (see \cite{Dijkgraaf:2000fq}).
%

\subsection{Generalized Attractor Formalism for $\CN=4$ and Leading Entropy}

Now, we would like to compute the black hole degeneracies from the
macroscopic side.  Since the attractor formalism is tailored for
$\CN=2$ supergravity, one should first decompose the spectrum
under an $\CN=2$ subalgebra. The $\CN=4$ supergravity multiplet
consists of the $\CN=2$ supergravity multiplet with its
graviphoton gauge field, two $\CN=2$ gravitino multiplets with 2
Abelian gauge fields each, and one $\CN=2$ vector multiplet. In
addition, each $\CN=4$ vector multiplet decomposes into one vector
and one hypermultiplet of $\CN=4$. The gauge fields from the
$\CN=2$ gravitino multiplets have different couplings from the
rest of the $\CN=2$ vectors and we will restrict to black holes
which are neutral with respect to them. In terms of $\CN=2$
multiplets, the spectrum of type IIA/$K3\times T^2$ has therefore
$n_V=24$ Abelian gauge fields. In order to evaluate the
generalized prepotential $F(X^A,W^2)$ which governs the $\CN=2$
supersymmetric couplings of these $24$ gauge fields, recall the
following:
\begin{itemize}
\item[i)]
The tree-level topological amplitude $F_0$ is fixed by the triple intersection
product on $H^2(K^3\times T^2)$. We choose a basis of two-cycles with
$\gamma_1=H^2(T^2)$ and $\gamma_{a=2,23}$ a basis of $H^2(K3)$. The
triple intersection product vanishes except between $\gamma_1$ and two
2-cycles $\gamma_a, \gamma_b$
in $H^2(K3)$, where it equals the signature $(3,19)$ intersection
product $C_{ab}$
\item[ii)]
The topological
amplitude $F_1$ has been computed in \cite{Harvey:1996ir},
and can be obtained as the holomorphic part of the
$R^2$ amplitude at one-loop,
\be \label{fr2} f_{R^2} = 24 \log ( T_2 |\eta(T)|^4 )
\ee
where $T,U$ denote the K\"ahler and complex
structure moduli of the torus $T^2$. From the heterotic point of
view, this result can be interpreted as NS5-brane instanton
corrections to the tree-level heterotic $R^2$ amplitude \cite{Harvey:1996ir}.
\item[iii)]
All higher topological
amplitudes $F_h$ for $h>1$  vanish for models with $\CN=4$ supersymmetry.
Indeed, the type II dilaton is part of the second factor in \eqref{modspace},
and a non-vanishing $F_h$ amplitude would be inconsistent with
$SO(6,n_V-2)$ duality.
\end{itemize}
We therefore obtain the generalized prepotential
\be
F(X^I,W^2)=  - \frac12 \sum_{a,b=2}^{23} C_{ab} \frac{X^a X^b X^1}{X^0}
-\frac{W^2}{128\pi i} \log\Delta(q) \ee
where $T=X^1/X^0$ and $q=e^{2\pi i T}$. The appearance of the
same discriminant function $\Delta(q)$ as in the heterotic result
\eqref{partition} is at this stage coincidental\footnote{The two are however
related by the following chain of arguments: the $R^2$ coupling is related
by mirror symmetry to a $(\nabla^2 S)^2$ coupling, where $S$ is the type IIA
axio-dilaton \cite{Antoniadis:1993ze}.
The latter can be computed from a 1-loop amplitude on the heterotic
side, which produces both a 1-loop $\log(U_2|\eta(U)|^4)$
contribution in type IIA,
and a series of D-instanton contributions on $K3\times S^1$; the latter are
governed by the Fourier coefficients of $1/\Delta(q)$, in agreement with the
partition function of the $D0-D4$ system \cite{Antoniadis:1997zt}.}.

We may now apply the $\CN=2$ attractor formalism summarized in Section
\ref{osvintro} to the heterotic DH states $(n,w)$, or equivalently
to bound states of  $p^1=w$ D4-branes wrapping $K3$
with $q_0=n$ D0-branes. Since this does not cause any
additional complications, we shall allow arbitrary electric charges
$q_0,q_{i=2..23}$, as long as $q_1=0$ and
the only non-vanishing magnetic charge
is $p^1$. Under these assumptions,
the black hole free energy \eqref{fbhw} reduces to
\be
\label{free622}
{\cal F}(\phi^I,p^I) = - \frac{\pi}{2} C_{ab} \frac{\phi^a \phi^b
p^1}{\phi^0} -\log|\Delta(q)|^2
\ee
where
\be
\label{qq}
q=\exp\left[ \frac{2 \pi}{\phi^0} \left( p^1 + i \phi^1 \right) \right].
\ee
According to \eqref{sbhw}, the Bekenstein-Hawking-Wald entropy
is simply obtained by performing a Legendre transform over
all electric potentials  $\phi^I, I=0,\dots23$. The Legendre
transform over $\phi^{a=2..23}$ sets $\phi^a=(\phi^0/p^1) C^{ab} q_b$,
where $C^{ab}$ is the inverse of the matrix $C_{ab}$. We will
check a posteriori that in the large charge
limit, it is consistent to approximate $\Delta(q)\sim q$,
whereby all dependence on $\phi^1$ disappears.
We thus obtain
\be
\label{sext}
S_{BHW} \sim  {\rm Extr}_{\phi^0}
\left[ - \frac{\pi}{2}  \frac{C^{ab} q_a q_b }{p^1} \phi^0
+ 4\pi \frac{p^1}{\phi^0} + \pi \phi^0 q_0  \right]
\ee
The extremum of the bracket lies at
\be
\label{phi0st}
\phi^0_*= \frac12\sqrt{-p^1/\hat q_0}\ ,\qquad
\hat q_0 := q_0 + \frac{1}{2p^1} C^{ab} q_a q_b
\ee
so that at the horizon the
K\" ahler class $\Im T \sim \sqrt{ - p^1 \hat q_0}$ is very large,
justifying our assumption. Evaluating \eqref{sext} at the extremum,
we find
\be
\label{sbhl}
S_{BH} \sim 4\pi \sqrt{Q^2/2} \ ,\quad
Q^2 = 2p^1 q_0 + C^{ab} q_a q_b
\ee
in agreement with the leading exponential behavior in
\eqref{boltzentropy}, including the precise numerical factor.
Note that this result is independent of  the OSV conjecture,
and relies only on the classical attractor mechanism in the presence of
higher-derivative corrections.
This observation, first made in \cite{Dabholkar:2004yr}, indicates
that the tree-level $R^2$ coupling in the effective action of the
heterotic string on $T^6$ (or, equivalently,
large volume limit of the 1-loop $R^2$ coupling in type IIA/$K3\times
T^2$) is sufficient to cloak the singularity of the small black hole
behind a smooth horizon. This is in fact confirmed by a study of the
corrected geometry \cite{Dabholkar:2004dq,Hubeny:2004ji}.
Furthermore, the fact that the
correct numerical factor is reproduced from $R^2$ interactions alone
indicates that, in contrast to general expectations based on the
form of the tree-level metric \cite{Sen:2004dp}, further higher-derivative
interactions do not correct the Bekenstein-Hawking-Wald entropy
(although they may still correct the actual solution). It would be
interesting to understand the origin of this non-renormalization.

\subsection{Testing the OSV Formula}

We are now ready to test the proposal \eqref{osvii} and evaluate
the inverse Laplace transform of $\exp({\cal F})$ with respect
to the electric potentials,
\be
\label{omosv1}
\Omega_{OSV}(p^\L,q^\L)
= \int d\phi^0~d\phi^1~d^{22}\phi^a
\frac{1}{\vert\Delta(q)\vert^2}
\exp\left[ - \frac{\pi}{2} C_{ab} \frac{\phi^a \phi^b p^1}{\phi^0}
+ \pi \phi^0 q_0+ \pi \phi^a q_a \right]
\ee
Due to the non-definite signature of $C_{ab}$,
the integral over $\phi^a$ diverges for real values.
This may be avoided by rotating the integration contour to $\eps+i \IR$ for all
$\phi$'s. The integral over $\phi^a$ is now a Gaussian, leading to
\be
\label{osvde}
\Omega_{OSV}(p^\L,q^\L)
= \int d\phi^0~d\phi^1 \left( \frac{\phi_0}{p^1} \right)^{11}
\frac1{\vert \Delta(q) \vert^2}
\exp\left( - \frac12  \frac{C^{ab} q_a q_b}{p^1} \phi^0 +  q_0 \phi^0 \right)
\ee
where we dropped numerical factors and used the fact that $\det C=1$.
Unfortunately, for imaginary $\phi^0,\phi^1$, $q$ is a pure phase,
and $\Delta(q)$ is ill-defined. The asymptotics of $\Omega$ is
independent of the details of the contour, as long as it selects
the correct classical saddle point \eqref{phi0st} at large charge.
Approximating again $\Delta(q)\sim q$, we find the quantum version
of \eqref{sext},
\be
\label{osvde2}
\Omega_{OSV}(p^\L,q^\L)
= \int d\phi^0~d\phi^1 \left( \frac{\phi_0}{p^1} \right)^{11}
\exp\left( - \frac12  \frac{C^{ab} q_a q_b}{p^1} \phi^0
-4\pi \frac{p^1}{\phi^0} +  q_0 \phi^0 \right)
\ee
The integral over $\phi^1$ superficially leads to an infinite result.
However, since the free energy is invariant under $\phi^1\to\phi^1+\phi^0$,
it is natural to restrict the integration to a single period
$[0,\phi^0]$, leading to an extra factor of $\phi^0$ in \eqref{osvde2}.
The integral over $\phi^0$ is now of Bessel type, leading to
\be
\label{osvfin}
\Omega_{OSV}(p^\L,q^\L) = (p^1)^2 \BesselI{13}{4}{Q^2/2}
\ee
in impressive agreement with the microscopic result \eqref{asymptotic2}
at all orders in $1/Q$.

While this result is encouraging, it however indicates that
\eqref{osvii} should interpreted with some care:

\begin{itemize}
\item The extra factor of $(p^1)^2$ in Equation \eqref{osvfin}
is inconsistent with $O(\Gamma_{6,22})$ duality, which requires
the exact degeneracies to be a function of $Q^2$ only. This indicates
that the integration measure implicit in \eqref{osvii} is not
the trivial Euclidean measure. Given the wave function
interpretation of $e^{F_{top}}$ \cite{Witten:1993ed}, one
attractive possibility would be to normalize it -- alas,
it appears to be severely non-normalizable. For lack of
a proper understanding of this integration measure, we are thus
forced to consider ratios $\Omega(p,q)/\Omega(p,q')$ only\footnote{The 
analysis of the $p^0\neq 0$ case in Section \ref{labhd6} indicates
that a proper duality-covariant measure will have to break holomorphicity.}.

\item In order to obtain the modified Bessel function
with the correct index, note that it was crucial to discard the
non-holomorphic correction proportional to $\log T_2$ in $F_1$
(keeping this correction would have resulted in an index $19$
rather than $13$, spoiling the agreement with the microscopic
result \eqref{asymptotic2}). In addition, it was important to
compare to the degeneracies of DH states with arbitrary
angular momentum $J$ (degeneracies of DH states
with $J=0$ are computed in Appendix \ref{appj}, and lead to
a Bessel function with index $29/2$ and a different intercept).

\item The ``all order'' result \eqref{osvfin} depends only on
the number of $\CN=2$ vector multiplets, as well as on the
leading large volume behavior of $F_1 \sim q /(128 \pi i)$.
By heterotic/type II duality, this term is mapped to a
tree-level $R^2$ interaction on the heterotic
side, which is in fact universal. We thus conclude that in all
$\CN=2$ models which admit a dual heterotic description, the
degeneracies of small black holes predicted by \eqref{osvii}
are given by
\be
\label{unideg}
\Omega_{OSV}(p^\L,q^\L) \propto \BesselI{\frac{n_V+2}{2}}{4}{Q^2/2}\ ,
\ee
provided it is justified to neglect higher genus $F_{h>1}$ and genus 0,1
Gromov-Witten instantons. We shall return to this point in
Section \ref{smbh2}.

\item In order to try and match \eqref{osvfin} and \eqref{asymptotic2} in
more detail, one may change variable $\beta=\pi/t$
in \eqref{density} and rewrite the exact microscopic
result as
\be
\Omega(n,w) = \int dt\ t^{-14} \ \frac{\exp\left( \frac{\pi n w}{t}
  \right)}{\Delta\left( e^{-4\pi t} \right)}
\ee On the other hand, it is convenient to change variables in the
OSV integral \eqref{osvde} to $\tau_1=\phi^1/\phi^0,
\tau_2=-p^1/\phi^0$, with Jacobian $d\phi^0 d\phi^1 = 8 (p^1)^2
d\tau_1 d\tau_2/\tau_2^3$, leading to
\be
\label{osvtry}
\Omega_{OSV}(p^\L,q^\L)
\sim \int d\tau_1 ~ d\tau_2 ~ \tau_2^{-14}
\frac{\exp\left(\frac{\pi(N-1)}{\tau_2} \right)} {\vert
\Delta\left( e^{-2\pi \tau_2 + 2\pi i \tau_1} \right) \vert^2} \ee
Despite obvious similarities, it appears unlikely that the two
results are equal non-perturbatively. Indeed, with any natural
interpretation of the integration contours consistent with the
quantum mechanics interpretation, the integral \eqref{osvtry}
diverges.

\item Just as the perturbative result \eqref{asymptotic2}, the
result \eqref{osvfin} misses subleading terms in the Rademacher
expansion \eqref{rade24}. It does not seem possible to interpret
any of the terms with $c>1$ as the contribution of a subleading
saddle point in either \eqref{asymptotic} or \eqref{osvde}. It would
be interesting to see if non-holomorphic Poincaré series can be used
to extract these contributions from \eqref{osvde}.
\end{itemize}

Despite these difficulties, we find it remarkable that the
black hole partition function in the OSV ensemble, obtained from
purely macroscopic considerations, reproduces the entire
asymptotic series exactly to all orders in inverse charge.

\subsection{Degeneracies vs. Helicity Supertrace}

If it is to satisfy the second law of thermodynamics,
the Bekenstein-Hawking-Wald entropy should be equal to the logarithm
of the total number of micro-states in the regime
where the black hole is formed. On the other hand, the degeneracies
of DH states have been computed at zero heterotic string coupling.
In general however, BPS states can
appear and disappear rather chaotically on various loci of the moduli
space, by (un)pairing up into longer multiplets.
If the absolute degeneracies at zero coupling
can be identified with a suitable index, it is then possible
to ensure that the total number of micro-states does not change
as the coupling is increased (barring the possible crossing of lines
of marginal stability).
The only such indices with a well-defined target space interpretation
are the helicity supertraces\footnote{ see \cite{Kiritsis:1997gu},
Appendices E and G for an extensive review of helicity supertraces.}
\be
\Omega_n = \Tr (-1)^F J_3^n
\ee
where $F$ is the target-space fermion number and $J_3$ is
a Cartan generator in the massive little group in 3+1 dimensions
(or, for massless states, the ordinary helicity), and $n$
is an even number ($\Omega_{2n+1}$ always vanishes by reason of
symmetry) . For a given
number $\CN$ of supersymmetry in 4 dimensions, $\Omega_{n<\CN}$
vanishes automatically in any multiplet, while $\Omega_{n\geq 2\CN}$
receives contributions from generic long multiplets. In the window
$\CN\leq n \leq 2\CN$, the helicity supertraces $\Omega_n$ receive
only contributions from short or intermediate multiplets, and are
therefore unaffected by recombination processes.

For the $\CN=4$ case of interest in this section, the first non-vanishing
supertrace is $\Omega_4$, which receives contributions only from the
supergravity multiplet, massless vector multiplet and short massive
multiplets\footnote{The superscript
$j$ indicates the spin $J_3$ of the middle state in the
short massive supermultiplet $S^j$},
\be
\Omega_4(sugra)=3, \quad \Omega_4(vect)=\frac32\ ,\quad \Omega_4(S^j)=\frac32(2j+1)(-1)^{2j}
\ee
while the intermediate and long $\CN=4$ multiplets cancel out. In particular,
$\Omega_4$ is unaffected by possible recombinations of four short multiplets
into a longer intermediate multiplet.
Similarly, the helicity supertrace $\Omega_6$ receives contributions from
short and intermediate multiplets only,
\be
\Omega_6(sugra)=\frac{13\cdot 15}{4},
\quad \Omega_6(vect)=\frac{15}{8}\ ,
\quad \Omega_6(S^j)=\frac{15}{8}(2j+1)^3(-1)^{2j}\ ,
\ee
\be
\quad \Omega_6(I^j)=\frac{45}{4}(2j+1)(-1)^{2j+1}
\ee
and is invariant under recombinations of four intermediate multiplets
into a longer one.

In order to compare with the absolute degeneracies \eqref{partition},
let us compute the helicity supertrace of the DH states
in the Het(4,24) model. Helicity supertraces are most easily computed
by introducing generating
parameters $v$ and $\bar v$ for the left and right moving components
of the space-time helicity $J_3$ \cite{Kiritsis:1997gu}
\be
Z(v,\bar v)=
\Tr (-1)^F e^{2\pi i v J_3^R}  e^{2\pi i \bar v J_3^L} q^{L_0} q^{\bar L_0}
\ee
and computing
\be
B_{n}(q,\bar q)
= \sum \Omega_n q^{L_0} \bar q^{\bar L_0}
= \left( \frac{\partial}{2\pi i \pa v}+ \frac{\partial}{2\pi i \pa \bar v}
\right)^n \vert_{v=\bar v=0} Z(v,\bar v)
\ee
The generating function for helicity supertraces
of the $E_8\times E_8$ heterotic string on $T^6$
is simply given by
\be
\label{z22h}
Z_{(4,24)}^H(v,\bar v) =
\frac{\xi(v)\bar\xi(\bar v)}{\tau_2 |\eta|^4}
\frac12 \sum_{\alpha,\beta} (-1)^{\alpha+\beta+\alpha\beta}
\frac{\bar\theta\ar{\alpha/2}{\beta/2}(\bar v)\
\bar\theta^3\ar{\alpha/2}{\beta/2}}{\bar\eta^4}
\frac{Z_{6,6}}{|\eta|^{12}} \left( \theta_{E_8[1]} \right)^2
\ee
where $\alpha,\beta=0,1$ label the four spin structures on the
superconformal side,
$\xi(v)$ incorporates the $U(1)$ charge of the bosons in the
two transverse directions,
\be
\xi(v)=\prod_{n=1}^{\infty} \frac{(1-q^n)^2}
{(1-q^n e^{2\pi i v})(1-q^n e^{-2\pi i v})}
= \frac{2\eta^3 \ \sin \pi v}{\theta_1(v)}
\ee
$\theta_{E_8[1]}$ is the numerator of the
character of the $E_8$ current algebra at level 1,
\be
\theta_{E_8[1]} = \frac12 \left( \theta_3^8 + \theta_4^8 + \theta_2^8 \right)
\ee
and $Z_{6,6}$ is the partition function of bosonic zero-modes on $T^6$.
By the Riemann identity, \eqref{z22h} can be converted into
\be
\label{z22hr}
Z_{(4,24)}^H(v,\bar v) =
\frac{\xi(v)\bar\xi(\bar v)}{\tau_2 |\eta|^4}\
\frac{\bar \theta^4\ar{1/2}{1/2}(\bar v/2)}{\bar \eta^4}\
\frac{Z_{6,6}}{|\eta|^{12}} \left(  \theta_{E_8[1]} \right)^2
\ee
which is recognized as a trace in the Ramond sector only, with an
insertion of $(-1)^{J_R}$. Since the Jacobi theta function
$\theta_1(z;\tau)$ has a single zero at $z=0$, a non-vanishing
supertrace is obtained only for $n\ge 4$.
Taking four $\bar v$-derivatives and
using $\theta\ar{1/2}{1/2}'(0)=\theta_1'(0)=2\pi \eta^3, \xi(0)=1$,
the first non-vanishing supertrace is easily computed:
\be
B_4 = \frac{1}{\tau_2}
Z_{6,6} \  \left(  \theta_{E_8[1]} \right)^2
\times  \frac32 \frac{1}{\eta^{24}}
\ee
where the factor $1/\tau_2$ corresponds to the contribution of the zero-mode
$p_2,p_3$ in the transverse directions. At a generic
point, the two factors in the numerator
combine into a lattice sum $Z_{(6,22)}$, leading to
\be
\label{b4622}
B_4 =
\frac{1}{\tau_2}
Z_{6,22} \times \frac32 \frac{1}{\eta^{24}}
\ee
The first factor simply corresponds to the continuous degeneracy due
to the momentum in 4 dimensions, while the second factor is just the
partition function of the lattice $\Gamma_{6,22}$ of electric
charges. For any vector $Q \in \Gamma_{6,22}$ , we
conclude that the helicity
supertrace of states with electric charges $Q$ is given by
\be
\Omega_4( Q )= \frac32 p_{24}(N) = \frac{3}{32} \Omega_{abs}(Q)
\ee
where $\Omega_{abs}$ is the absolute degeneracy computed in
\eqref{partition} up to an overall
numerical factor. This suggests that, in the case
of $\CN=4$ backgrounds, the OSV integral \eqref{osvii} may compute
the fourth helicity supertrace of the black hole micro-states.

An immediate problem with this proposal is that it
implies that the OSV prescription should vanish in the case of
``large'' black holes, which form intermediate (1/4-BPS) multiplets of
$\CN=4$ supersymmetry. These states cancel from $\Omega_4$ and contribute
to sixth helicity supertrace $\Omega_6$ onward.
In the case of the $Het(4,24)$ model, $\Omega_6$ may
be obtained straightforwardly by taking either 6 $\bar v$-derivatives, or
4 $\bar v$-derivatives and 2 $v$-derivatives, leading
to \cite{Kiritsis:1997gu}
\be
\label{b6622}
B_6 = \frac{1}{\tau_2}
Z_{6,22} \times \frac{15}{8} \frac{2-E_2}{\eta^{24}}
\ee
Since the perturbative heterotic spectrum contains no intermediate multiplets,
this result arises from the contributions of the
same DH states which contributed
to \eqref{b4622}. While the Rademacher formula does not apply to the
non-modular invariant Eisenstein series $E_2$, one may simply use the
identity
\be
\frac{E_2}{\eta^{24}} = - q \frac{d}{dq} \frac{1}{\eta^{24}}
\ee
to obtain the asymptotic behavior of the Fourier coefficients of
$B_6$ to all orders in $1/N$,
\be
\label{rd6}
\Omega_6 (N) \sim  \frac{15}{8}
(N+1) \BesselI{13}{4}{N-1}
\ee
where
\be
B_6 = \frac{1}{\tau_2} Z_{6,22} \sum_{N=0}^{\infty} \Omega_6(N) q^{N-1}
\ee
In particular, the extra factor of $N+1$ in \eqref{rd6} makes it
impossible to include a contribution from $\Omega_6$ to the index
relevant for the OSV proposal \eqref{osvii} for
half-BPS states, since one would have to modify the integration measure
by a $q$ dependent factor. On the other hand, $\Omega_4$ is clearly
inadequate for $1/4$ BPS states. We conclude that the index computed
by \eqref{osvii} must depend on the number of supersymmetries preserved
by the BPS states under consideration.

Before closing this section, let us briefly comment on the case
with $\CN=2$ supersymmetry originally envisaged in \cite{Ooguri:2004zv}.
In this case, the only index protected by supersymmetry is
the second helicity supertrace $\Omega_2$, to which only 1/2 BPS states
contribute:
\be
\Omega_2(sugra)=\Omega_2(vect)=1\ , \quad \Omega_2(hyper)=-1\ ,
\quad \Omega_2(S^j)=(2j+1)(-1)^{2j+1}
\ee
This is the space-time interpretation of the ``vectors minus hypers'' index
introduced from a world-sheet point of view in \cite{Harvey:1995fq}, since
short multiplets with integer (resp. half-integer) spin $j$ are the massive
generalization of the massless hypermultiplet (resp. vector multiplet).
In particular, $\Omega_2$ is invariant under the recombination of a hyper and
a vector multiplet into a long multiplet of $\CN=2$. Note however that
$\Omega_2$ may change at lines of marginal stability in moduli space.
Since we do not have the freedom to add higher helicity supertraces,
we conjecture that the OSV prescription computes the second helicity
supertrace of the $\CN=2$ black hole micro-states. Evidence for this claim
will be given in Section 5.

\subsection{DH states in type II/$K3\times T^2$ \label{dhii}}

In addition to the heterotic DH states, the $(4,24)$ model also
admits DH states on the type IIA side, corresponding to
fundamental type II strings with momentum $n_i$ and winding $w^i$
along $T^2$ ($i=5,6$). These can have either left-moving or
right-moving excitations, depending on the sign of $n_i w^i$.
Since there are now 8 bosonic and 8 fermionic oscillators, with
total central charge $c=12$, the degeneracies grow as
\be S_{DH}^{IIA} \sim 2\pi \sqrt{2|n_i w^i|}
\ee
In contrast to the heterotic DH states, these states preserve only
1/4 of the supersymmetries, unless $n_i w^i=0$. According to
\eqref{ppqqpq}, they have $P^2=Q^2=P\cdot Q=0$, hence zero
tree-level entropy. Their helicity supertraces have been computed
in \cite{Kiritsis:1997gu} (eqs. $(G.24)$ and $(G.25)$), and vanish
identically except for $n_i w^i=0$: \bea\label{typestii}
\Omega_4(Q) & = & 36\ \delta_{n_i w^i,0} \\
\Omega_6(Q) & = & 90\ \delta_{n_i w^i,0}
\eea
This indicates that these intermediate multiplets come in pairs and
may combine into longer multiplets and leave the spectrum.

Since the type II DH states are
charged under the four $\CN=2$ gravitino multiplets,
the $\CN=2$ attractor formalism does not apply directly. Nevertheless,
by a $O(6,22)$ duality, they may be mapped to a D0-D2/$T^2$ state with
charges $(q_0,q_1)$.

More generally, we may try and apply the OSV formula \eqref{osvii} to
purely electrically charged states in the type II polarization, with arbitrary
electric charges $(q_0,q_1,q_a)$. The perturbative part of the free-energy
\eqref{fbhw} vanishes, leaving only the Gromov-Witten instanton series,
evaluated at real $X^A=\phi^A/\phi^0$, where it is no longer convergent.
The integral \eqref{osvii} is therefore highly singular. Nevertheless,
discarding the Gromov-Witten contribution, \eqref{osvii} produces
a delta function of the electric charges, in qualitative agreement with
the helicity supertraces above.

It should be noticed that similar DH states occur in Type IIA/$T^6$,
with $\CN=8$ supersymmetry. The first non-trivial helicity
supertraces occur at order $\Omega_{12}, \Omega_{14}$, but they
are given by modular forms with positive weight, so that the indexed
degeneracies of intermediate multiplets grow as a power-law rather
than exponentially.

\section{Small Black Holes in $\CN=4$ Models with Reduced Rank}\label{N=4}
In this section, we proceed to compare the macroscopic and microscopic
entropy of small black holes in a variety of string vacua with $\CN=4$
supersymmetry. While the (4,24) model discussed in the previous section
has been the most studied one in the literature, a large
number of $\CN=4$ vacua can be obtained  using  fermionic
\cite{Ferrara:1989nm,Chaudhuri:1995fk} or orbifold constructions
\cite{Chaudhuri:1995bf,Gregori:1997hi,Dabholkar:1998kv}.
The latter has the advantage that a dual description can often
be found by using six-dimensional heterotic/type II duality
and adiabatic arguments \cite{Schwarz:1995bj,Chaudhuri:1995dj,Gregori:1997hi}.
Each of these models has a moduli space of the form \eqref{modspace},
where the first factor corresponds to the heterotic dilaton and
$n_V$ denotes the number of massless Abelian gauge fields (including
the graviphoton, but discarding the gauge fields from the two
$\CN=2$ gravitino multiplets).
We will denote such vacua as $Het(4,n_V)$ or $II(4,n_V)$, assuming
that all models with the same number of vector multiplets belong
to the same moduli space. As in the (4,24) case, the only
non-vanishing F-term $F_1$ can be extracted from the one-loop amplitude
$R^2$ amplitude in the type II model, while the exact degeneracies
of small black holes are most easily determined in the heterotic dual.

\subsection{$F_1$ in Reduced Rank Type II Models}
\label{f1ii}

The topological amplitude $F_1$ has been computed in a number of
$(4,n_V)$ type II models in \cite{Gregori:1997hi}. In general, it
is given by the holomorphic, $T$-dependent part of the integral of
the ``new'' supersymmetric index on the fundamental domain of the
upper half plane \cite{ Harvey:1996ir},
\be
\label{fr24} f_{R^2} =
\int_{\cal F} \frac{d^2\tau}{\tau_2} ~\Tr_{RR} (-1)^{J_L+J_R} J_L
J_R q^{L_0+\bar L_0} = -\frac23
\int_{\cal F} \frac{d^2\tau}{\tau_2} \
B_4
\ee
As indicated in the second equality, the
supersymmetric index is proportional to the generating function
$B_4$ of the helicity supertraces $\Omega_4$
of the perturbative type II spectrum
\cite{Gregori:1997hi}. For completeness, we briefly review the CFT
construction of these models\footnote{While the inclusion of
discrete RR fluxes on $K3$ is required non-perturbatively for
level matching \cite{Schwarz:1995bj}, this does not affect the
perturbative computation of $F_1$ in these models. Such fluxes
do however affect the BPS spectrum \cite{distlermoore}. } and list the
corresponding supertrace and $R^2$ amplitudes:
\begin{itemize}
\item The   $(4,16)$ model is obtained by starting from
Type IIA on $K3\times T^2$ at the $T^4/\Zint_2$ orbifold point of $K3$,
and performing a further $Z_2$
orbifold which acts as $(-1)$ on half of the twisted sectors, and
shifts one of the coordinates of $T^2$ by a half-period. The generating
function of the 4-th helicity supertraces is
\be
B_4 = 18\ Z_{2,2} + 6 \sum_{(h,g)\neq (0,0)}
Z_{2,2}^{\delta_1}\ar{h/2}{g/2}
\ee
where
\be
Z^\delta_{2,2}\ar{h/2}{g/2}(T,U;\tau,\bar\tau) =
\sum_{p \in \Gamma_{2,2}+\frac{h}{2} \delta} e^{-i\pi g (p,\delta)}
q^{\frac12 \Pi_L^2(p)} \bar q^{\frac12 \Pi_R^2(p)}
\ee
is the shifted lattice sum for the Narain lattice of the torus $T^2$.
We choose a symmetric shift vector $\delta_1=(1,1)/2$
along the first circle,
so as to entertain a geometric description.

\item A $(4,12)$ model may be obtained by performing a
further $\Zint_2$ orbifold of the $(4,16)$ model,
which acts as $(-1)$ on a different half
of the 16 twisted states, together with a shift by half a period on
the remaining circle in $T^2$. The helicity supertrace generating
function is
\be
B_4 = 9\ Z_{2,2} + 3 \sum_{(h,g)\neq (0,0)}
\left( Z_{2,2}^{\delta_1}\ar{h/2}{g/2}
+Z_{2,2}^{\delta_2}\ar{h/2}{g/2}
+Z_{2,2}^{\delta_1+\delta_2}\ar{h/2}{g/2}
\right)
\ee

%\item A $(4,10)$ model may be obtained by performing
%a further $\IZ_2$ orbifold  acting  as $(-1)$ on yet a different half
%of the 16 twisted states, together with yet another shift $\delta_3$ on
%$T^2$. Similarly, a further $\IZ_2$ orbifold will yield a $(4,9)$ model.
%Unfortunately, $F_1$ has been computed for neither of these models.%

\item A $(4,8)$ model can be obtained % in one go,
by returning to the $II(4,24)$ model at the $T^4/\IZ_2$ orbifold
point, and by orbifolding by a further $\IZ_2$
which acts as $(-1)$ on all twisted sectors. The result is
\be
B_4 = 18\ Z_{2,2} - 6 \sum_{(h,g)\neq (0,0)}
Z_{2,2}^\delta\ar{h/2}{g/2}
\ee
\end{itemize}

In each of these cases, the modular integral \eqref{fr24} can be reduced to
the $(4,24)$ case by  making use of the following identities,
\bea
\label{allsum2}
\frac{1}{2} \left(
Z_{2,2}\ar{00}{00}+Z_{2,2}\ar{00}{\half 0}+Z_{2,2}\ar{\half 0}{00}+Z_{2,2}\ar{\half 0}{\half 0} \right)
&=&Z_{2,2}(T/2,2U) \\
\label{allsum3}
\frac{1}{2} \left(
Z_{2,2}\ar{00}{00}+Z_{2,2}\ar{00}{0\half }+Z_{2,2}\ar{0\half }{00}+Z_{2,2}\ar{0\half }{0\half } \right)
&=&Z_{2,2}(T/2,U/2) \\
\label{allsum4}
\frac{1}{2} \left(
Z_{2,2}\ar{00}{00}+Z_{2,2}\ar{00}{\half \half }+Z_{2,2}\ar{\half \half }{00}+Z_{2,2}\ar{\half \half }{\half \half } \right)
&=&Z_{2,2} (T/2,(U+1)/2) \ ,
\eea
where, on the left hand side, all partition functions are evaluated at $(T,U)$.
We thus obtain \cite{Gregori:1997hi}
\be
\label{fr2all}
\begin{array}{ccclllcl}
(4,24)&:& f_{R^2} &=&24 \log T_2 |\eta(T)^4|
&\sim& 24 \log T_2 - 8 \pi T_2+\dots\\
(4,16)&:& f_{R^2}&=&16  \log T_2 | \eta^3(T) \theta_4(T)|
&\sim& 16\log T_2 -4 \pi T_2 + \dots\\
(4,12)&:& f_{R^2}&=&12 \log T_2 | \eta^2(T) \theta_4^2(T)|
&\sim& 12\log
T_2-2 \pi T_2 + \dots\\
(4,8) &:& f_{R^2}&=&8 \log T_2 | \eta(T)^6 / \theta_4(T)^2 |
&\sim& 8 \log T_2 -4 \pi T_2 + \dots
\end{array}
\ee
where $T$ is the K\"ahler modulus of the $T^2$ covering of the
base of the $K3$ fibration. We have also indicated the large
volume expansion. The leading linear
term is proportional to the
size $A$ of the base of the $K3$ fibration, which differs
from $T$ by a power of two.  The logarithmic divergence is proportional to the
helicity supertrace $\Omega_4 = 3 + (3/2) (n_V-2)$ of the massless
spectrum.  The dots correspond to a finite term, dependent on the
details of the IR cut-off, and a sum
of worldsheet instantons. In general, we therefore
have
\be
f_{R^2} = n_V \log A - 8 \pi A_2 + \dots
\ee
with $A=(T,T/2,T/4,T/2)$ for the four models above.
In the heterotic dual, $A$ becomes
the heterotic dilaton $S=\theta+ i V_6/g_s^2$. where $V_6$
is the volume of the 6-torus. The term linear in $T_2$ is therefore
a tree-level term, coming from the compactification of the $R^2$
interaction in the 10-dimensional heterotic string. The type II worldsheet
instantons are interpreted on the heterotic side
as Euclidean NS5-branes wrapping $T^6$.

\subsection{Heterotic Duals and Exact Counting of DH States}
Heterotic $\CN=4$ models with reduced rank can be obtained by
orbifolding the $E_8\times E_8$ or $SO(32)$ ten-dimensional
heterotic strings at an enhanced symmetry point, by a
symmetry leaving the right-moving superconformal algebra
untouched. In particular, we consider the following models:
\begin{itemize}
\item A $(4,16)$ model obtained by
orbifolding the $E_8\times E_8$ heterotic string on $T^6$ by the exchange
of the two $E_8$, combined with a translation on one of the directions
of the torus $T^6$. Equivalently, one may orbifold the $SO(32)$
heterotic string at an $SO(16)\times SO(16)$ point by the exchange of the
two $SO(16)$ factors.
\item A $(4,12)$ model obtained by
orbifolding the $SO(32)$ heterotic string at a $SO(8)^4$ point
by the group $\IZ_4$ permuting the four $SO(8)$
factors circularly\footnote{It is also possible
to orbifold by the full permutation group $S_4$,
or the alternate subgroup $A_4$, but the required action on $T^2$
is more complicated.} combined with
a translation of order 4 on the torus.
\item A $(4,10)$ model obtained by
orbifolding the $SO(32)$ heterotic string at a $SO(4)^8$ point
by the group $\IZ_8$ permuting the
eight $SO(4)$ factors circularly. Viewing $SO(4)^8$ as $SU(2)^{16}$, one
may also orbifold by $\IZ_{16}$ and get a $(4,9)$ model.
\end{itemize}
In each of these models, it is important to include a translation
on one of the directions of the torus $T^6$ so as to give a mass
to the twisted sectors, and ensure that the rank of the gauge group
is effectively reduced.

The common property of these models is that they give rise to an enhanced
gauge symmetry with a current algebra at level $k>1$.
However, in order to have a type II dual
with a smooth geometry, one should further break the gauge symmetry
to an Abelian group $U(1)^{n_V+4}$.

\subsection{A Detailed Analysis of the $Het(4, 16)$ Model}
Let  us now discuss in detail the degeneracies of DH states in the
$Het(4,16)$ model obtained by orbifolding the $Het(4,24)$ model at
a point of enhanced gauge symmetry $E_8\times E_8$.
The Narain lattice of the $Het(4,24)$ model may be decomposed as
\be
\Gamma_{6,22} = E_8(-1) \oplus E_8(-1) \oplus II^{1,1} \oplus II^{5,5}
\ee
where $II^{1,1} \oplus II^{5,5}$ describe the momenta and winding
numbers on the 6-torus $S^1\times T^5$. Accordingly, we shall denote the
momentum eigenstates as $P=(P_1,P_2,P_3,P_4)$. At any point in the
moduli space \eqref{modspace}, this vector may be
projected into a sum of a left-moving and a right-moving part,
\be
P= \Pi_L(P) + \Pi_R(P)
\ee
where $\Pi_R(P) \in \IR^6$ are the 6 central charges of $\CN=4$ supersymmetry,
and $\Pi_L(P)  \in \IR^{22}$ are the 22 electric charges under the vector
multiplets of the $Het(4,24)$ model. While the charge vector $P$
takes quantized values independent of the moduli, the projections
$\Pi_L(P)$ and $\Pi_R(P)$ are real numbers depending continuously on
the moduli.

\subsubsection*{Untwisted sector}

Now, the $Het(4,16)$ model can be obtained as a $\IZ_2$ orbifold acting
on momentum eigenstates as
\be
g \vert P_1,P_2,P_3,P_4\rangle
= e^{2\pi i \delta \cdot P_3} \vert P_2,P_1,P_3,P_4 \rangle
\ee
where $2\delta$ is the vector $(1,1)\in II^{1,1}$ corresponding to
the translation by half a period along the circle. The action on
the oscillators is most easily described by diagonalizing the action
of $g$: $8$
left-moving
oscillators obtain a negative parity under $g$, while the
remaining left-moving and all right-moving oscillators have positive parity.
Let $\CP^\pm(\alpha)$ denote a generic monomial
in left-moving creation oscillators, with definite parity $\pm$ under $g$.

DH states in the untwisted sector of the $Het(4,16)$ model
can be constructed as invariant combinations of the DH states of the
$Het(4,24)$ model under the orbifold action,
\be
\label{cands}
\CP^\pm(\alpha) \biggl( \vert P_1,P_2,P_3,P_4\rangle
\pm  e^{2\pi i \delta \cdot P_3} \vert P_2,P_1,P_3,P_4\rangle\biggr) \otimes
\vert \tilde s \rangle
\ee
where the parity of the oscillators is correlated with that of the
zero-modes,  and $\vert \tilde s \rangle$ is a right-moving
groundstate\footnote{In the following, we omit the factor of
$2^4$ due to the degeneracy of the  right-moving
groundstate.}.
The level matching conditions identifies the level $N$
of the oscillator state $\CP(\alpha)$ with
\be
\label{lm2}
 N  -1 = \half q_R^2 - \half q_L^2 = \half P^2
\ee
The DH states \eqref{cands} are thus enumerated by the partition functions
\be
\label{candsc}
\half \biggl( {1\over \eta^{24} } \pm   {2^4\over \eta^{12} \vartheta_2^4}
\biggr)\sum_{II^{22,6}}  q^{\half q_L^2} \bar q^{\half q_R^2}
\frac{1 \pm \Theta(P)}{2}
\ee
where $\Theta(P) = e^{2\pi i \delta P_3} \delta_{P_1,P_2}$. The last
factor in \eqref{candsc} guarantees that states with $P_1\neq P_2$
are counted twice with $1/2$ multiplicity, while states with $P_1=P_2$
and $e^{2\pi i \delta P_3} = \mp 1$ are dropped out, in agreement with
Equation \eqref{cands}.

This is not the final answer however, since we need to extract from
\eqref{candsc} the contribution of states with a given electric charge.
Due to the orbifold projection, the only massless vector multiplets
are the linear combinations of the $E_8\times E_8$ gauge bosons of the
$(4,24)$ model which are symmetric under exchange of the two factors.
Therefore, a state of the form $\CP(\alpha) \vert P_1, P_2, P_3, P_4\rangle$
has electric charge
\be
Q(P) := (P_1 + P_2; P_3, P_4)
\ee
taking values in the (non self-dual) lattice\footnote{In this
expression, $(-1/2)$ indicates that the norm of the
$E_8$ is multiplied by $1/2$, in order to keep the canonical
normalization for the gauge fields.}
\be
M_0 =  E_8(-\half) \oplus II^{1,1} \oplus II^{5,5}
\ee
In particular, the momentum eigenstates
$(P_1-Q, P_2 + Q, P_3, P_4)$
have the same electric charge $Q(P)$ as
$(P_1 , P_2  , P_3, P_4)$, for {\it any} $Q$ in the $E_8$
root lattice.
It can be checked that all these states have the same central
charges $\Pi_R$ on the subspace $SO(6,14)/SO(6)\times SO(14)$
of the moduli space of the $Het(4,24)$ model \eqref{modspace}
invariant under the orbifold projections. They therefore have the
same mass and electric charge, but differ by the excitation
level $N$ of the oscillators.

In order to extract the exact degeneracy of DH states for a
given electric charge $Q$, it is appropriate to change the
basis and decompose the two $E_8(-1)$ charge vectors into their
sum and difference,
\bea
P_1 + P_2 &=& 2 \Sigma + \wp \\
P_1 - P_2 &=& 2 \Delta - \wp
\eea
where $\Sigma,\Delta$ both take
values in the $E_8$ root lattice, and ${\cal P}$ is an element of
the finite group $Z=\Lambda_r(E_8)/2 \Lambda_r(E_8)$, of index
$2^8$. Expressing the square of the left-moving momentum as \be
\Pi_L^2(\Sigma+\Delta, \Sigma-\Delta+ \wp, P_3, P_4) =
\Pi_L^2\left(\Sigma + \frac12 \wp, \Sigma + \frac12 \wp, P_3,
P_4\right) + 2\left( \Delta - \frac12 \wp \right)^2 \ee we can
carry out the sum over the ``unphysical charges'' $\Delta$ by
introducing $E_8$ theta functions with characteristics: \be
\label{thp} \Theta_{E_8[2], \wp}(\tau) := \sum_{\Delta\in E_8(1)}
e^{2\pi i \tau( \Delta - \frac12 \wp)^2 } \ee This allows to
decompose the $E_8(1)\oplus E_8(1)$ lattice as a sum of products
of shifted $E_8(2)$ lattices, \be \label{thth}
\theta^2_{E_8[1]}(\tau) = \sum_{{\cal P}\in E_8/2E_8}
\theta_{E_8[2],{\cal P}}(\tau) \theta_{E_8[2],{\cal P}}(\tau) \ee
Note that $\Theta_{E_8[2], \wp}$ depends only on the orbit of
${\cal P}$ under the Weyl group of $E_8$. It may be checked that
the finite group $Z$ decomposes into three orbits only,
corresponding to the orbit of the fundamental weight of the
trivial, adjoint and 3875 representations, of respective length 1,
120 and 135, respectively. The theta series \eqref{thp} are thus
the numerator of affine characters of $E_8$ at level 2, and can be
computed explicitly using free fermion representations, \be
\begin{array}{lcccc}
\theta_{E_8[2],1} &=& \theta_{E_8[1]}(2\tau) &=& 2^{-4}
\left( \theta_3^8 + \theta_4^8 + 14\ \theta_3^4 \theta_4^4 \right) \\
\theta_{E_8[2],248} &=& \frac12 \left( \th_3^6 \th_2^2 + \th_2^6 \th_3^2
\right)(2\tau) &=& 2^{-4} \left( \th_3^8 - \th_4^8 \right) \\
\theta_{E_8[2],3875} &=& \th_3^4 \th_2^4 (2\tau) &=& 2^{-4} \th_2^8
\end{array}
\ee
where we used the duplication identities \eqref{doubling}.
One may indeed check
that \eqref{thth} holds thanks to the modular
identity
\be
\label{ththe}
\theta_{E_8[1]}^2 = \theta^2_{E_8[2],1} +
120\ \theta^2_{E_8[2],248} + 135\ \theta^2_{E_8[2],3875}
\ee

For a fixed electric charge vector $2\Sigma+\wp$, the untwisted DH states
(irrespective of their oscillator level) are thus
enumerated by
\be
\label{zu}
\half \frac{\Theta_{E_8[2], \wp}(\tau)}{\eta^{24} }
+ \half  \delta_{\wp,0}   e^{2\pi i \delta \cdot P_3}
{2^4\over \eta^{12} \vartheta_2^4}  := q^{\Delta_\wp}
\sum_{N=0}^\infty \Omega^u_{\wp}(N) q^{N}
\ee
where $N+\Delta_\wp=\frac12 Q^2$. Notice that the second term
on the left-hand side corresponds
to states with charges $P_1=P_2$, hence $\wp=\Delta=0$.

\subsubsection*{Twisted sector}

Let us now analyze the DH states in the twisted sectors. Many details are
easily obtained by taking the modular transform of the partition function
with boundary conditions $(1,g)$. Unlike the untwisted sector,
twisted states automatically have $P_1=P_2$, however their charges now
take values in
\be
M_1 = E_8(-\half) \oplus (II^{1,1} + \delta) \oplus II^{5,5}
\ee
This is not a lattice since a sum of two vectors in $M_1$ ends up
in $M_0$. DH states take the form
\be
\CP^\pm(\alpha) \left( 1 \mp   e^{i \pi (\half P^2 + (P_3 + \delta)^2)} \right)
\vert P; P_3+\delta , P_4\rangle\otimes \vert t\rangle
\otimes \vert \tilde s \rangle
\ee
where $\vert t\rangle$ is the twisted left-moving
ground state,
and 8 of the bosonic oscillators in $\CP^\pm(\alpha)$ are half-integer
modded. DH states with electric charges $Q_e=
(P; P_3+\delta , P_4)\in M_1$ are
now enumerated by the partition function
\be
\label{dt14}
\frac{1}{2}
\biggl(  {1\over \eta^{12} \vartheta_4^4}
\pm {1\over \eta^{12} \vartheta_3^4 } \biggr)
:=  q^{\Delta_\pm} \sum_{N=0}^{\infty} \Omega^t_{\pm}(N) q^{N}
\ee
where the sign is that of $-e^{i \pi (\half P^2 + (P_3 + \delta)^2)}$,
and $\Delta_+ = -\half, \Delta_- = 0$.
By the level matching condition \eqref{lm2}, $N+\Delta_\pm$ is equated
to the square of the electric charge $Q^2/2$.

\subsubsection*{Comparison with macroscopic prediction}

Having obtained the exact degeneracies in the untwisted and twisted
sectors, we may now extract their asymptotics using the Rademacher
formula \eqref{radi},
\be
\label{deg414}
\Omega_{abs}(Q) =
\frac12 \BesselI{9}{4}{Q^2/2}
+  2^{-6} \BesselI{9}{4}{Q^2/4}
\left\{ \begin{matrix}
15 + 16  e^{2\pi i P_3\cdot \delta }\ ,  & \wp \in {\cal O}_1\\
1\ , & \wp \in {\cal O}_{248}\\
-1 \ ,  & \wp \in {\cal O}_{3875} \\
- e^{i\pi Q^2}\ ,  & Q \in M_1
\end{matrix} \right\}
+ \dots \nn
\ee
where $Q\in M_0$ in the first three cases.
Comparing to the general prediction \eqref{unideg} for a $\CN=2$ theory
with $n_V=16$ vectors, we see that
the microscopic counting \eqref{deg414} matches the macroscopic entropy
to all orders in $1/N$, in all sectors. However, the subleading correction
depends on the fine details of the charge vector
in the lattice $M_0 \oplus M_1$.

\subsubsection*{Helicity supertraces}
Finally, it is useful to check the analysis above against a direct
computation of the helicity supertraces. The partition function of
the $E_8\times E_8$ heterotic string on $T^6$ with
an insertion of $e^{i\pi v J_3^R}  e^{i\pi \bar v J_3^L}$
is given by
\be
\label{z14h}
Z_{(4,16)}^H(v,\bar v) =
\frac{\xi(v)\bar\xi(\bar v)}{4 \tau_2 |\eta|^4}
\sum_{h,g}
\sum_{\alpha,\beta} (-1)^{\alpha+\beta+\alpha\beta}
\frac{\bar\theta\ar{\alpha/2}{\beta/2}(\bar v)\
\bar\theta^3\ar{\alpha/2}{\beta/2}(0)}{\bar\eta^4}
\frac{Z_{6,6}}{|\eta|^{12}}  Z_{cur}\ar{h/2}{g/2}
\ee
where $\alpha,\beta=0,1$ run over the four spin structures
and $h,g=0,1$ run over the four (untwisted/twisted,
unprojected/unprojected) sectors of the orbifold. In the above
expression,
\be
Z_{6,6}\ar{h/2}{g/2} = \sum_{p \in \Gamma_{6,6}}
(-1)^{g (\delta,p)}
q^{\frac12 \Pi^2_L(p+ \frac{h}{2} \delta)}
\bar q^{\frac12 \Pi_R^2(p+ \frac{h}{2} \delta)}
\ee
is the partition function for the shifted $\Gamma_{6,6}$ lattice, and
\bea
\label{zcure8}
Z_{cur}\ar{0}{0} &=& \frac{\theta^2_{E_8[1]}}{\eta^{16}}(\tau) \ ,\quad
Z_{cur}\ar{0}{\half} = \frac{\theta_{E_8[1]}}{\eta^{8}}(2\tau) \\
Z_{cur}\ar{\half}{0} &=& \frac{\theta_{E_8[1]}}{\eta^{8}}
\left(\frac{\tau}{2}\right) \ ,\quad
Z_{cur}\ar{\half}{\half} = e^{-2i\pi/3}
\frac{\theta_{E_8[1]}}{\eta^{8}} \left(\frac{\tau+1}{2} \right)
\eea
are the orbifold blocks corresponding to the
exchange of the two $E_8$ factors.
Since the orbifold acts purely on the right-moving part, the
helicity partition function is obtained just as in the $(6,22)$
case, leading to the helicity supertraces
\bea
B_4 &=&
\frac{3}{2\tau_2 \eta^{8}}\times
\frac12 \sum_{h,g} Z_{6,6}\ar{h/2}{g/2} \  Z_{cur}\ar{h/2}{g/2} \\
B_6 &=&
\frac{15(2-E_2)}{8\tau_2\eta^{8}} \times
\frac12 \sum_{h,g} Z_{6,6}\ar{h/2}{g/2} \  Z_{cur}\ar{h/2}{g/2}
\eea
Using the duplication identities \eqref{doubling}, we obtain
\bea
\label{b466}
B_4 &=&
\frac{3}{2\tau_2}\times \frac12 \left[
\frac{\theta^2_{E_8[1]}}{\eta^{24}}
Z_{6,6}\ar{0}{0}
+ 2^4 \frac{\theta_{E_8[1]}(2\tau)}{\theta_2^4\ \eta^{12}}
Z_{6,6}\ar{0}{\half} \right.\\
&& \left.
+ \frac{\theta_{E_8[1]}(\frac{\tau}{2})}{\theta_4^4\ \eta^{12}}
Z_{6,6}\ar{\half}{0}
-
\frac{\theta_{E_8[1]}(\frac{\tau+1}{2})}{\theta_3^4\ \eta^{12}}
Z_{6,6}\ar{\half}{\half}
\right]
\eea
where the theta series in the numerator can also be written as
\bea
\theta_{E_8[1]}(\tau) &=& \frac12 \left( \theta_3^8 + \theta_4^8
+ \theta_2^8 \right) \\
\theta_{E_8[1]}(2\tau) &=& 2^{-4}
\left( \theta_3^8 + \theta_4^8 + 14\ \theta_3^4 \theta_4^4 \right) \\
\theta_{E_8[1]}(\tau/2) &=&
\theta_3^8 + \theta_2^8 + 14\ \theta_3^4 \theta_2^4 \\
\theta_{E_8[1]}((\tau+1)/2) &=&
\theta_4^8 + \theta_2^8 - 14\ \theta_4^4 \theta_2^4
\eea
Using \eqref{ththe} above, the untwisted contribution ($h=0$) may be
rewritten as
\bea
\frac{3}{2\tau_2}\times \sum_{\eps=\pm1}
\frac{Z_{6,6}\ar{0}{0} + \eps Z_{6,6}\ar{0}{\half} }{2}
\left[ \theta_{E_8[2],1} \times
\frac12 \left( \frac{\theta_{E_8[2],1}}{\eta^{24}}
+ \eps \frac{2^4  \vartheta_2^4}{\eta^{12}} \right)
\right.  \\ \left.
+ 120\  \theta_{E_8[2],248} \times
\left(\frac{\theta_{E_8[2],248}}{2\eta^{24}}\right)
+135\ \theta_{E_8[2],3875} \times
\left(\frac{\theta_{E_8[2],3875}}{2\eta^{24}} \right)
 \right]
\eea
Each term in round brackets can now be interpreted as
the multiplicity for the DH states in the conjugacy class ${\cal O}_\wp$
of $M_0$ indicated by the $E_8$ character
which multiplies it. Similarly, in the twisted sector we have
\bea
\frac{3}{2\tau_2}\times \sum_{\eps=\pm1}
\frac{Z_{6,6}\ar{\half}{0} + \eps Z_{6,6}\ar{\half}{\half} }{2}
\frac12 \left[  \frac{1 }{\theta_4^4\ \eta^{12}}
\theta_{E_8[1]} \left(\frac{\tau}{2} \right)
- \eps
 \frac{1}{\theta_3^4\ \eta^{12}} \theta_{E_8[1]}
\left(\frac{\tau+1}{2}\right)
\right]
\eea
This indeed reproduces the result \eqref{dt14} above.
It is also clear the generating function of the 6-th
helicity supertrace $B_6$ is given
by the same partition functions as before, up to a factor $5(2-E_2)/4$.

\subsection{General Reduced Rank Models}
The agreement found for the (4,16) model of the previous section
and the (4,24) model of Section 2 can in fact be easily seen to
generalize to all freely acting $\CN=4$ orbifolds of the heterotic
string compactified on $T^6$ by the following reasoning. In these
models, DH states can always be constructed in the untwisted sector,
by taking an arbitrary excitation of the left-moving 24 bosons, with
appropriate momenta and winding, and ensuring invariance under the
discrete symmetry. If $k=24-n_V$ is the number of vector fields which are
projected out by the orbifold, the generating function for the
absolute degeneracies (or, equivalently the helicity supertraces
$\Omega_4$) of DH states in the untwisted sector will take the form
\be
\label{chla}
\frac{1}{|G|} \left( \frac{Z}{\eta^{24}} + \sum_{g\in G \backslash
  \{1\}}
\Theta_g \right)
\ee
where $|G|$ is the order of the orbifold group,
$Z$ is the partition of the lattice of charges which have
been projected out, and $\Theta_g$ are the partition function
with an insertion of the generator $g\in G$. Indeed, the $k$ charges
are not physical and correspond to internal degrees of freedom.
The first term in \eqref{chla} is a modular form of weight $k/2-12
=-n_V/2$ and, provided the left-moving ground state is invariant
under the orbifold, has leading term $1/q$. The Rademacher formula
gives a Bessel function of the required order
$1-w=(n_V+2)/2$,
\be
\Omega^{u}(Q) \sim \hat I_{(n_V+2)/2}\left( 4\pi \sqrt{Q^2/2} \right)
\ee
in agreement with the prediction \eqref{unideg}.
The other terms have the same modular weight, but mix with
twisted sectors under modular transformation, and as a result
are exponentially suppressed.
In the twisted sectors, the generating functions can be obtained by modular
invariance, hence have the same modular weight. Their mixing with
the untwisted terms $\Theta_g$ implies that the leading term in the
Rademacher expansion is controlled by the same pole with $\Delta=1$.
Thus, the agreement with the OSV prediction \eqref{unideg} is expected
to hold for all $\CN=4$ reduced rank models. This is confirmed by the
analysis of other Het(4,$n_V$) models in Appendix D. As we shall see
in Section 5, the situation is quite different for $\CN=2$ models,
where the leading term in \eqref{chla} is absent in the case of the
helicity supertrace $\Omega_2$, or moduli dependent for absolute
degeneracies $\Omega_{abs}$.

\subsection{A Type II (2,2)/(0,4) Dual Pair \label{0422}}
Let us now turn to a different type II model, where the degeneracies of DH
states can be computed exactly by using a type II dual, albeit with unusual
(0,4) worldsheet supersymmetry \cite{Sen:1995ff}.

Consider type IIA compactified on the orbifold $(T^4 \times T^2)
/\Zint_2$, where the orbifold acts by a reversal of the coordinates
on $T^4$, times a translation along one circle in $T^2$. Since the
16 twisted sectors obtain a mass due to the shift, the
massless spectrum consists of 6 vector multiplets of $\CN=4$, together with
the gravity multiplet. The moduli space is thus given by
\eqref{modspace} with $n_V=6$, where the $Sl(2)$ factor
corresponds to the K\"ahler modulus of $T^2$. This orbifold
can be viewed as a variant of a $K3$
compactification. We shall denote this model by $(2,2)$, reflecting
the fact that the $\CN=4$ supersymmetries in target
space arise from the world-sheet supersymmetry symmetrically
between the left and right-movers.

This model was argued to be U-dual to a (4,0) type IIA model,
constructed as the different orbifold $(T^4 \times T^2)
/\Zint_2$ by $(-1)^{F_L}$ (where
$F_L$ is the left-moving world-sheet fermion number) times
a translation along on circle in $T^2$ \cite{Sen:1995ff}.
The orbifold gives a mass to all Ramond-Ramond fields, leaving
again 6+6 vector multiplets of $\CN=4$. While it has the same
$\CN=4$ supersymmetry, the latter now comes entirely from
the right-moving supercharges on the world-sheet.
Just as in the heterotic string, the $Sl(2)$ factor  in \eqref{modspace}
now parameterizes the axio-dilaton. The duality between these
$(0,4)$ and $(2,2)$ models is thus very similar to
the usual heterotic/$T^6$- type IIA/$K3 \times T^2$ duality.

Just as in the heterotic/$T^6$ case, DH states of the (0,4) model
can be constructed by exciting the left-movers only,
combined with appropriate momenta and winding along $T^6$.
Their helicity supertraces have been computed in \cite{Gregori:1997hi}
(Eq. 6.11):
\bea
\label{om440}
B_4 &=& \frac32 \sum_{(h,g)\neq (0,0)}
H_4\ar{h/2}{g/2}  Z_{2,2}\ar{h/2}{g/2} \ ,\quad \\
B_6 &=& \frac{15}{8}\sum_{(h,g)\neq (0,0)}
\left( H_4\ar{h/2}{g/2}
+ H_6\ar{h/2}{g/2} \right) Z_{2,2}\ar{h/2}{g/2} Z_{4,4}
\eea
where
\be
H_4\ar{h/2}{g/2} = e^{i\pi g} \frac{\theta^4\ar{h}{g}}{\eta^{12}}\ ,\quad
H_6\ar{h/2}{g/2} = \begin{cases}
 \frac{\th_3^8-\th_4^8}{2\eta^{12}}\ ,\quad     (h,g)=(0,1)\\
 \frac{\th_2^8-\th_3^8}{2\eta^{12}}\ ,\quad     (h,g)=(1,0)\\
 \frac{\th_4^8-\th_2^8}{2\eta^{12}}\ ,\quad     (h,g)=(1,1)
\end{cases}
\ee
Note that the contribution of the $(h,g)=(0,0)$ sector vanishes as it
has $N=8$ supersymmetry. From these expressions it easy to disentangle
the contributions of the various sectors: the degeneracies of DH
states in the untwisted sector are generated by
\be
\label{t4et12}
\frac12 \frac{\theta_3^4 - \theta_4^4}{\eta^{12}}
\quad \mbox{or} \quad
\frac12 \frac{\theta_2^4}{\eta^{12}}
\ee
depending whether the momentum along the shifted circle in $T^2$ is even
or odd, respectively. Those in the twisted sector are given by the
same expressions for odd and even momentum, respectively. In either
case, the degeneracies grow as
\be
\label{om440a}
\Omega_4 = \frac32 \Omega_{abs} \sim 3\cdot 2^5\
\hat I_5 (2\pi \sqrt{Q^2/2} )\ ,
\ee
hence have half the entropy of the DH states in the heterotic  $(4,24)$ model.
As in that model, the  helicity supertrace $\Omega_6$ originates entirely
from 1/2-BPS DH states, and the perturbative string spectrum contains no
intermediate multiplets.

Let us now turn to the type II $(2,2)$ side, and see if this entropy may
be accounted for by higher derivative interactions.
The $R^2$ amplitude in the $(2,2)$ model has been obtained
by a one-loop computation in \cite{Gregori:1997hi} (Sec. 6.1):
\be
f_{R^2} = 8 \log T_2 |\theta_4(T)|^4
\ee
where we use the same normalization as in \eqref{fr2all}.
In contrast to the other $\CN=4$ type II model considered in this
section, this amplitude contains only worldsheet instantons
(except for the logarithmic term), and vanishes in the large volume
limit $T\to i\infty$. This is in agreement with the fact that, on
the (0,4) side, the tree-level higher-derivative corrections start
at order $R^4$ corrections, as expected
for an orbifold of type IIA. In particular, the geometry
remains singular, and the OSV formula appears to be unable
to reproduce the microscopic entropy in \eqref{om440a}. It would be
interesting to see if $R^4$ corrections can resolve the singularity.

Finally, let us note that the (2,2) model also has
purely electric DH states, analogous to the states discussed in
Section \ref{dhii}. Their helicity supertraces have been computed in
\cite{Gregori:1997hi} (Eq. 6.4). In contrast to the type $IIA/K^3\times
T^2$ case, the 1/4-BPS states do not entirely cancel from the helicity
supertraces, instead the latter are given by modular forms of positive
weight,
\bea
\label{h6dh}
B_4 &=& 12 \sum_{(h,g)\neq (0,0)} Z_{2,2}\ar{h/2}{g/2}\ ,\\
B_6 &=& \frac{15}{2} \sum_{(h,g)\neq (0,0)} \left( 4 +
H_2\ar{h/2}{g/2}
+ \bar H_2\ar{h/2}{g/2}
\right)Z_{2,2}\ar{h/2}{g/2}
\eea
where
\bea
H_2\ar{h/2}{g/2} = \begin{cases}
 \th_3^4+\th_4^4  \ ,\quad   &  (h,g)=(0,1)\\
 -\th_2^4-\th_3^4 \ ,\quad   &  (h,g)=(1,0)\\
 \th_2^4-\th_4^4  \ ,\quad   &  (h,g)=(1,1)
\end{cases}
\eea
Depending on the sign of $Q^2$, the helicity supertrace
$\Omega_6$ of 1/4-BPS states
is generated by either $H_2$ or $\bar H_2$ in \eqref{h6dh}. Since the
modular weight of the counting function is positive,
the helicity supertrace $\Omega_6$ grows as a power of the charges,
rather than exponentially. In contrast, absolute degeneracies are counted
by the same functions as in \eqref{t4et12},
hence have an entropy of order $2\pi\sqrt{Q^2/2}$.
Just as in the type II/$K3 \times T^2$ case, it would be interesting
to understand how these states acquire a smooth horizon.

%%% pasted from I V3 may 4

\section{Macroscopic Predictions for Extremal Black Holes Degeneracies
in $\CN=2$ Models}

In this section, we return to the realm of $\CN=2$ supersymmetry,
where the OSV conjecture was originally formulated, and
extract the degeneracies of extremal black
holes as predicted by the conjectural relation \eqref{osvii}.
We start in Subsection \ref{gratop}
by reviewing the relation between the generalized
prepotential and the topological string amplitude. We then
evaluate \eqref{osvii} for large black holes with no D6-brane
charge ($p^0=0$), in particular scaling limits of the charges. The case
of small black holes in $K3$-fibrations is discussed in
Subsection \ref{smbh2}. Finally, in Subsection \ref{labhd6} we compute the
integral \eqref{osvii} for arbitrary D6-brane charge, for tree-level
prepotentials of the form $F=X^1 X^a C_{ab} X^b/X^0$. This is a special
example of the Legendre invariant prepotentials discussed in 
\cite{Pioline:2005vi}.

\subsection{Generalized Prepotential and
the Topological String Amplitude \label{gratop}}

As we recalled in the introduction, $\CN=2$ supergravity admits
an infinite series of higher-derivative corrections which can
be written as integral of a chiral density in $\CN=2$ superspace,
\bea
\int d^4\theta~ F(X^I,W^2) &=&
\int d^4\theta~ \sum_{h=0}^{\infty}  F_h(X^A) W^{2h} \\
&=& {\cal L}_{tree} + \sum_{h=1}^{\infty}
F_h(X^A) (^- C^-)^2 (T^-)^{2h-2} + \dots
\eea
where $X^I$ ($I=0..n_V-1$) are the homogeneous superfields for the vector
multiplets, $W$ is the $N=2$ Weyl superfield, with
$W^2=(T^-)^2 + \dots + \theta^4 (^- C^-)^2$, and the ellipses denote
other interactions related by supersymmetry (see e.g.
\cite{Mohaupt:2000mj} for a review of this formalism).
In the above expression,
$^- C^-$ denotes the anti-self-dual part of the Weyl tensor,
$T^-$ the anti-self-dual part of the graviphoton field-strength.
For $h=0$, one recovers the two-derivative $N=2$
Lagrangian controlled by the prepotential $F_0(X^I)$

For $\CN=2$ models obtained by compactifying type IIA string theory
on a Calabi-Yau three-fold, it can be shown that the only contribution
to the $(^- C^-)^2 (T^-)^{2h-2}$ coupling (or its on-shell equivalent
$(^- R^-)^2 (T^-)^{2h-2}$)  occurs at genus $h$, and
reduces to a vacuum amplitude in the A-model topological
string, obtained from the $(2,2)$ superconformal sigma model on
$\CX$ by a topological twist
\cite{Bershadsky:1993cx,Antoniadis:1993ze}.
In general, it includes non-holormophic contributions
from massless states propagating in the loops.
The holomorphic topological string amplitude is defined
as an asymptotic expansion in the topological string coupling near
some large radius limit
(i.e. in a   neighborhood of a point of maximal unipotent
monodromy). It includes  perturbative
contributions\footnote{In general one should allow for an extra
quadratic polynomial in $t^a$ with
real coefficients. These terms can be reabsorbed by a change of
variable and do not play any role in our discussion.}
at genus $0$ and $1$, together with
an infinite sum of world-sheet instanton
contributions at arbitrary genera,
\be
\label{Ftop}
  F_{\rm top} = - {i (2\pi)^3 \over 6 \lambda^2} C_{ABC} t^A t^B t^C
- {i \pi \over  12} c_{2A} t^A + F_{GW}(\lambda,q)
\ee
where $\lambda$ is the topological string coupling\footnote{
In the notations of \cite{Ooguri:2004zv}, $\lambda^2
  = - g_{top}^2$.} , $t^A = \theta^A +
i r^A$ with  $r^A>0$ are the  complexified K\"ahler moduli on a
basis $\gamma^A$ of $H_2(\CX,\IZ)$ ($A=1,\dots, n_V-1$),  $C_{ABC}$ are the
triple intersection numbers $C_{ABC} = \int_\CX  J_A J_B J_C $,
$c_{2A} = \int_\CX J_A c_2(T^{1,0} \CX)$,  and
\begin{subequations}
\bea
\label{gwdef}
  F_{GW}(\lambda,q) &=& \sum_{h\geq 0,\beta} N_{h,\beta} \, q^{\beta} \,
  \lambda^{2h-2}\\
\label{gvdef}
 &=& \sum_{h\geq 0 ,\beta,d\geq 1}
  n^h_\beta {1 \over d} \biggl( 2 \sin {d \lambda \over 2} \biggr)^{2h-2}
  q^{d\beta}.
\eea
\end{subequations}
is the Gromov-Witten instanton sum.
Here $\beta= \beta^A \gamma_A$ runs over effective curves with
$\beta^A\in \IZ^+$, $q^{\beta} := e^{2 \pi i \beta_A t^A}$, and
$N_{h,\beta}$ are the (rational) Gromov-Witten invariants. In the second
line we have used the identity of Gopakumar and Vafa to rewrite
$F_{GW}$ in terms of integral BPS invariants  $n^h_{\beta}$.

The precise relation between the topological string amplitude
and the generalized prepotential is
\be
F_{top}(t^A,\lambda) = \frac{i\pi}{2} F_{SUGRA}(X^A, W^2) \ , \quad
t^A = \frac{X^A}{X^0}\ ,\quad
\lambda^2  = \left( \frac{\pi}{4} \frac{W}{X^0} \right)^2
\ee
leading to the standard supergravity normalization\footnote{The
factor of proportionality relating $\lambda$ and $W/X^0$ can be
obtained by demanding the correct automorphic result for IIA/$K3\times
  T^2$.}
\be
\label{Fsugra}
F_{\rm sugra} = - {1 \over 6} C_{ABC} {X^A X^B X^C \over X^0} - {W^2 \over
 64} {c_{2A} \over 24} {X^A \over X^0} - {{X^0}^2 \over (2\pi i)^3}
 \sum_{h,\beta} N_{h,\beta} q^{\beta} \biggl({\pi W \over 4
 X^0}\biggr)^{2h}
\ee

It is important to note that the sum in \eqref{gvdef} contains
degenerate instanton contributions, with $\beta=0$. Those occur
only at genus 0, and are controlled by the single BPS
invariant $n^0_0=-(1/2) \chi(\CX)$, where $\chi$ is the Euler number
of $\CX$:
\be
\label{deffls0}
F_{GW}^{\rm deg} (\lambda) = -\half \chi(\CX) f(\lambda)
:= -\half \chi(\CX)
\sum_{d=1}^\infty {1\over d} {1\over (2\sin {d\la \over 2})^{2}}
\ee
where the second equality defines the Mac-Mahon function $f(\lambda)$.
$F_{GW}^{\rm deg}$ admits an asymptotic expansion at weak topological coupling, \be
\label{hzergw}
F_{GW}^{deg} =
   -\half \chi(\CX) \Biggl[\lambda^{-2} \zeta(3)
+ K
- \sum_{n=0}^\infty \lambda^{2n+2} {\vert B_{2n+4} \vert \over (2n+4)!}
 {(2n+3)  \over (2n+2)  }    B_{2n+2}  \Biggr]
\ee
where the ``constant'' $K$ is computed in Appendix \ref{mcmah},
\be
\label{klogl}
K =  {1\over 12} \log{2\pi i \over \la} - {1\over 2 \pi^2} \zeta'(2)
+ \frac{1}{12} \gamma_E
\ee
In equation \eqref{hzergw} above, the $O(1/\lambda^2)$ term corresponds to
the famous contribution to the prepotential coming from the
reduction of the tree-level $R^4$ coupling in 10 dimensions
\cite{Candelas:1990rm}, and the coefficient of $\lambda^{2n+2}$ is
the Euler character of the moduli space of genus $n+2$, as computed
in \cite{faber}. The ``constant'' $K$ depends logarithmically on
$\lambda$, hence cannot be attributed to any order in the genus
expansion. Nevertheless, it follows from a careful analysis of the
weak coupling behavior of $f(\lambda)$, which is analytic for  $\Im \lambda\not=0$.
This term is usually dropped in the
topological string literature, but will play an important role
in the analysis of the black hole degeneracies below.

Instead, for $N=2$ backgrounds obtained by compactifying the heterotic
string on $K3\times T^2$, the higher-derivative coupling
$(^- C^-)^2 (T^-)^{2h-2}$ for any $h$ receives contributions at 1-loop
already \cite{Antoniadis:1995zn} (as well as tree-level for $h=0,1$).
In fact, using heterotic-type II duality, this is a powerful way
to compute the Gromov-Witten invariants of compact K3-fibered
Calabi-Yau three-folds, at least for effective curves $\beta$ lying
only in the K3 fiber \cite{Marino:1998pg} (see
 \cite{Klemm:2004km} for recent progress).

Finally, let us note that by expanding the parenthesis in
\eqref{gvdef} in binomial series and summing term by term over $d$,
we may rewrite
\be
\label{gopva}
\begin{split}
F_{GW} =& \sum_{\beta}\sum_{k=1}^{\infty} k\ n_\beta^0
\log( 1 - e^{i k \lambda }) - \sum_{\beta} n^1_\beta \log ( 1 - q^{\beta})\\
&+ \sum_{h\geq 2} \sum_{\beta} \sum_{l=0}^{2h-2}
(-1)^{h+l} {2h-2 \choose \ell} n^h_{\beta}
\log\left( 1 - q^\beta e^{i(h-1-l)\lambda} \right)
\end{split}
\ee
hence obtaining $\exp(F_{GW})$ as an infinite
product \cite{Gopakumar:1998ii,Gopakumar:1998jq}.
Unfortunately,
for $h\geq 2$ the infinite product is in general divergent, falling
short of providing a non-perturbative definition of the topological
string amplitude.

\subsection{Large Black Holes with $p^0=0$ \label{labh2}}

Let us now turn to the evaluation of the integral \eqref{osvii},
for large black holes, with non-zero entropy at the classical level.
Since their entropy at large charges is already well reproduced by the
tree-level prepotential, it is natural to expect that Gromov-Witten
instantons can be neglected, at least in some large charge regime.
Under this assumption (to which we shall return below), and
restricting to $p^0=0$ for simplicity (see \cite{Pioline:2005vi} for
a discussion of the $p^0\neq 0$ case), the free energy \eqref{fbhw}
reads
\be\label{cfhptv} \CF^{\rm pert} =
-  {\pi \over 6} {\hat C(p)\over \phi^0 } + {\pi \over 2}{ C_{AB}(p) \phi^A
\phi^B\over \phi^0 }
\ee
where we use the standard notation
\be\label{dfsn}
C_{AB}(p)  = C_{ABC} p^C, \quad C(p)
= C_{ABC} p^A p^B p^C , \quad \hat C(p) = C(p) + c_{2A}p^A.
\ee
Note in particular that, in this limit,  the only effect of
higher derivative corrections is to replace $C(p)\to \hat C(p)$.

We further assume that the measure $[d \phi]$ is
the standard Euclidean measure, extending over the infinite real axis or
some deformation thereof. The integral over $\phi^A$ is
therefore Gaussian, with a peak at
\be
\phi_*^A = -C^{AB}(p) q_B \phi^0
\ee
Due to the indefinite signature of the quadratic form $C_{AB}(p)$,
it is well defined only upon rotating the contour of integration
so that $\phi^A/\sqrt{\phi^0} \sim e^{\pm i\pi/4}$. Proceeding
formally, we find
\be
\label{omint}
\Omega(p^A, q_A) \sim
\int d\phi^0\ \frac{(2\phi^0)^{\frac{n_V-1}{2}}}{|\det C_{AB}(p)|^{1/2}}
\exp\left( -  {\pi \over 6} {\hat C(p)\over \phi^0 } + \pi \phi^0 \hat
q^0 \right)
\ee
where $C^{AB}(p)$ is the inverse matrix of $C_{AB}(p)$ and
\be\label{dfnqh} \hat q_0 = q_0 - \half q_A C^{AB}(p)  q_B
\ee
is invariant under unipotent monodromies.
The integral over $\phi^0$ is now of Bessel type, with a
saddle point at
\be
\label{sdlepoint}
\phi^0_* = \pm \sqrt{ - \hat C(p) \over 6 \hat q_0   }
\ee
When $\hat q_0<0$, the action at the saddle point is real, and
equal to
\be
S_0 = 2\pi \sqrt{-  \hat C(p) \hat q_0  /6}
\ee
Provided the saddle point is actually selected by the contour
integral, we thus find that the formula \eqref{osvii} predicts
\be
\label{ibessl}
\Omega(p^A, q_A) \sim
\pm \half \vert \det C_{ab}(p)\vert^{-1/2}
\Bigl({\hat C(p) / 6} \Bigr)^{\nu}\ \times \
\hat I_{\nu } \Biggl(2\pi \sqrt{- \hat C(p) \hat q_0/ 6} \Biggr)
\ee
where
\be
\label{nuind}
\nu = \half(n_V+1)
\ee
Using the asymptotic expansion \eqref{besselasymp}, we thus
find
\be
\label{logo}
\log \Omega(p^A, q_A) \sim  S_0 -  \half(n_V+1) \log (S_0/4\pi)
- \log \CN(p) + \dots
\ee
where $\CN(p)$ is the $p$-dependent prefactor in \eqref{ibessl},
and the ellipses denote an infinite number of calculable
power-suppressed contributions. The first term in this equation
reproduces the classic result of \cite{Maldacena:1997de} (generalized
to $q^A \neq 0$), which was successfully matched to the microscopic
counting based on M5-branes wrapping a 4-cycle in $\CX$.

Let us now discuss the validity of our assumptions. Since this has
already been discussed in \cite{Dabholkar:2005by}, we shall be brief:
\begin{itemize}
\item Upon scaling all electric and magnetic charges to infinity
(but keeping $p^0=0$), the topological coupling $\lambda = 4\pi
/(i \phi^0_*)$ at the saddle point goes to zero, hence all higher
derivative corrections can be neglected. However, the K\"ahler
classes at the saddle point $\Im t^A =  p^A/\phi^0_*$ stay of
order 1, so it is not legitimate to drop the Gromov-Witten
instantons. \item If all $p^A\neq 0$ (but $p^0=0$), it is possible
to stay at weak topological coupling and get rid of the
Gromov-Witten instantons by scaling $\hat q^0$ faster than $p^A$.
In this case, the leading correction to the entropy comes from the
tree-level $\zeta(3)$ term in \eqref{hzergw}, which perturbs the
saddle point. This predicts a correction\footnote{A similar
correction was computed in \cite{LopesCardoso:1999ur}, without
taking into account the contrubution from the measure.} to
the Bekenstein-Hawking entropy 
\be S(p^A, q_A) =  2\pi \sqrt{-
\hat C(p) \hat q_0  /6} + \frac{\zeta(3)\chi(\CX)}{96\pi^2}
\frac{\hat C(p)}{ \hat q_0 } + \dots 
\ee 
which still grows like a
power of the charges. \item On the microscopic side, the leading
entropy is well reproduced from the M5-brane conformal field
theory when the Ramanujan-Hardy formula is applicable, i.e. when
$\hat q^0 \gg \hat C(p)$. In this regime, the topological coupling
at the saddle point is strong, although the K\"ahler classes can
still be taken to be large. This means that non-degenerate
Gromov-Witten instantons could be neglected, provided the
BPS  invariants grow sufficiently slowly. However, the series
of degenerate instantons is strongly coupled, and one should
instead use the Gopakumar-Vafa representation in terms of the
Mac-Mahon function, which is exponentially suppressed at large
coupling. The $\log \lambda$ term in \eqref{klogl} implies an
extra factor $(\phi^0)^{\chi(\CX)/24}$ in \eqref{omint}, which
would affect the index of the Bessel function in \eqref{nuind}.
Since \eqref{omint} will be further supported by the microscopic
analysis, we propose to modify by hand the definition of the
topological string amplitude $\Psi_{top}$ into
\be
\label{modosv}
\tilde\Psi_{top}:=\lambda^{\chi/24} \Phi_{top}
\ee
More generally,
it would be interesting to have a better understanding of the
integration measure in \eqref{osvii}.
\end{itemize}

To summarize, provided the OSV conjecture \eqref{osvii} holds,
the infinite number of power-suppressed corrections encapsulated
in the Bessel function \eqref{ibessl} can be trusted in the
strong coupling regime $\hat q^0 \gg \hat C(p)$, provided the
Gopakumar-Vafa infinite product is convergent.

%\subsubsection*{Large black holes and the $(0,4)$ CFT dual}

\vskip 5mm

Regrettably\footnote{The remainder of this section is excerpted 
from \cite{Dabholkar:2005by}.}, 
there are no examples where the degeneracies of large black holes
are known exactly. In principle the index $\Omega_2$ should be
computable from a $(0,4)$ sigma model described in
\cite{Maldacena:1997de,Minasian:1999qn}, presumably from the elliptic genus
of this model. While the sigma model is rather complicated,
and has not been well investigated we should note that
from the Rademacher expansion it is clear that the
leading exponential asymptotics of negative weight modular forms
depends on very little
data. Essentially all that enters is the order of the pole and 
the negative modular weight. There are $c_L =   C(p) + c_2\cdot p= \hat C(p)$ real left-moving
bosons. Since the sigma model is unitary, the
relevant modular form has the expansion
$q^{-c_L/24} + \cdots $. This gives the order of the pole, and thus 
we need only know the modular weight. This in turn depends on 
the  number of left-moving noncompact bosons. Each noncompact boson 
contributes $w=-\half $ to the modular weight. Now, the sigma model 
of \cite{Maldacena:1997de} splits into a product of a relatively simple 
``universal factor'' and a rather complicated ``entropic factor,'' 
as described in \cite{Minasian:1999qn}. Little is known about the entropic 
factor other than that it is a $(0,4)$ conformal theory with 
$c_R = 6 k$, where $k= {1\over 6} C(p) + {1\over 12} c_2\cdot p-1$, 
where $p\in H^2(\CX,\IZ)$. 
The local geometry of the target space was worked out 
in \cite{Minasian:1999qn}. 
Based on this picture we will  assume the target space is 
compact and does not contribute to the modular weight. 
(Quite possibly the model is a ``singular conformal field theory'' 
in the sense of \cite{Seiberg:1999xz} because the surface in the linear 
system $\vert p\vert$ can degenerate along the discriminant locus. 
It is reasonable to model this degeneration using a Liouville 
theory, as in \cite{Seiberg:1999xz}. 
If this is the case we expect   the entropic 
factor to contribute order one modular weight.) 
The universal factor is much more explicit. The target is $\IR^3 \times S^1$, 
it has $(0,4)$ supersymmetry with $k=1$ and there are $h-1$ (where $h=h_{1,1}$)
compact leftmoving bosons which are $N=4$ singlets. They have momentum in the 
anti-self-dual part of $H^{1,1}(\CX,\IZ)$ (anti-self-duality is 
defined by the surface in $\vert p \vert$). Since we fix these 
momenta we obtain $w=-\half (h-1)$. Finally there are 3 
noncompact left-moving bosons 
describing the center of mass of the black hole in $\IR^3$. Thus, the 
net left-moving modular weight is $-(h+2)/2$. Now, applying the 
Rademacher   expansion in the region $\vert \hat q_0 \vert \gg \hat C(p)$
we find the elliptic genus is proportional to 
\be
\label{radmsw}
\hat I_{\nu}\biggl(2\pi \sqrt{{\vert \hat q_0 \vert \hat C(p)\over 6}}\biggr)
\ee
with $\nu =  {h+4\over 2}$. This is remarkably close to  \eqref{ibessl} !  
Clearly, further work is needed here since it is likely there are 
a number of important subtleties in the entropic factor. Nevertheless,  
 our argument suggests that a deeper investigation of the elliptic 
genus in this model will lead to an interesting test of  
 \eqref{osvii} (or rather \eqref{modosv}, since it must be
done at strong topological string coupling) for the case of large black holes.

\subsection{Small Black Holes \label{smbh2}}

We now turn to the case of small black holes with $C(p)=0$
but $\hat C(p)\not=0$: these are singular solutions of the tree-level
$\CN=2$ supergravity Lagrangian, but it is expected that quantum
corrections will smooth out the singularity and lead to a bona fide
black hole. For such charges, the matrix $C_{AB}(p)$ is not
invertible and some of the manipulations in the previous section
need to be rethought.

We are particularly interested in the case when $\CX$ is a $K3$
fibration over $\IP^1$ admitting a heterotic dual. In this case,
we can divide up the special coordinates so that $X^1/X^0$ is the
volume of the base and $X^a/X^0$, $a=2, \dots n_V -1$ are
associated with the (invariant part of the) Picard lattice of the
fiber. The cubic intersection form becomes
\be
- {1 \over 6} C_{ABC} {X^A X^B X^C} =
-\frac12 C_{ab} X^1 X^a X^b
- \frac16 C_{abc} X^a X^b X^c
\ee
where the indices $a,b$ run from $2$ to $n_V-1$, and $C_{ab}$
is the intersection form of the (invariant part of the)  Picard lattice of the
fiber\footnote{Notice that $C_{abc}=0$ at tree-level on the heterotic
side, but not on the type II side in general.}
The matrix $C_{AB}(p)$ thus takes the form
\be\label{dabmatrx}
C_{AB}(p) = \begin{pmatrix} 0 & C_{ab} p^b  \\ C_{ab} p^b  &
p^1 \tilde C_{ab} + C_{abc} p^c\end{pmatrix}
\ee
We now specialize to heterotic DH states, with charges
$p^0=0$, $ p^a=0, a=2,\dots, n_V-1$, and
$q_1=0$, with $p^1q_0 \not=0$ and $q_a\not=0$ for $a=2,\dots, n_V-1$.
Using $c_1=24$ for the K3 fiber, the integral \eqref{osvii} now becomes
\be
\label{smosv}
\Omega( p^1, q_0, q_a) =
\int d\phi^0 d\phi^1 d\phi^a \ \exp\left( - 4\pi
\frac{p^1}{\phi^0} + \frac{\pi}{2} \frac{p^1 C_{ab} \phi^a
  \phi^b}{\phi^0}
+ \pi q_0 \phi^0 \right)
\ee
The $\phi^1$ dependence disappears from the
integrand and one must  make a discrete identification on $\theta
= \phi^1/\phi^0$. As in the benchmark case in Section 2,
we find that \eqref{osvii} gives
\be\label{iresult}
(p^1)^2 \ \hat I_{\nu}\bigl( 4\pi \sqrt{\vert p^1
      q_0
- \half
q_{a}\tilde{ C}^{ab} q_{b} \vert }\bigr)
\ee
where the index of the Bessel function is now
\be\label{iresulti} \nu = \half (n_V+2)
\ee
Let us now re-discuss the validity of our assumption that
Gromov-Witten instantons could be neglected in the small
black hole case. Since $C(p)=0$ and $\Im t^a = - p^a/\phi^0_*$
at the saddle point, the attractor values of the K\"ahler
moduli are necessarily at the boundary of the K\"ahler cone.
In principle, one must retain the full worldsheet instanton series
(or rather, its analytic continuation, should it exist.)

Remarkably\footnote{This paragraph is again excerpted 
from \cite{Dabholkar:2005by}.}, 
for $\CN=4$ compactifications this is not a problem. In
this case, due to the decoupling between the two factors in \eqref{modspace},
$F_{\rm top}$ is only a function of a single K\"ahler modulus $t^1$,
and moreover $\chi(\CX)=0$. Hence, at the saddle-point,
\be\label{sdlsbh}
\phi^0_* = - \sqrt{4 p^1 \over   \vert \hat q_0 \vert} \qquad  \Im t^1 = \half \sqrt{p^1 \vert \hat q_0 \vert}
\ee
Thus, whether or not the topological string coupling is strong ($\vert \hat q_0 \vert \gg p^1$) or
weak ($p^1\gg \vert \hat q_0 \vert $) the relevant K\"ahler class is
large
and the Bessel asymptotics \eqref{iresult} are justified.

The situation is rather different for $\CN=2$ compactifications. In this case
$F_{top}$ is in general a function of $t^1$ as well as $t^a$ for $a\geq 2$.
Thus the computation in \eqref{smosv} is {\it not} justified.
We stress that the problem is not  that the
topological string is strongly coupled. Indeed,  for $\chi=0$
examples such as the FHSV example discussed in Section 5.3 below,
the saddlepoint value \eqref{sdlsbh} can be taken in the weak coupling regime
by taking $p^1\gg \vert \hat q_0 \vert $.
In fact, the difficulty appears to be with the formulation of the
integral \eqref{osvii} itself
for the case of charges of small black holes.
Recall that we must evaluate
\be\label{curlyff} \CF_{top} := - \pi ~ \Im F_{top}(p^I + i \phi^I, 256)
\ee
Since $X^a/X^0 = \phi^a/\phi^0$ is {\it real}, for $a>1$,  one must
evaluate the worldsheet instanton sum for real values $t^a = \phi^a/\phi^0$.
For some Calabi-Yau manifolds it is possible to analytically continue
$F_0$ from  large radius to small values of $\Im t^a$. However we may
use  the
explicit results of \cite{Harvey:1996ts,Henningson:1996jz}, which express
 $F_1 \sim \log \Phi$, where $\Phi$ is an automorphic form for $SO(2,n;\IZ)$.
It appears that $\Im t^a =0$ constitutes a natural boundary of the
automorphic form  $\Phi$. Thus, in the case of $K3$ fibrations
with heterotic duals the formalism of \cite{Ooguri:2004zv}
becomes singular for these charges, even at weak topological
string coupling.

Remarkably, if we ignore these subtleties, the formula \eqref{iresult}
turns out
to  match perfectly with the asymptotic expansions  of twisted sector
DH states, as we show below. For untwisted sector DH states the asymptotics
do not match with either $\Omega_{abs}$ nor with $\Omega_2$.

\subsection{Large Black Holes with $p^0\neq 0$ \label{labhd6}}
Finally, let us evaluate the integral \eqref{osvii} for large black 
holes with non-zero D6-brane charge. For simplicity, we restrict ourselves 
to $K3$ fibrations with $C_{abc}=0$. and $c_{2A}=0$
\footnote{The case $c_{2,1}\neq 0, c_{2,a}=0$ can be obtained
by shifting $\vec p^2 \to \vec p^2 + \frac13 c_{2,1}$ in
the equations below.}, and, as in previous cases, 
disregard the Gromov-Witten instanton series. For convenience,
we drop inessential numerical factors. The computation
in this section is a special case of the analysis in \cite{Pioline:2005vi},
which applies for cubic prepotentials  $F=I_3(X)/X^0$
which are invariant under Legendre transform in all variables.
When this is not the case, such as in the $STU+U^3$ model, the 
attractor mechanism is significantly more involved.

{}From \eqref{fbhw}, one computes the black hole free energy
in the mixed ensemble,
\be
{\cal F} = \frac{p^0 \phi^1 - p^1 \phi^0}{(p^0)^2+(\phi^0)^2}
\left( \frac12 \vec \phi^2 - \frac12 \vec p^2  \right)
- \frac{p^0 p^1 + \phi^0 \phi^1}{(p^0)^2+(\phi^0)^2}
\left( \vec p \vec \phi \right)
\ee
where $\vec \phi^2 =  \phi^a \phi^b C_{ab},
\vec p^2 = p^a p^b C_{ab}, \vec p \vec \phi=  p^a C_{ab} \phi^b$,
and determines the microcanonical degeneracies via \eqref{osvii}.
The integral over the potentials $\phi^a$ is still
Gaussian, leading to
\be
\label{omj}
\begin{split}
\Omega_{OSV}(p,q)& =  \int d\phi^0 d\phi^1 
\left( \frac{(p^0)^2+(\phi^0)^2}
{p^0 \phi^1 - p^1 \phi^0}\right)^{\frac{n_V-2}{2}} \\
&\exp\left[
\frac{ [(p^1)^2+(\phi^1)^2] \vec p^2 +
[(p^0)^2+(\phi^0)^2] \vec q^2
- 2( p^0 p^1 + \phi^0 \phi^1) \vec p \vec q
}{2(p_1 \phi^0 - p^0 \phi^1)}
+ q_0 \phi^0 + q_1 \phi^1\right]
\end{split}
\ee
where $\vec q^2 = q_a C^{ab} q_b$ and $\vec p \vec q = p^a q_a$.
In order to compute the integral over $\phi^0,\phi^1$,
let us change variables to
\bea
p^0 \cosh x &=& \sqrt{(p^0)^2 + (\phi^0)^2}  \\
(p^0)^2 y &=& (p^1 \phi^0 - p^0 \phi^1) (\vec p^2 - p^0 q_1)
\eea
with Jacobian $d \phi^0 d\phi^1 / (dx dy) = (p^0)^2 \cosh x /
(\vec p^2 - 2 p^0 q_1)/2$.
The argument of the exponential in \eqref{omj} becomes
\be
y + \frac{B^2 \cosh^2 x}{4(p^0)^2 y} + \frac{A}{p^0} \sinh x
\ee
where
\bea
\label{d6ab}
A &=& -p^1 \vec p^2 + p^0 (p^0 q_0 + p^1 q_1 + \vec p \vec q) \\
B &=& \sqrt{(\vec p^2  - 2 p^0 q_1)
\left[ (p^1)^2 \vec p^2 + (p^0)^2 \vec q^2 - 2 p^0 p^1 \vec p \vec q \right]}
\eea
Together with the above det, this gives
\be
\label{d6i}
(p^0)^2
(\vec p^2 - 2 p^0 q_1)^{\frac{n_V-4}{2}}
\int (\cosh x)^h\ y^{\frac{2-n_V}{2}} \exp \left(
y + \frac{B^2 \cosh^2 x}{4(p^0)^2 y} + \frac{A}{p^0} \sinh x \right) \ dx\ dy
\ee
The integral over $y$ is of Bessel type, leading to
\be
(p^0)^{\frac{n_V}{2}}
\left(\frac{\vec p^2 - 2 p^0 q_1}{B}\right)^{\frac{n_V-4}{2}}
\int (\cosh x)^{n_V-1}
\exp\left[  \frac{A}{p^0} \sinh x \right]
I_{\frac{n_V-4}{2}} \left( \frac{B}{p^0} \cosh x \right)\ dx
\ee
In the limit where all charges are scaled to infinity at the same
rate, the integral \eqref{d6i} may be evaluated by saddle point 
approximation: the saddle lies at
\be
\phi^0 = \frac{A}{S_0}\ ,\quad
\phi^1 = \frac{1}{p^0 S_0} \left( A p^1 + \frac{B^2}{\vec p^2 - 2 p^0
  q_1} \right)
\ee
where 
\be
\label{d6s0}
S_{0} = \frac{1}{p^0} \sqrt{ B^2 - A^2}
\ee
In particular, the K\"ahler moduli at the saddle point are given by
\bea
\label{t1a}
\Im t^1 &=&
= \frac{p^0 \phi^1- p^1 \phi^0}{(p^0)^2+(\phi^0)^2}
= \frac{2S_0}{\vec p^2  - 2 p^0 q_1}  \\
\Im t^a &=&
= \frac{p^0 \phi^a- p^a \phi^0}{(p^0)^2+(\phi^0)^2}
= \frac{S_0}{B^2} \left(\vec p^2  - 2 p^0 q_1 \right)
\left( p^0 C^{ab} q_b - p^1 p^a \right)
\eea
Including the fluctuation determinant, we obtain 
\be
\label{d6fin}
\Omega_{OSV}(p,q) \sim B^2  ( \vec p^2  - 2 p^0 q_1)^{(n_V-4)/2}
S_0^{-(n_V+2)/2}
\exp( S_0 )
\ee
The leading entropy $S_0$ in \eqref{d6s0} agrees with the
general result in \cite{Shmakova:1996nz}. Using \eqref{d6ab}, it may be
rewritten as 
\be
S = \sqrt{(p^0)^2 q_0^2 + 2 p^0 q_1 \vec q^2 +2 p^0q_0(p^1 q_1+ \vec p \vec q)
+(p^1 q_1 - \vec p \vec q)^2 - 2 p^1 q_0 \vec p^2 - \vec p^2 \vec q^2}
\ee
where $\vec p^2 =p^a C_{ab} p^b, \vec q^2 =  q_a C^{ab} q_b$
and $\vec p \vec q = p^a q_a$.
Defining $Q=(q_0,p^1,q_a)$ and $P=(p^0,-q_1,p^a)$, this is 
recognized as the familiar discriminant
\be
S = \sqrt{(P\cdot P)(Q\cdot Q) - (P\cdot Q)^2}
\ee
One may check that the result \eqref{d6fin} agrees with \eqref{logo}
in the limit $p^0\to 0$, using the fact that $\det(C_{AB}(p)) 
= (p^1)^{h-2} \vec p^2$, $C(p)=3 p^1 \vec p^2$.

On the other hand, it is important to note that the prefactors in 
\eqref{d6fin}, which follow from using a trivial integration measure
for the electric potentials $\phi^I$ in \eqref{osvii}, are not 
consistent with T-duality. This problem may be cured by using an 
appropriate integration measure such as
\be
\label{tosv}
\tilde \Omega_{OSV}(p,q) = \int 
\frac{d\phi^0 d\phi^1 d\phi^a}
{|X^0|^{n_V+2} (\Im t^1)^2 (\Im t^a C_{ab} \Im t^b )^{n_V/2}}
\ e^{{\cal F} + \pi q^A \phi_A}
\ee
where, as usual, $X^I=p^I+i\phi^I$ and $t^A= X^A/X^0$.
To 1-loop order, this does not change the location of 
the saddle point \eqref{t1a}, but simply removes the offending
factors in \eqref{d6fin}, leading to
\be
\label{d6fin2}
\tilde\Omega_{OSV}(p,q) \sim S_0^{-(n_V+2)/2} \exp( S_0 )
\ee
For $p^0=0$, the measure in \eqref{tosv} reduces to the flat integration
measure used in \eqref{omint}, up to an overall factor $[C(p)]^2$ which depends
on the magnetic charges only. However, there is no guarantee that this
prescription will be consistent with T-duality at higher orders. 

The measure \eqref{d6fin2} is obviously not the only choice which removes
the non-duality invariant 
factors in \eqref{d6fin}. In particular, as shown in 
\cite{Pioline:2005vi} the following measure
\be
\label{tosv2}
\hat\Omega_{OSV}(p,q) = \int d\phi^0 d\phi^1 d\phi^a \ 
|X^0|^{-2} \ (\Im t^1)^{(n_V-4)/2} \ e^{{\cal F} + \pi q^A \phi_A}
\ee
has the remarkable effect of rendering the one-loop approximation 
to the integral exact, leading to the manifestly duality invariant
result
\be
\label{d6fin3}
\hat\Omega_{OSV}(p,q) = \hat I_{1/2}(S_0) \sim S_0^{-1} \exp( S_0 )
\ee
Note however that it does not reduce to the constant measure when $p^0=0$,
and it would therefore spoil agreement with the microscopic counting
of DH states. At any rate, irrespective of the choice of measure, it
is clear that a duality-invariant measure can no longer be holomorphic 
for $p^0\neq 0$.  It would be very desirable to have a deeper 
understanding of the integration measure implicit in \eqref{osvii}.

Finally, let us discuss the validity of the assumption that Gromov-Witten
instantons can be neglected. If we scale all electric and magnetic
charges uniformly by $s$, the entropy $S_0$ scales as $s^2$, 
the topological coupling $\lambda \sim 1/|X^0|$ as $1/s$  while the
K\"ahler classes $\Im t_A$ are fixed. The $\zeta(3) (X^0)^2$ term 
in \eqref{hzergw} is however comparable to the leading entropy $S_0$, so
that its effect cannot be neglected. It is therefore necessary to
scale the charges $(p^0,q_0)$ and $(p^A,q_A)$ differently if one is
to neglect the Gromov-Witten instanton contributions. One option is
to take $q_A\gg p^0 \gg (q_0, p^A)$. In this regime, the K\"ahler classes 
$\Im t_A$ grow to infinity as $\sqrt{q_A/p^0}$, while the coupling 
$\lambda = \Im (1/X^0)$ can be made arbitrarily small (in fact zero
when $q_0=p^A=0$), so that Gromov-Witten instantons can indeed
be neglected.

\section{Microscopic Counting of DH States in $\CN=2$ Models}\label{N=2}

In this section, we compute the microscopic degeneracies of perturbative
DH states in heterotic models with $\CN=2$ supersymmetry in four dimensions,
which are dual to small black holes in type II string theory compactified
on a Calabi-Yau three-fold $\CX$. In section 5.1 and 5.2, we discuss the
$E_8\times E_8$ heterotic string compactified on $K_3$ with standard,
respectively symmetric embedding of the spin connection in the gauge group.
In section 5.3, we turn to the FHSV model, which can be viewed as a $\CN=2$
analogue of the $\CN=4$ models with reduced rank discussed in Section 3.
In section 5.4, we obtain a formula which applies to all asymmetric
orbifolds of the heterotic string, with $\CN=2$ or $\CN=4$ supersymmetry.

\subsection{Het$/K3\times T^2$ with Standard Embedding}
A simple class of heterotic models with $\CN=2$ supersymmetry can be
obtained by compactifying the $E_8\times E_8$ heterotic string on
$K3$, and identifying the spin connection on $K3$ with the gauge
connection for one of the $E_8$ factors. The corresponding
conformal field theory is most easily constructed
at the $\IZ_2$ orbifold point of $K3$, where the orbifold generator
acts as $-1$ on the four coordinates of $T^4$ (as well
as their right-moving superpartners), and as a shift $(\frac12,\frac12,0^6)$
in the charge lattice of one of the $E_8$ factors. This gives a
$\CN=2$ model with 628 hypermultiplets transforming as a
\be
 4(1,1,1)+8(1,56,1)+(1,56,2)+32(1,1,2)
\ee
representation of the $E_8 \times E_7\times SU(2) \times U(1)^4$
gauge symmetry. In particular, $N_V-N_H=388-628=-240$.
This model is part of a large network of $\CN=2$ vacua which can be reached
by a sequence of fundamental or adjoint Higgsing
transitions \cite{Kachru:1995wm}. Of particular interest are the vacua
with Abelian gauge symmetry, which can be dual to compactifications of
type II string theory on a smooth Calabi-Yau threefold.
At a generic point in the moduli
space of $K3$, the $SU(2)$ factor is Higgsed, leaving 10 charged hypers in the
$56$ of $E_7$ and $65$ neutral hypermultiplets, for the same value of
the index $N_V-N_H=-240$.
Going to the Coulomb branch of $E_8$ reduces the gauge
symmetry to $E_7 \times U(1)^{12}$, with index $N_V-N_H=-480$.
Further higgsing the $E_7$ factor reduces the gauge symmetry to
$U(1)^{12}$ with $492$ neutral hypers, a $(12,492)$ model
in the notation of \cite{Kachru:1995wm}.
This model has been argued to be dual to type II on an hypersurface
in $WP^4_{1,1,12,28,42}$ \cite{Kachru:1995wm}.
Instead, one may go to the Coulomb branch of $E_7$
and obtain a $(19,65)$ model,
with 19 vector multiplets and 65 neutral hypers.
However, we could also consider going to the Coulomb branch of the original
$E_8 \times E_7\times SU(2) \times U(1)^4$ gauge symmetry, leading
to a (20,4) model with 20 Abelian vectors and 4 neutral hypers.

Let us now consider the degeneracies of DH states in the original
model with unbroken $E_8\times E_7\times SU(2)\times U(1)^4$ gauge symmetry.
The helicity generating partition function is obtained straightforwardly as
\bea
 \label{393}
Z(v,\bar v)
&=&\frac{\xi(v)\bar \xi(v)}{\tau_2 |\eta|^4} \frac12 \sum_{h,g=0}^1~
\frac12 \sum_{a,b=0}^1~(-1)^{a+b+ab}~{\bar\th[^\frac{a}{2}_\frac{b}{2}](v)
\bar\th[^\frac{a}{2}_\frac{b}{2}]
\bar\th[^{\frac{a+h}{2}}_{\frac{b+g}{2}}]
\bar\th[^{\frac{a-h}{2}}_{\frac{b-g}{2}}]\over \bar \eta^4} \\
&& \frac{Z_{2,2}}{|\eta|^4}
\frac{ Z_{(4,4)}^{orb}[^\frac{h}{2}_\frac{g}{2}]}{|\eta|^8}
\times {1\over 2}\sum_{\g,\d=0}^1~{\th[^{\frac{\g+h}{2}}_{\frac{\d+g}{2}}]
\th[^{\frac{\g-h}{2}}_{\frac{\d-g}{2}}]
\th^6[^{\frac{\g}{2}}_{\frac{\d}{2}}]\over \eta^8}\times
\theta_{E_8[1]}
\eea
where $Z_{(4,4)}^{orb}$ are the orbifold blocks of the $T^4/\IZ_2$
orbifold,
\be
Z_{(4,4)}\ar{0}{0} = Z_{4,4}\ ,\quad
Z_{(4,4)}\ar{h/2}{g/2} =
\frac{2^4}{|\theta\ar{1-\frac{h}{2}}{1-\frac{g}{2}}
\theta\ar{1+\frac{h}{2}}{1+\frac{g}{2}}|^2}
\ee
The sum over spin structures $a,b$ can as usual be performed by
using the Riemann identity \eqref{tt14}. Taking two $\bar v$ derivatives
and setting $v=\bar v=0$, the generating function for the
second helicity supertraces is thus
\be
B_2= \frac12 \sum'_{h,g}
\frac{Z_{2,2} \theta_{E_8[1]}}
{\tau_2 \eta^{18}\th[^{\frac{1+h}{2}}_{\frac{1+g}{2}}]
 \th[^{\frac{1-h}{2}}_{\frac{1-g}{2}}]}
\times {1\over 2}\sum_{\g,\d=0}^1~
\th[^{\frac{\g+h}{2}}_{\frac{\d+g}{2}}]
\th[^{\frac{\g-h}{2}}_{\frac{\d-g}{2}}]
\th^6[^{\frac{\g}{2}}_{\frac{\d}{2}}]
\ee
where the prime indicates that the untwisted, unprojected sector $h=g=0$
has to be omitted.

In order to read off the degeneracies of DH states with prescribed electric
charges from this expression, it is convenient to go to a general point in
the vector multiplet moduli space. This depends on the phase under
consideration:
\begin{itemize}
\item in the $(12,492)$ model above, where the gauge symmetry is broken
to $U(1)^{12}$, the two factors $Z_{2,2}$ and $\theta_{E_8[1]}$ combine into
a the partition function $Z_{2,10}$ of the charge lattice
$II_{2,2} \oplus E_8$ at a general point in the $SO(2,10)/SO(2)\times SO(10)$
moduli space. Using \eqref{E3} and \eqref{E4}, the sum over $h,g$
simplifies to
\be
\label{om2k3}
B_2= \frac{Z_{2,10}}{\tau_2} \frac{E_6}{\eta^{24}}
\ee
We thus deduce that the indexed degeneracies of DH states in this phase
are given by the coefficients of
\be
\frac{E_6}{\eta^{24}} = \sum_{N=0}^{\infty} \Omega(N) q^{N-1}
= \frac{1}{q} - 480 - \dots
\ee
where $N-1=Q^2$. By the Rademacher formula, the degeneracies grow as
\be
\Omega_2(Q) \sim  \BesselI{7}{4}{Q^2/2}
\ee
in agreement with the general prediction \eqref{iresult} with $n_V=12$.

\item in the (20,4)  model above,  the two factors $Z_{2,2}$
and $\theta_{E_8[1]}$
combine with the eight theta series in the numerator into a
a vector of partition functions $Z_{2,18}\th[^h_g]$ of a lattice
\be
\label{218s}
II_{2,2} \oplus \left( E_8 \cup (E_8 + \delta) \right) \oplus E_8
\ee
at a general point in its moduli space.
The helicity supertrace can be decomposed into four sectors,
\bea
\tau_2 B_2&=&
   {Z_{2,18}[^0_0]+Z_{2,18}[^0_1]\over 2} F_u
  -{Z_{2,18}[^0_0]-Z_{2,18}[^0_1]\over 2} F_u \nn\\
&&-{Z_{2,18}[^1_0]+Z_{2,18}[^1_0]\over 2} F_+
  -{Z_{2,18}[^1_0]-Z_{2,18}[^1_0]\over 2} F_-
\label{460}
\eea
with
\be
 F_u={\th^2_3\th_4^2\over \eta^{24}}\;\;\;,\;\;\;
 F_{\pm}={\th^2_2(\th_3^2\pm\th_4^2)\over
\eta^{24}}
\label{461}
\ee
We thus find that the second helicity supertraces of DH states are
enumerated by a different generating function in each
conjugacy class of the lattice \eqref{218s}. The asymptotics are given by
\bea
\Omega_2^u(Q)   &\sim& \BesselI{11}{4}{\frac{3}{8}Q^2} \\
\Omega_2^{\pm} (Q) &\sim& \BesselI{11}{4}{\frac{1}{2}Q^2}
\eea
In particular, the indexed degeneracies in the untwisted sector are
exponentially smaller than in the twisted sector. Only the latter
coincide with the macroscopic prediction \eqref{iresult} with $n_V=20$.
As we shall see, this is in fact a generic feature of $\CN=2$ orbifolds
where twisted states can be distinguished from untwisted ones by
their charges.

\item Similarly, in the (19,65) model,
the $Z_{2,2}$ and $\theta_{E_8[1]}$ combine with 7 out of the 8 theta
functions in the numerator, into the partition function of a
signature $(2,17)$ lattice. The second helicity supertraces
in the various sectors are generated by
\be
 F_u={\th^2_3\th_4^2 (\th_3 + \th_4) \over \eta^{24}}\;\;\;,\;\;\;
 F_{\pm}={\th^2_2(\th_3^2 (\th_3 + \th_2) \pm \th_4^2 (\th_4 + \th_2) )\over
\eta^{24}}
\label{462}
\ee
Again, using the Rademacher formula, we find agreement
with the  macroscopic prediction \eqref{iresult} with $n_V=19$
in the twisted sectors, but not in the untwisted one.
\end{itemize}

{}From the above discussion, it is thus clear that the degeneracies of DH
states depend on the phase under consideration: as a vector field
become massive, black holes which used to carry different charges
under this field are no longer distinguishable, leading to an increase
of the entropy at fixed charges under massless charges. The total
number of states is however conserved. In particular, the same argument
as in Section 3.4 shows that the modular weight of the
generating function of the second helicity supertrace at fixed charges
is directly correlated to the rank of the charge lattice, in agreement
with the relation $1-w = (n_V+2)/2$. The numerical factor in the leading
entropy however depends on the sector of consideration, and is typically
smaller in the untwisted sector. As we shall discuss in more detail in
Section 5.3 in the context of the FHSV model,
the absolute degeneracies are however much larger, as the
result of large cancellations between massive vector and hypermultiplets.

\subsection{Het$/K3\times T^2$ with Symmetric Embedding}
In general, one may construct $\CN=2$ heterotic backgrounds by
embedding the spin connection into the sum of two rank 2 bundles
with $c_2=12$ in each $E_8$ factor. This admits a simple conformal
field theory description as a $\IZ_2 \times \IZ_2$ orbifold, where
the first generator acts as in the standard embedding case, and
the second acts purely by a shift along one direction of $T^4$
as well as a vector $(\frac12,\frac12,0^6)$ in the other $E_8$
factor \cite{Aldazabal:1995yw}. This results in a model
with $E_7 \times SU(2) \times E_7 \times SU(2) \times U(1)^4$
gauge symmetry and hypermultiplets in
\be
4(56,1;1,1)+4(1,1;56,1)+16(1,2;1,1)+16(1,1;1,2)
\ee
This model has $N_V-N_H=-244$ and can be completely Higgsed into a
(4,244) model, dual to type II string theory on $WP_{24}^{1,1,2,8,12}$
with Euler number $\chi=-480$ \cite{Kachru:1995wm}.
The helicity partition function at the orbifold point reads
\bea
 \label{394}
Z&=&\frac{\xi(v)\bar \xi(v)}{\tau_2 |\eta|^4} \frac12 \sum_{h,g=0}^1~
\frac12 \sum_{h',g'=0}^1~
\frac12 \sum_{a,b=0}^1~(-1)^{a+b+ab}~{\bar\th[^\frac{a}{2}_\frac{b}{2}](v)
\bar\th[^\frac{a}{2}_\frac{b}{2}]
\bar\th[^{\frac{a+h}{2}}_{\frac{b+g}{2}}]
\bar\th[^{\frac{a-h}{2}}_{\frac{b-g}{2}}]\over \bar \eta^4} \\
&& \frac{Z_{2,2}}{|\eta|^4} \frac{
Z_{(4,4)}^{orb}[^{\frac{h}{2};\frac{h'}{2}}
_{\frac{g}{2};\frac{g'}{2}}]}{|\eta|^8} \times {1\over
2}\sum_{\g,\d=0}^1~{\th[^{\frac{\g+h}{2}}_{\frac{\d+g}{2}}]
\th[^{\frac{\g-h}{2}}_{\frac{\d-g}{2}}]
\th^6[^{\frac{\g}{2}}_{\frac{\d}{2}}]\over \eta^8} \times {1\over
2}\sum_{\g',\d'=0}^1~ {\th[^{\frac{\g'+h'}{2}}_{\frac{\d'+g'}{2}}]
\th[^{\frac{\g'-h'}{2}}_{\frac{\d'-g'}{2}}]
\th^6[^{\frac{\g'}{2}}_{\frac{\d'}{2}}]\over \eta^8} \nn \eea In
this expression, $ Z_{(4,4)}^{orb}[^{\frac{h}{2};\frac{h'}{2}}
_{\frac{g}{2};\frac{g'}{2}}]$ denotes the orbifold block
corresponding to a torus $T^4$ with twist $(h,g)$ on the 4
directions and shift $(h',g')$ along, say, the first circle. It is
non-vanishing only for $(h',g')=(0,0)$ or $(h,g)=(0,0)$ or
$(h,g)=(h',g')$. In the latter case, it reduces to the orbifold
block $ Z_{(4,4)}^{orb}[^{\frac{h}{2}} _{\frac{g}{2}}]$ with twist
only. In particular, despite appearances, one may check that
the construction is
symmetric under exchange of the two $E_8$. By using the Riemann
identity and \eqref{E3},\eqref{E4}, it is again possible to
simplify the helicity supertrace into \be
B_2 =
\frac{Z_{2,2}}{\tau_2} \frac{E_4 E_6}{\eta^{24}} \ee
Degeneracies of DH states from this equation can be extracted in
the same way as before. The result is simplest in the ``maximally
Higgsed'' phase of the (4,244) model,
where the 4 $U(1)$ charges correspond to the
$T^2$ lattice: the generating function for second helicity
supertraces of DH states is
simply
\be \label{E4E6} \frac{E_4 E_6}{\eta^{24}} =
\sum_{N=0}^{\infty} \Omega_2(N) q^{N-1} = \frac{1}{q} -240 + \dots \ee
with asymptotics \be \Omega_2(Q) \sim  - \BesselI{3}{4}{\frac{1}{2}Q^2}
\ee in full agreement with \eqref{iresult} for $n_V=4$. As before,
one may unhiggs this model and increase the rank of the gauge
group: in all cases the indexed degeneracies are counted by
modular forms of weight $w=-n_V/2$, and agree with \eqref{iresult}
in the twisted sectors only.

\subsection{The (2,12) FHSV Model \label{sec:FHSV}}
The FHSV model introduced in  \cite{Ferrara:1995yx}
is one of the simplest and best understood examples of
heterotic/type II duality with $\CN=2$ symmetry. On the type II side,
it consists of an orbifold of Type IIA string theory on $K3 \times
T^2$ by the Enriques involution on $K3$ times a reversal of $T^2$ --
a close cousin of the $(4,16)$ model. Its dual description may
be formulated as a $\IZ_2$ orbifold of the $E_8\times E_8$ heterotic string
on $T^4 \times T^2$, where the orbifold acts by exchanging the two
$E_8$ factors\footnote{We slightly deviate from the
action in \cite{Ferrara:1995yx}, reversing the coordinates on $T^4$,
and translating
one of the circles in $T^2$, which exchanges two $\Gamma_{9,1}$
and reverses a $T^3$; the two constructions are expected to be on
the same moduli space.}. In terms of the momentum lattice
\be
\label{g622fh}
\Gamma_{6,22} = E_8(-1) \oplus E_8(-1) \oplus II^{2,2} \oplus II^{4,4}
\ee
the action on the momenta is therefore
\be
g \vert P_1,P_2,P_3,P_4\rangle
= e^{2\pi i \delta \cdot P_3} \vert P_2,P_1,P_3,-P_4 \rangle
\ee
where $2\delta$ is the vector $(1,0,1,0)\in II^{2,2}$ corresponding to
the translation by half a period along the first circle.

Diagonalizing the action of $g$ on the oscillators,  there are $12$
untwisted and $12$ twisted left-moving bosons, and
$4$ twisted and $4$ untwisted right-moving $\CN=1$ multiplets.
Denoting by $(\eps_L,\eps_R)$ the parity of the left and right moving
oscillators under the orbifold action, massless states with parity
$(+,+)$ correspond to hypermultiplets, while massless states with
parity $(-,-)$ correspond to vector multiplets as well as the
graviphoton. The massless spectrum therefore consists of 12 hypermultiplets,
11 vectors multiplets and the gravity multiplet, with tree-level
moduli space
\be
\label{fhsvm}
\frac{SO(4,12,\IR)}{SO(4)\times SO(12)} \times
\frac{SO(2,10,\IR)}{SO(2)\times SO(10)}
\ee
where the first (resp. second) factor is parameterized by the
scalar fields in the hypermultiplets (resp. vector multiplets).
In fact, it can be shown that there are no quantum corrections
to the moduli space metric, and that \eqref{fhsvm} is the exact
quantum moduli space, up to global identifications  \cite{Ferrara:1995yx}.
At any point
on the vector multiplet moduli space, a vector $P$ of the lattice
\eqref{g622fh} may be projected into a sum $\Pi_L(P)+\Pi_R(P)$
in $\IR^{22} \oplus \IR^{6}$. The linear combination
\be
Z = \Pi_{R}^1(P) + i \Pi_R^2(P)
\ee
is the complex
central charge $Z$ of the $\CN=2$ algebra, while the remaining
components $\Pi_R^{3,4,5,6}(P)$ are the remnants of the
central charges of the $\CN=4$ supersymmetry, which is broken by the twist
on $T^4$. By the same reasoning as in Section 3.3, the 22
left-moving charges $\Pi_L^i(P)$ decompose into 12 electric charges
\be
Q(P) = (P_1+P_2; P_3)
\ee
under the gauge fields in the vector multiplets, taking values
in the signature $(2,10)$ non-self dual lattice
\be
\Lambda_{0} = E_8(-\half ) \oplus II^{2,2}
\ee
while the remaining 10 are ``unphysical charges'' under gauge fields which
have been projected out.

\subsubsection*{Untwisted sector}

Now, candidate DH states in the untwisted sector can be
constructed as \be\label{untwist} \CP^\pm ( \alpha )\cdot
\biggl(\vert P_1, P_2, P_3,P_4\rangle \pm e^{2\pi i P_3\delta}
\vert P_2, P_1, P_3,-P_4\rangle\biggr) \otimes \vert\tilde I
\rangle^\pm \ee where  $\CP^\pm(\alpha)$ denotes a generic
monomial in the left-moving creation oscillators, with definite
parity $\pm$ under the orbifold action $g$, and $|\tilde
I\rangle^\pm$ denotes the right-moving ground states transforming
as $8_v \oplus 8_s$ under the transverse $so(8)$ rotations in
ten-dimensions, with definite parity under $g$. Unlike the
$(4,16)$ model, the states \eqref{untwist} are BPS only if they
saturate the BPS bound $M^2=|Z|^2$, i.e. $\Pi_R^{i}(P)=0$ for
$i=3,4,5,6$. More formally, this condition may be written as \be
\label{bpscond} \Pi_R(P)^2 = \Pi_R(Q(P))^2 \ee Note that this
condition explicitly depends on the values of the vector multiplet
moduli space. For $P_4\neq 0$, it is only obeyed on a codimension
one submanifold of the vector moduli space, providing an example
of the ``chaotic BPS states'' mentioned in the introduction. As we
shall see shortly, these states always come in a vector multiplet
/ hypermultiplet pair and cancel from the helicity supertrace
$\Omega_2$. On the other hand, states \eqref{untwist} with $P_4=0$
are always BPS. In order to enumerate the DH states
\eqref{untwist}, let us introduce the partition function
\be\label{bpm} A_\pm := {\Tr}_{\CH_{P}} \half (1\pm g) =
\sum_{II^{22,6}} q^{\half \Pi_L(P)^2} \bar q^{\half\Pi_R(P)^2}
\Pi_{bps}(P) \half (1 \pm \Theta(P)) \ee where the projection
operator $\Pi_{bps}(P)$ is $=1$ when (\ref{bpscond}) is satisfied,
and $=0$ otherwise, and \be\label{thetap} \Theta(P)=\delta_{P_1,
P_2} e^{2\pi i P_3 \cdot\delta} \delta_{P_4,0} \ee incorporates
the fact that states with $P_1=P_2, P_4=0$ and  $e^{2\pi i \delta
P_3} = \mp 1$ are dropped out, while those we $P_1\neq P_2$ or
$P_4\neq 0$ are counted twice with 1/2 multiplicity, just as in
\eqref{candsc}. Note that $\Pi_{bps}(P)\Theta(P) = \Theta(P)$.

In addition, let us introduce the
partition functions of the  left-moving oscillator excitations
$\CP^\pm ( \alpha )$,
\be\label{ii}
B_\pm := {\Tr}_{\CH_{osc}} \half (1\pm g) q^{L_0} \bar q^{\bar L_0} =
\half \biggl( {1\over \eta^{24} } \pm  {2^6\over \eta^{6}
\vartheta_2^6} \biggr) := q^{-1} \sum_{N=0}^\infty  d^u_{\pm}(N) q^N
\ee
The partition function for DH states \eqref{untwist} with positive
parity for the right-moving ground state is thus
\be\label{hypers}
Z_H=A_+ B_+  + A_- B_- = \half {1\over \eta^{24} } \sum_{P\in II^{22,6}}
q^{\half q_L^2} \bar q^{\half q_R^2} \Pi_{bps}(P)
+  {2^5\over \eta^{6} \vartheta_2^6} \sum_{P\in II^{22,6}}
q^{\half q_L^2} \bar q^{\half q_R^2} \Theta(P)
\ee
while, for DH states with negative right-moving parity, it is
\be\label{vectors}
Z_V=A_+ B_- + A_- B_+=\half {1\over \eta^{24} } \sum_{P\in II^{22,6}}
q^{\half q_L^2} \bar q^{\half q_R^2} \Pi_{bps}(P)
-  {2^5\over \eta^{6} \vartheta_2^6} \sum_{P\in II^{22,6}}
q^{\half q_L^2} \bar q^{\half q_R^2} \Theta(P)
\ee
Generalizing the terminology from the massless sectors, and
consistent with the definition in \cite{Harvey:1995fq}, we shall
refer to the states of the first type \eqref{hypers}
as ``massive hypermultiplets'', and states of the second type
\eqref{vectors} as ``massive vector multiplets''. Taking
the difference, we find the index
\be
B_2 = Z_H - Z_V =   {2^6\over \eta^{6} \vartheta_2^6} \sum_{P\in II^{22,6}}
q^{\half q_L^2} \bar q^{\half q_R^2} \Theta(P)
\ee
The notation anticipates the fact, to be demonstrated shortly, that
this index indeed coincides with second helicity supertrace. The
chaotic BPS states thus cancel out from $\Omega_2$, leaving only states
with $P_1=P_2$ and $P_4=0$. For these states, the indexed degeneracies are
thus counted by
\be
\label{untfh}
  \frac{2^6}{\eta^{6} \vartheta_2^6} :=  \sum d^u_{\pm}(N) q^{N-1}
\ee
Using the Rademacher formula, this is given asymptotically by
\be
d^u(N) \sim 2^{-7}
\BesselI{7}{2}{N-1}
\ee
Note that the argument of the Bessel function is one half of its usual
value, in agreement with the fact that unbroken $\CN=4$ supersymmetry
in the untwisted sector leads to drastic cancellations in the index $\Omega_2$.

\subsubsection*{Chaotic BPS states}
While the BPS states cancel from the index $\Omega_2$, it is nevertheless of
interest to investigate their degeneracies, and exhibit their dependence
on the moduli. Let us therefore consider the sum
\be
\label{zh+zv}
Z_H+Z_V= {1\over \eta^{24} } \sum_{P\in II^{22,6}}
q^{\half q_L^2} \bar q^{\half q_R^2} \Pi_{bps}(P)
\ee
Now, as in the (4,16) case, we need to rewrite \eqref{zh+zv}
as a partition for the physical charges $Q=(P_1+P_2;P_3)$.
Let us therefore change basis to
\bea
P_1 + P_2 &=& 2 S + \wp \\
P_1 - P_2 &=& 2 \Delta - \wp
\eea
where $S,\Delta$ both take values in the $E_8$ root lattice, and ${\cal P}$ is
an element of the finite group $Z=\Lambda_r(E_8)/2 \Lambda_r(E_8)$.
When  $\Pi_{bps}(P)=1$, it is easy to check that
\be\label{conds}
\Pi_L(P)^2 - \Pi_L(Q(P))^2 = -2(\Delta - \half \wp)^2 - P_4^2
\ee
This allows to rewrite \eqref{zh+zv} into
\be
Z_H+Z_V= {1\over \eta^{24} }
\sum_{Q\in \Lambda_0} q^{\half \Pi_L(Q)^2} \bar q^{\half \Pi_R(Q)^2}
\CF_Q(q)
\ee
where $\CF_Q(q)$ is a sum over ``unphysical charges'',
\be\label{dfnshe}
\CF_Q(q) = \sum_{\Delta\in E_8[1], P_4\in II^{4,4}} q^{-(\Delta -
\half \wp)^2 -\half P_4^2} \
\Pi_{bps} \left( S+\Delta, S-\Delta+ \wp, P_3, P_4  \right)
\ee
Now, for generic moduli $\Pi_{bps}\neq 0$ only for $\Delta
=0, \wp=0, P_4=0$, so that
\be\label{simplefq}
\CF_Q(q) =\delta_{\wp,0}
\ee
At special moduli however, $\CF_Q(q)$ will be a non-trivial theta series.
E.g., at the $E_8\times E_8$ enhanced symmetry point with
generic (non-rational) moduli for $II^{4,4}$, the BPS condition
 puts $P_4=0$, however allows $\Delta,
\wp$ to be purely leftmoving, leading to
\be\label{bigeffq} \CF_{Q}(q) = \sum_{\Delta\in E_8(+1)}
q^{(\Delta - \half \wp)^2} = \Theta_{E_8[2], \wp}(\tau) \ee The
absolute degeneracies of the DH states, counted by \be {1\over
\eta^{24} } \CF_{Q}(q) = :=  \sum d^{abs}_{\pm}(N) q^{N-\Delta}
\ee will thus have different asymptotics at different points in
moduli space, \be d^{abs}(N) \sim \BesselI{\nu}{4}{N-\Delta} \ee
where the index of the Bessel function will be $\nu=13$ for
generic moduli, $\nu=9$ at moduli where \eqref{bigeffq} is valid,
and may take other values at different loci. Since the index $\nu$
controls the logarithmic correction to the entropy, the latter
would in general depend on the moduli. Note that in all cases, the
index $\Omega_2$ is exponentially suppressed with respect to the absolute
number of BPS states in the untwisted sector.

\subsubsection*{Twisted sectors}
Let us now briefly turn to the BPS states in the twisted sector
of the FHSV model. By a modular transformation, it is easy to see
that the electric charges for twisted sectors is
\be
\Lambda_1 = E_8(-\half) \oplus (II^{2,2} + \delta)
\ee
 DH states take the form
\be
\CP^\pm(\alpha) \left( 1 \mp   e^{i \pi (\half P^2 + (P_3 + \delta)^2)} \right)
\vert P; P_3+\delta \rangle\otimes \vert t\rangle
\otimes \vert \tilde s \rangle
\ee
where $\vert t\rangle$ denotes one of the $2^{6}$ twisted left-moving
ground states, and $ \vert \tilde s \rangle $ one of the $2^6\times 2^3$
twisted right-moving ground states, in the Neveu-Schwarz or Ramond sector.
DH states with electric charges $Q_e=
(P; P_4)\in \Lambda_1$ are
now enumerated by the partition function
\be
\label{dt15}
\frac12
\biggl(  {2^6\over \eta^{6} \vartheta_4^6}
\pm
 {2^6\over \eta^{6} \vartheta_3^6 } \biggr)
:=  \sum \Omega^t_{\pm}(N) q^{N-\Delta_\pm}
\ee
where the sign is that of $-e^{i \pi (\half P^2 + (P_3 + \delta)^2)}$,
and $\Delta_\pm = \pm 1/4$. Using the Rademacher formula we obtain
the asymptotics
\be
\label{fhsvdt}
\Omega^t_{\pm}(N) = 2
\BesselI{7}{4}{N-\Delta_\pm} + \cdots
\ee
Comparing with the macroscopic prediction \eqref{unideg}
with $n_V=12$, we find agreement to all orders in inverse
charges. As in previous cases, the prescription \eqref{osv1}
however fails to reproduce the ``non-perturbative'' corrections
in \eqref{fhsvdt}.

\subsubsection*{Degeneracies vs. helicity supertraces}
Finally, let us rederive the above results using the formalism of helicity
partition functions. By the same reasoning as in \eqref{z22h},
the helicity partition function of the FHSV model reads
\be
\label{fhsv1}
\begin{split}
Z_{\rm FHSV}^H(v,\bar v) =&
\frac{\xi(v)\xi(\bar v)}{\tau_2 |\eta|^4} \frac12 \sum_{h,g}
\frac{Z^{orb}_{4,4}\art{h}{g}}{|\eta|^{8}}
\frac{Z_{2,2}\art{h}{g}}{|\eta|^{4}}
\bar Z_{cur}\art{h}{g}
\\
&\times\frac12 \sum_{\alpha,\beta} (-1)^{\alpha+\beta+\alpha\beta}
\left(\frac{
\bar\theta\art{\alpha}{\beta}(\bar v)
\bar\theta\art{\alpha}{\beta}(0)
\bar\theta\art{\alpha-h}{\beta-g}(0)
\bar\theta\art{\alpha-h}{\beta-g}(0)
}{\bar\eta^4} \right)
\end{split}
\ee
where
\be
Z_{2,2}\art{h}{g} = \sum_{p \in II^{2,2}}
(-1)^{g (\delta,p))}
q^{\frac12 \Pi^2_L(p+ h \delta)} \bar q^{\frac12 \Pi_R^2(p+ h \delta)}
\ee
is the partition function for $T^2$ orbifolded by a translation
by the order 2 vector $\delta$,
\bea
Z_{4,4}^{orb}\ar{0}{0} &=& \sum_{p \in II^{4,4}}
q^{\frac12 \Pi_L^2(p)} \bar q^{\frac12 \Pi_R^2(p)} \\
Z_{4,4}^{orb}\art{h}{g} &=& 16\frac{|\eta|^{12}}
{|\theta\ar{\half+\frac{h}{2}}{\half+\frac{g}{2}}
\theta\ar{\half-\frac{h}{2}}{\half-\frac{g}{2}}|^2}\ ,
\quad(h,g)\neq (0,0)
\eea
are the partition functions of the orbifold $T^4/\Zint_2$,
and $Z_{cur}$ is the same as in \eqref{zcure8}.
The sum over spin structures can be performed using the
Riemann identity, leaving
\be
\label{fhsv2}
\begin{split}
Z_{\rm FHSV}^H(v,\bar v) =\frac12 \sum_{h,g}
\frac{\xi(v)\xi(\bar v)}{\tau_2 |\eta|^4}\
&\frac{\bar\theta_1^2 (\frac{\bar v}{2})
\ \bar \theta\ar{\half-\frac{h}{2}}{\half-\frac{g}{2}}(\frac{\bar v}{2})
\ \bar \theta\ar{\half+\frac{h}{2}}{\half+\frac{g}{2}}(\frac{\bar
  v}{2})} {\bar\eta^4}\\
&\times \frac{Z^{orb}_{4,4}\art{h}{g}}{|\eta|^{8}}
\frac{Z_{2,2}\art{h}{g}}{|\eta|^{4}}
Z_{cur}\art{h}{g}
\end{split}
\ee
The leading trace comes at order $v^2$, and does not receive any
contribution from the $(h,g)=(0,0)$ sector, which has $\CN=4$ supersymmetry:
\be
B_2 = \frac{1}{2\tau_2\ \eta^2}
\sum_{(h,g)\neq(0,0)}
\frac{16}{ \theta\ar{\half-\frac{h}{2}}{\half-\frac{g}{2}}
\ \theta\ar{\half+\frac{h}{2}}{\half+\frac{g}{2}}}
Z_{2,2}\art{h}{g}
\bar Z_{cur}\art{h}{g}
\ee
or, equivalently,
\be
B_2 = \frac{16}{2\tau_2}
\left[
16 \frac{\theta_{E_8[1]}(2\tau) Z_{2,2}\ar{0}{\half}}{\eta^6 \theta_2^6}
+ \frac{\theta_{E_8[1]}(\frac{\tau}{2}) Z_{2,2}\ar{\half}{0}}{\eta^6 \theta_4^6}
- \frac{\theta_{E_8[1]}(\frac{\tau+1}{2}) Z_{2,2}\ar{\half}{\half}}
{\eta^6 \theta_3^6}
\right]
\ee
Identifying the numerators as the partition functions for the lattice
$\Lambda_0$ and $\Lambda_1$, we directly obtain the degeneracies
\eqref{untfh} and \eqref{dt15}. The contribution of the chaotic states
can be exhibited by looking at the fourth helicity supertrace $\Omega_4$.

\subsection{General $\CN=2$ Asymmetric Orbifolds}
Having described the FHSV model in detail, it is not too difficult
to compute the degeneracies of DH states for
arbitrary asymmetric orbifolds of the heterotic
string on the torus $T^6$ by a discrete group $\Gamma$. We assume
that the constraints level matching and anomaly cancellation
are satisfied, which still leaves a large class of possibilities.
For simplicity we will focus on  the index of DH states in
the untwisted sector. We discuss the twisted sector states briefly
at the end of this section.

Let $\Gamma$ is a discrete group, with an
embedding $R: \Gamma \to O(22)\times O(6)$. The orbifold
group acts by shifts so that the action on momentum vectors is
\be\label{momact}
g \vert P \rangle = e^{2\pi i \delta(g)\cdot P} \vert R(g) P \rangle
\ee
In $\IR^{22,6}$, with metric $Diag(+1^{22},-1^{6})$  we can diagonalize
the rotational part of $R(g)$ as
\be
\label{diagg}
R(g) = R(\theta_1(g)) \oplus \cdots \oplus R(\theta_{11}(g) ) \oplus R(2\tilde \theta_1(g) )\oplus
R(\tilde \theta_2(g) )\oplus R(\tilde \theta_3(g) )
\ee
where $R(\theta)$ is the usual $2\times 2$ rotation matrix
\be
R( \theta)= \begin{pmatrix} \cos(2\pi \theta)& \sin(2\pi \theta)\\
-\sin(2\pi \theta)& \cos(2\pi \theta)\ \end{pmatrix}
\ee
We will sometimes denote
$\theta_j(g) = r_j(g)/N$ where $N = \vert \Gamma \vert$.
$\CN=2$ supersymmetry requires
that $\tilde\theta_1+\tilde\theta_2+\tilde\theta_3\equiv
0 \mod 1$ so that their exists a complex combination $Z$
of the charges $\Pi_R(p)$ which is invariant under $\Gamma$,
and which can be identified as the $\CN=2$ central charge.

The moduli are the boosts in $O(22,6)$ commuting with the image $R(\Gamma)$.
We consider embeddings $\Lambda\subset \IR^{22,6}$ of $II^{22,6}$,
and let $\Lambda(g)$ denote the sublattice of vectors
fixed by the group element $g$.

DH states in the untwisted sector  are contained in the subspace
of the 1-string Hilbert space of the form \be\label{thens}
\CH_{osc,L} \otimes \CH_{mom} \otimes \tilde \CH_{gnd} \ee As
already stressed in the FHSV model, even after imposing the level
matching constraints, it is still necessary to insert a projection
$\Pi_{bps}$ on states which satisfy the BPS condition $M^2=|Z|^2$.
The DH states can therefore be enumerated by introducing the
partition function for the momenta, \be \label{smontg}
{\Tr}_{\CH_{mom}} \biggl(U_2(g) q^{H}\bar q^{\tilde H}  \biggr) =
\sum_{P\in \Lambda(g)} q^{\half P_L^2} \bar q^{\half P_R^2}
e^{2\pi i \delta(g)P} \Pi_{BPS}(P) \ee where $U_i(g)$ is the
representation of $g$ in each of the factor spaces, and for the
left-moving oscillators in the 22 internal directions, \be
\label{oscint} {\Tr}_{\CH_{osc,L}} \biggl(U_1(g) q^{H} y^{2J_3}
\biggr) = \prod_{j=1}^{11} \frac{-2 \eta (\tau) \sin \pi
\theta_j(g)} {\theta\ar{\half}{\half + \theta_j(g)}(0;\tau)} \ee
where we understand that \be \label{engins} \frac{-2 \eta (\tau)
\sin \pi \theta_j(g)} {\theta\ar{\half}{\half +
\theta_j(g)}(0;\tau)} \rightarrow {1\over \eta^2} \ee if
$\theta_j(g)=0$. The contribution of the right-movers as well as
the left-moving bosons in the transverse directions can be written
as \be \frac12 \sum_{a,b} (-1)^{a+b+ab}
\bar\theta\ar{\frac{a}{2}}{\frac{b}{2}}(\bar v) \prod_{i=1,3}
\bar\theta\ar{\frac{a}{2}}{\frac{b}{2}- \tilde\theta_i(g)}(0)
\frac{\sin \pi \tilde \theta_i}{\bar\theta\ar{ \half}{\half-
\tilde \theta_i}(0)} \ee The sum over spin structures can be
carried out by using the generalized Riemann identity \eqref{t15}.
In the supersymmetric case, this reduces to \be
\bar\theta\ar{\half}{\half}(\bar v/2) \prod_{i=1,2,3}
\bar\theta\ar{\half}{\half- \tilde\theta_i(g)}(\bar v/2)
\frac{\sin \pi \tilde \theta_i}{\bar\theta\ar{\half}{\half- \tilde
\theta_i(g)} (\bar v/2)} \ee The ground state contribution is
therefore \be \left( \sqrt{y} - \frac1{\sqrt{y}} \right) \times
\prod_{i=1,2,3}  \left(\sqrt{y} e^{i\pi \tilde\theta_i} -
\frac{1}{\sqrt{y}} e^{-i \pi \tilde\theta_i} \right) \ee In
particular, the zeroth and first helicity supertraces vanish,
while \be \Omega_2 = 2 (\sin \pi \tilde \theta(g))^2 \ee where
$\tilde \theta_3=0, \tilde \theta_1 = - \tilde \theta_2 := \tilde
\theta \mod 1$.

Now let us discuss the charge lattice. Suppose that $k$ pairs of
left-moving bosons are fixed for all $g\in \Gamma$. Together with
the 2 right-moving  directions in the plane of $\tilde \theta_3$ we
have a plane $\CQ \subset \IR^{22,6}$ of signature $(2k,2)$. The
vector-multiplet moduli come from the $SO(2k,2)$ rotations in this
plane. The number of $U(1)$ vector fields is $n_V = 2k+2$. The
projection of $\Lambda$ into the plane $\CQ$ defines the charge
lattice (in the untwisted sector) $M_0$. Let $\rho: \Lambda \to
M_0$ be the projection. States in the untwisted sector are
labelled by $P \in II^{22,6}$ but we only want to discuss
degeneracies at a fixed $Q\in M_0$.
Using the BPS condition $P_R^2 = Q_R^2$, we may rewrite:
\be \label{lattsum} \sum_{P\in \Lambda(g)} q^{\half P_L^2} \bar
q^{\half P_R^2} e^{2\pi i \delta(g)P} \Pi_{BPS}(P) = \sum_{Q\in
M_0} q^{\half Q_L^2} \bar q^{\half Q_R^2} \CF_{g,Q}(q)
\ee
where
\be \label{extraeff} \CF_{g,Q}(q) = \sum_{P\in \Lambda(g),
\rho(P)=Q} q^{\half (P_L^2- Q_L^2)} e^{2\pi i \delta(g)P}
\ee
The function \eqref{extraeff}\ is actually very simple in many important cases.
For example if $\Lambda(g) \subset M_0$, which is typical if the fixed
space under the group element $g$ coincides with $\CQ$ then we
simply have $\CF_{g,Q}(q) = e^{2\pi i \delta(g)\cdot Q} $. For
this reason it is useful to distinguish between ``{\it minimal
twists}'', which leave only the subspace $\CQ$ invariant (i.e. $0
< \theta_j(g)<1$ for $j>k$) and nonminimal twists. For nonminimal
twists the kernel of $Q_{el}$ will be nontrivial and
$\CF_{g,Q}(q)$ will be a theta function.

Putting all this together we find that the degeneracies of untwisted sector BPS
states are given by
\be\label{bsnttn}
 \Omega_{\rm w}(Q) =  e^{4\pi Q_R^2} \int d\tau_1\ q^{\half Q_L^2} \bar
q^{\half Q_R^2} \CZ_\omega
\ee
\be\label{andthcz} \CZ_w = {1\over  N} \sum_{g\in \Gamma} {1\over
\eta^{2+2k} }   \Biggl[ \prod_{j=1}^{11-k} (-2\sin \pi
\theta_j(g)) {\eta \over \vt{\half}{\half + \theta_j(g)}{} }
\Biggr] w(g) \CF_{g,Q}(q)
\ee
where $w(g)$ is given by
\be
\label{omegacases}
 w(g) = \begin{cases}
 16 \cos \pi \tilde \theta_1(g)
\cos \pi \tilde \theta_2(g)\cos \pi \tilde \theta_3(g)  & $w=abs$
\\ 2 (\sin \pi \tilde \theta(g))^2 & $w=2$ \\
 {3\over 2} & $w=4$\\
{15\over 8}(2-E_2(\tau)) & $w=6$
\end{cases}
\ee

This formula is exact. Now let us determine its asymptotics.
The general counting function appearing in \eqref{bsnttn} is
\be
\label{kayfunc}
K(\tau) = {1\over \eta^{24-3t}} \prod_{j=1}^t
{1\over \vt{\half}{\half+ r_j/N }{0} }
= q^{-1} \sum_{n\geq 0} K_g(n) q^n
\ee
Together with the functions
\be
\label{bigspace}
  {1\over \eta^{24-3t}} \prod_{j=1}^t
{\half\over \vt{\half+ a_j/N}{\half + b_j/N}{0} }
\ee
with $0 \leq a_j, b_j < N$, the function $K$ transforms as a
matrix of dimension $N\times N$ and modular weight $w=t-12 $
under the congruence subgroup
$\Gamma_0(N,\IZ)$ of $Sl(2,\IZ)$.
In order to apply the Rademacher formula, one  must
diagonalize the $T$ operator in the space spanned by \eqref{bigspace}.
After some computation, we find:
\be
\label{kenns}
K(n)
= \vert \Delta_g\vert^{1-w} e^{-i \pi \sum_j (\half -
  \theta_j)} \BesselI{1-w}{4}{(n-1) \vert \Delta_g\vert }
+ \cdots
\ee
and
\be
\label{deltagee} \Delta_g := -1 + \half \sum_{j=1}^{11-k}
\theta_j(g)(1-\theta_j(g)) , \qquad\qquad 0< \theta_j(g) < 1
\ee
is the ground-state energy in the left-moving sector twisted by $g$.
We only get contributions from $g$ such that $\Delta_g <0$. In
addition, there are non-perturbative corrections of order
\be
\BesselI{1-w}{4}{ (n-1) \vert \Delta_g+ {\ell \over N} \vert }
\ee
for $\ell$ such that  $\Delta_g+ {\ell \over N} <0$,
and of order
\be
\label{conrths}
\BesselI{1-w}{\frac{4}{c}}{(n-1) \vert \Delta_g+ {\ell \over N} \vert  }
\ee
for $c>1$. We conclude that the leading asymptotics for the
degeneracies of untwisted DH states from the minimal twists is ($w\not=6$
here):
\be
\label{asympts}
   {1\over  4N} \sum_{g\in \Gamma, minimal }' w(g) h(g)
\prod_{j=1}^{11-k}(-2\sin \pi \theta_j(g)) \vert
\Delta_g\vert^{k+2}  \hat I_{k+2 }(4\pi \sqrt{\vert
\Delta_g\vert \half Q^2  })
\ee
where
\be\label{hgee} h(g) =
\begin{cases} (-1)^{(12-k)/2} \sin\bigl(2\pi \delta(g)
Q + \pi \sum_j \theta_j(g)\bigr)  & $k$ even \\
(-1)^{(11-k)/2}
\cos\bigl(2\pi \delta(g) Q + \pi \sum_j \theta_j(g)\bigr)  & $k$
odd \end{cases}
\ee
The prime\footnote{The rest of this section is excerpted from
\cite{Dabholkar:2005by}.}
on the sum indicates we only get contributions from
$g$ such that $\Delta_g <0$. For nonminimal twists there will be
similar contributions as described above. In particular the index
on the Bessel function will be the same, but \eqref{deltagee} receives
an extra nonnegative contribution from the shift $\delta$, and the
coefficient $ \vert \Delta_g\vert^{k+2 }$ is modified (and
still positive). In some  examples the leading asymptotics is
provided by the minimal twists alone.

It is interesting to compare this with the twisted sectors. Since
the sector $(1,g)$ always mixes with $(g,1)$ under modular
transformation, and since the oscillator groundstate energy is
$-1$ in the untwisted sector,
 it is clear that for charges $Q$ corresponding to states in the
twisted sector the asymptotics will grow like
\be
\label{twisted} \hat I_{k+2 } (4\pi \sqrt{\half Q^2} )
\ee
This is true both for the absolute number of BPS states and for
the supertraces.
Recall that $k+2 = \half (n_V +2)$ for $\CN=2$ compactifications,
so we have   agreement with \eqref{iresulti}.

There are some interesting general lessons we can draw from
the result \eqref{asympts}.  Due to the factor $h(g)$   it is
possible that the leading $I$-Bessel functions cancel for certain
directions of $Q$. Moreover, a general feature of $\CN=2$
compactifications is that $g=1$ does not contribute to $\Omega_2$
in \eqref{asympts}. Then, since $\vert \Delta_g \vert <1$ the
degeneracies are exponentially smaller in the untwisted sector
compared to those of the twisted sector.
 We have seen explicit examples of this above.
In contrast, for $\CN=4$ compactifications, the $g=1$ term
{\it does} contribute to $\Omega_4$, which thus has the
same growth as in the twisted sector.

One general lesson seems
to be that the degeneracies, and even their leading asymptotics
can be sensitive functions of the ``direction'' of $Q$ in charge
space. In general it is quite possible that    the exact BPS degeneracies
and their asymptotics will be subtle arithmetic functions of the
charge vector $Q$\footnote{Such a phenomenon was conjectured based on
other considerations in \cite{Miller:1999ag}.}.
In the physics literature it is taken for granted that there
is a smooth function $S: H^{\rm even}(\CX,\IR) \to \IR$ so that
$S(sQ ) \sim \log \Omega_w(s Q)$ for $s\to \infty$, but the
true situation might actually be much more subtle.
The Rademacher expansion shows that Fourier coefficients of negative
weight modular forms have well-defined asymptotics governed by
Bessel functions. On the other hand, by contrast,
 the Fourier coefficients $a_n$ of   cusp forms of positive
weight $k$ have a lot of ``scatter'' and
can only be described by a probability distribution
for $a_n/n^{(k-1)/2}$ (see e.g.  \cite{millergelbart} for
an introduction to this subject).
As we have remarked above, certain supertraces {\it do} in fact
have expressions in terms of positive weight forms and we may
expect the asymptotics to be expressed in terms of such probability
distributions.
It would be very interesting to explore further this dichotomy
for the  functions
$\Omega_{\rm w}(Q)$.

\section{The Black Hole Partition Function}

In this section we reconsider the black hole partition function,
starting with what is known about degeneracies of BPS states, and
try to reproduce the structure of the topological string free
energy.

Since BPS degeneracies, even counted with signs, depend on the
background vector multiplet moduli $t^A$ (due to jumps at marginal
stability), one should specify the background to properly define the
partition sum. Furthermore, because the original OSV partition
function \eqref{zosv} (henceforth denoted as $Z_0$) 
%\begin{equation} \label{Zdef}
% Z_0 := \sum_q \Omega(p,q) \, e^{\pi q_i \phi^i}
%\end{equation}
does not converge, a regularization needs to be introduced. We will
consider
\begin{equation} \label{Zalphadef}
 Z_{\alpha} := \sum_q \Omega(p,q) \, e^{\pi q_i \phi^i - \pi \alpha
 H(p,q;t)}.
\end{equation}
As we will see below, a suitable and natural choice for $H(p,q;t)$
is the BPS energy. This introduces additional explicit dependence on
$t$, which formally disappears when $\alpha \to 0$.

For definitiveness we will work in the IIA picture. Since the
topological string wave function is defined as an expansion around
an infinite radius point, a natural guess is that we should take
$\Omega(p,q)$ to count the degeneracies in the corresponding large
radius limit. More precisely, we tentatively define
\begin{equation}
  \Omega_{\infty}(p,q;u):=\lim_{R \to \infty} \Omega(p,q;t=i \, R \, u),
\end{equation}
where $\Omega(p,q;t)$ is an appropriate index counting the number of
BPS states with charge $(p,q)$ on a Calabi-Yau with complexified
K\"ahler form $B+i \omega = t^A J_A$ and $u$ is a fixed real vector
inside the K\"ahler cone. Note that this definition of the
degeneracies still depends on the chosen direction $u$ in the
K\"ahler cone.

For simplicity, we will again mainly consider the case $p^0=0$ in
what follows. In the $R \to \infty$ limit, IIA BPS states are then
described at vanishing string coupling $g_s$ by D4 branes wrapping a
divisor $S$, with D2 and D0 branes dissolved into it. For $r$ D4
branes on a rigid divisor $S$, the moduli space $\CM$ of this system
is the moduli space of semistable rank $r$ coherent sheaves on $S$,
with fixed Chern classes $c_i$.\footnote{In the case $r=1$,
instantons are always pointlike, and $\CM$ is simply the Hilbert
scheme of $N=c_2$ points on $S$. Alternatively, one can turn on a
B-field and consider noncommutative instantons, which are smooth
even if $r=1$.} If the divisor is not rigid, $\CM$ also includes
deformations of $S$. At $g_s=0$, BPS ground states are in one to one
correspondence with cohomology classes on $\CM$. At finite $g_s$,
some of these may be lifted, but the hyper$-$vector index $\Omega_2$
will remain invariant.

\subsection{Rigid Divisors} \label{sec:rigdiv}

We first consider a class of examples for which the counting is
under good control, namely rigid divisors $S$, i.e.\
$h^{2,0}=h^{1,0}=0$, wrapped by a single D4-brane (so $r=1$), with
$N$ D0-branes bound to it. We can always construct at least a
noncompact Calabi-Yau $\CX$ containing $S$, namely the canonical
line bundle over $S$. The simplest example is given by $S=\IP^2$, in
which case $\CX = O(-3) \to \IP^2$. A compact example is given by a
D4 wrapping an Enriques surface $S=K3/\IZ_2$ in the FHSV Calabi-Yau
threefold $\CX = (T^2 \times K3)/\IZ_2$. These branes are dual to
the twisted sector DH states in the dual heterotic model described
in section \ref{sec:FHSV}.

For a rigid divisor $S$, the moduli space $\CM_N$ is simply the
Hilbert scheme of $N$ points on $S$. The number of BPS states
$d_N:=\dim H^*(\CM_N) = \chi(\CM_N)$ is given by the generating
function
\begin{equation} \label{calZdefeta}
 {\cal Z}(q) := q^{-\chi/24} \sum_N d_N q^{N} = \frac{1}{\eta(q)^\chi},
\end{equation}
where $\chi=h^{1,1}(S)+2$ is the Euler characteristic of $S$. We can
also turn on $U(1)$ gauge flux $F$ on $S$, which will induce D2 and
D0 brane charge but will otherwise not affect the moduli space.
Choosing a basis $C_I$ of $H^2(S,\IZ)={\rm Pic}(S)$, which pulls
back to a basis of $H_{\rm cpct}^{1,1}(\CX)$ if $\CX$ is the
canonical line bundle over $S$, we get the following net D2 and D0
brane charges
\begin{eqnarray}
 q_I &=& \int_S J_I \wedge F \\
 q_0 &=& - \left(N - \int_S \frac{1}{2} F \wedge F - \frac{\chi}{24}\right) \\
 &=& - \left(N - \frac{1}{2} C^{IJ} q_I q_J -
 \frac{\chi}{24}\right). \label{q0Nrel}
\end{eqnarray}
Here $C^{IJ}:=(C^{-1})^{IJ}$ with $C_{IJ}:=C_I \cdot C_J$. The
electric charges can in general have nonintegral shifts:
\begin{equation} \label{chargequantshift}
q_I \in \frac{c_{s,I}}{2} + \IZ, \qquad q_0 \in -\frac{c_2(X) \cdot
S}{24} + \IZ.
\end{equation}
The class  $c_s \in H^2(S,\IZ)$  defines a spin$^c$ structure, and
is equivalent, modulo two, to the second Steifel-Whitney class. This
charge quantization law follows from the K-theoretic formulation of
RR charges and is needed to cancel anomalies, both on the brane
worldvolume \cite{Minasian:1997mm} and on the  fundamental string
worldsheet  \cite{Freed:1999vc}. The magnetic charges are given by
the homology class of $S$. The Euler characteristic $\chi(S)$ is
determined in terms of these magnetic charges only:
\begin{equation} \label{eulerS}
 \chi = S^3 + c_2(X) \cdot S.
\end{equation}

Using (\ref{q0Nrel}), we get:
\begin{equation}
 \Omega_\infty(p,q;u) = d_{N=-q_0 + \frac{1}{2} C^{IJ} q_I q_J +
 \frac{\chi}{24}}.
\end{equation}
Note that in this case, the degeneracies are in fact independent of
the choice of $u$. The partition function (\ref{zosv}) becomes
\begin{eqnarray}
 Z_0 &=& \sum_{N,q_I} d_N \, e^{-\pi \phi^0(-N + \frac{1}{2} C^{IJ} q_I q_J + \frac{\chi}{24}) - \pi \phi^I
 q_I} \\
 &=& {\cal Z}(e^{\pi \phi^0}) \, \Theta_0(\phi^0,\phi^I)
\end{eqnarray}
where ${\cal Z}=1/\eta^\chi$ as in (\ref{calZdefeta}) and
\begin{equation}
 \Theta_0(\phi^0,\phi^I) := \sum_{q_I} e^{-\pi \phi^0 \frac{1}{2} C^{IJ} q_I q_J - \pi \phi^I
 q_I}.
\end{equation}
Convergence of ${\cal Z}$ requires ${\rm Re}\, \phi^0 < 0$. On the
other hand $C_{IJ}$ has signature $(1,h^{1,1}-1)$. In particular the
direction $q_I \sim C_{IJ} u^J$ has positive norm squared. Therefore
$\Theta$ is divergent. This signals an instability of the ensemble.

A physically natural way to regularize the partition function is to
add an energy dependent Boltzmann factor as in (\ref{Zalphadef}).
More precisely we will take
\begin{equation}
 H(p,q;u) := \lim_{R \to \infty} \bigl( M(p,q;i \, R \, u) - M(p,0;i \, R \, u)
 \bigr),
\end{equation}
with $M(p,q;t)$ the mass in string units of a BPS state with charges
$p,q$ at the point $t$ in moduli space. We subtracted the $q=0$
energy to get a finite result in the limit $R \to \infty$.
Normalizing $u$ for convenience such that $C_{IJ} u^I u^J := 1$, we
get
\begin{eqnarray}
 H &=& \lim_{R \to \infty} \left( \left|q_0 + i \, R \, q_I u^I - \frac{R^2}{2} \right|
 - \frac{R^2}{2} \right) \\
 &=& -q_0 + (q_I u^I)^2.
\end{eqnarray}
Alternatively we could have obtained this by simply evaluating the
$U(1)$ Yang-Mills action on $S$ coupled to D0-branes, to which the
DBI action reduces in the limit $R \to \infty$.

For $-{\rm Re} \, \alpha < {\rm Re} \, \phi^0 < {\rm Re} \, \alpha$,
the modified partition sum (\ref{Zalphadef}) is convergent:
\begin{equation}
 Z_\alpha = {\cal Z}(e^{\pi (\phi^0 - \alpha)}) \, \Theta_\alpha(\phi^0,\phi^I)
\end{equation}
with
\begin{eqnarray}
 \Theta_\alpha(\phi^0,\phi^I)&:=&\sum_{q_I} e^{-\frac{\pi}{2} g_\alpha^{IJ} q_I q_J - \pi \phi^I
 q_I} \\
 g^{IJ}_\alpha &:=& (\phi^0 - \alpha) C^{IJ} + 2 \, \alpha \, u^I u^J.
\end{eqnarray}
The quadratic form $g^{IJ}_\alpha$ has positive definite real part
in the range of $\phi^0$ specified above. In particular, the
previously problematic direction $q_I = C_{IJ} u^J$ now gives
$g^{IJ}_\alpha q_I q_J = \phi^0 + \alpha$, which has positive real
part. Furthermore, using $\det C^{IJ}=(-1)^{h^{1,1}-1}$, we get
$\det g^{IJ}_\alpha = (\alpha-\phi^0)^{h^{1,1}-1} (\alpha+\phi^0)$
and similarly for $\det {\rm Re}\, g^{IJ}_\alpha$ by replacing the
factors by their real parts. Note that this is indeed positive.

Now that we have a convergent expression, we can perform a Poisson
resummation on $\Theta$:
\begin{equation} \label{Thetaresummed}
 \Theta_\alpha(\phi^0,\phi^I)=2^{\frac{h^{1,1}}{2}}(\alpha-\phi^0)^{-\frac{h^{1,1}-1}{2}}
 (\alpha+\phi^0)^{-\frac{1}{2}}
 \sum_{k^I} e^{\frac{\pi}{2} g_{IJ}^\alpha (\phi^I + 2 i k^I)(\phi^J + 2 i
 k^J) + 2 \pi i k^I \frac{c_{s,I}}{2}},
\end{equation}
with $c_{s,I}$ as in (\ref{chargequantshift}) and
\begin{equation}
 g_{IJ}^\alpha = \frac{1}{\phi^0-\alpha} C_{IJ} + \frac{2 \alpha}{\alpha^2 -
 {\phi^0}^2} u_I u_J
\end{equation}
with $u_I:=C_{IJ} u^J$.

Finally, we do a modular transformation on ${\cal Z}=1/\eta^\chi$:
\begin{equation}
 {\cal Z}(e^{\pi (\phi^0 - \alpha)})=2^{-\frac{\chi}{2}}(\alpha-\phi^0)^{\frac{\chi}{2}}
 {\cal Z}(e^{\frac{4 \pi}{\phi^0-\alpha}}).
\end{equation}
Combining this with (\ref{Thetaresummed}) and using $\chi=h^{1,1}+2$
and the product formula for $\eta$ gives
\begin{eqnarray}
 Z_\alpha&=&\frac{\alpha-\phi^0}{2} \left( \frac{\alpha-\phi^0}{\alpha+\phi^0}
 \right)^{1/2} \prod_n \left(1-e^{\frac{4 \pi
 n}{\phi^0-\alpha}}\right)^{-\chi} \nonumber\\
 &&\times
 \sum_{k_I} \exp\left( - \frac{\pi \, \chi}{6 (\phi^0-\alpha)} +
 \frac{\pi}{2} g_{IJ}^\alpha (\phi^I + 2 i k^I)(\phi^J + 2 i k^J)
 + 2 \pi i k^I \frac{c_{s,I}}{2} \right)  \label{Zalphaformula}
\end{eqnarray}
Inverting (\ref{Zalphadef}), we thus get
\begin{equation}
 \Omega_\infty(p,q) = \int_{-i}^{i} d\phi^0 \int_{-i}^{i} d\phi^I \, e^{\pi q_i \phi^i + \alpha H(p,q;u)} \,
 Z_\alpha(\phi^0,\phi^I;u).
\end{equation}
Note that the sum over $k^I$ in (\ref{Zalphaformula}) can be dropped
by extending the domain of the integrals over $\phi^I$ to
$(-i\infty,+i\infty)$. Furthermore, by definition, the expression is
independent of $\alpha$ (and $u$), so we can take the limit $\alpha
\to 0$, which formally gives
\begin{eqnarray}
 \Omega_\infty(p,q)&=& \int_{-i}^i d\phi^0 \int_{-i\infty}^{i \infty} d\phi^I \,
 f(\phi^0)
  \,
 e^{\pi \bigl( - \frac{S^3 + c_2 \cdot S}{6 \phi^0} +
 \frac{1}{2 \phi^0} C_{IJ} \phi^I \phi^J + q_0 \phi^0 + q_I \phi^I \bigr)}, \label{osvlike}
\end{eqnarray}
where we used (\ref{eulerS}) to express $\chi$ in terms of the
magnetic charge given by $S$, and we defined
\begin{equation} \label{fdef}
 f(\phi^0):=\frac{\phi^0}{2 i} \prod_n \left(1-e^{\frac{4 \pi
 n}{\phi^0}}\right)^{-\chi(S)}.
\end{equation}
The integral (\ref{osvlike}) is somewhat formal, because of the
oscillatory Gaussian integral and the infinite product in
(\ref{fdef}) which is not well behaved on the imaginary axis. From
the above we know however that it is unambiguously defined as the
limit $\alpha \to 0+$ of the same integral with replacements
\begin{equation}
 \phi^0 \to \phi^0 - \alpha, \qquad C_{IJ} \to C_{IJ} - \frac{2 \alpha}{\alpha
 + \phi^0} \, u_I u_J.
\end{equation}

\subsubsection*{Comparison with the topological string}

Comparing to (\ref{cfhptv}), we see that the quantity in the
exponential in (\ref{osvlike}) is exactly the \emph{perturbative}
part of the free energy derived from the topological string
amplitude.\footnote{Note that our basis of charges indeed has a
cubic prepotential, as assumed in (\ref{cfhptv}). In such a basis
the electric charges of D4-brane states will in general have
nonintegral shifts. By substituting $q_0 \to n_0 - (c_2 \cdot S)/24$
and $q_I \to n_I + c_{s,I}/2$ in accordance with the quantization
shifts (\ref{chargequantshift}), we get the free energy in an
integral basis. The additional terms proportional to the charge
shifts correspond to the linear resp.\ quadratic terms in the
prepotential which indeed generally appear for such a basis.} Let us
elaborate a bit on the term quadratic in $\phi$. On the topological
string side, it corresponds to the term $C_{AB} \phi^A \phi^B$ with
$C_{AB}:=C_{ABC} p^C$ and $A,B:1,\ldots,b_2(\CX)$. The integers
$C_{AB}$ give the intersection products of the pullbacks of a basis
of $H^{1,1}(\CX)$ to $S$. Since $h^{2,0}(S)=0$, these pullbacks span
all of $H^2(S)$, and $C_{AB}$ has rank $b_2(S)$. Note that in
general this can be smaller than $b_2(\CX)$. After a suitable change
of $\phi$ variables, we can thus rewrite the term $C_{AB} \phi^A
\phi^B$ as $C_{IJ} \phi^I \phi^J$, with $I=1,\ldots,b_2(S)$ and
$C_{IJ}$ the intersection form on $H^2(S)$ as defined before. The
remaining $\phi^A$ with $A=b_2(S)+1,\ldots,b_2(\CX)$ no longer
appear in the perturbative part of the supergravity free energy, and
the latter thus reduces to the ``$S$-local'' expression in the
exponential in (\ref{osvlike}).

Clearly however, at least for this simple class of wrapped D-branes,
the Gromov-Witten part of the topological string free energy is not
generated by the BPS partition sum. In particular there is no
$\phi^I$ dependence apart from the quadratic term, whereas typically
the Gromov-Witten series is a very complicated function of the
$\phi^I$. The infinite product in $f(\phi^0)$ looks somewhat like
the infinite products appearing in the Gopakumar-Vafa formula \eqref{gopva}
for the topological string wave function
but actually does not seem to have any
obvious interpretation in this context. It depends only on $\phi^0$,
so it would have to come from the homologically trivial worldsheet
sector, which however has a quite different form.

At large $|q_0|$, the integral is well approximated by a saddle
point evaluation, and at the saddle point, $\phi^0$ will be small
and negative, so the infinite product in $f(\phi^0)$ will be
exponentially close to 1. Dropping this factor will therefore merely
give exponentially small deviations from the exact answer. This is
not so however for the additional $\phi^0$ factor, which does not
appear on the topological string side of the conjecture. On the
other hand, we just saw that $\Delta b := b_2(\CX)-b_2(S)$ of the
$\phi^A$ decouple from the perturbative part of the free energy on
the topological string side. Moreover these $\phi^A$ have a natural
periodicity $\phi^0$, so integrating them out would naturally give
an additional factor $(\phi^0)^{\Delta b}$. In the FHSV example with
$S$ an Enriques surface, we have $\Delta b = 1$, hence for large
$q_0$ this procedure leads to complete agreement between microscopic
and perturbative macroscopic answers, up to exponentially suppressed
terms. This agrees with what we found in section \ref{sec:FHSV}, and
will hold similarly for more general $K3$-fibered examples.

However, more generally, it need not be true that $\Delta b = 1$.
One can easily imagine simple divisors $S$ of low $b_2(S)$ embedded
in a Calabi-Yau $\CX$ with large $b_2(\CX)$. In those cases the
discrepancy by a factor of $\phi^0$ cannot be compensated by taking
into account the decoupled integrals. Perhaps a better prescription
would therefore be to simply discard all decoupled integrals,
restricting only to the ``local'' variables, and adding a factor
$\phi^0$ by hand as a universal measure contribution.

It should be noted though that the rigid divisors we are considering
here are not ample (ample divisors have typically many moduli and
always give a nondegenerate $D_{AB}$). This implies that the
attractor point computed from the perturbative part of the
prepotential will \emph{not} lie inside the K\"ahler cone, so there
is a priori no reason whatsoever to expect any agreement between the
microscopic degeneracies and the macroscopic prediction computed
with only the perturbative part of the prepotential. The fact that
(modulo the small issue of the $\phi^0$ factor) there \emph{is}
nevertheless agreement to all orders in $1/|q_0|$ is therefore very
remarkable.

\subsubsection*{A remark on $k$-shifts}

The expression (\ref{Zalphaformula}) for the partition function
contains a sum over shifts labelled by $k$. This gives $Z$ the
required periodicity in $\phi^A$. It is easy to see that this sum
over $k$-shifts will be a general feature of the partition function
if one assumes the integral form of the conjecture,
\begin{equation}
 \Omega(p,q) \sim \int d\phi \, e^{\pi q \cdot \phi + \CF(\phi)},
\end{equation}
where the integrations are over the imaginary axis. Indeed,
substituting this in (\ref{zosv}) gives
\begin{equation}
 Z_0 \sim \int d\phi' \, e^{\CF(\phi')} \sum_q e^{2 \pi i
 q(\frac{\phi-\phi'}{2i})}.
\end{equation}
Assuming $q$ is quantized as $q = \bar{q} + s$ with $\bar{q} \in
\IZ$, and using $\sum_{\bar q} e^{2 \pi i \bar{q} x} = \sum_k
\delta(x-k)$, this gives
\begin{equation}
 Z_0 \sim \sum_{k \in \IZ} \, e^{\CF(\phi + 2 i k) + 2 \pi i s \cdot
 k},
\end{equation}
which is precisely the $k$-shift structure found above. Note however
that (\ref{Zalphaformula}) does not contain a sum over $k^0$. This
is related to the fact that the integrand in (\ref{osvlike}) is
periodic in $\phi^0$ and that the $\phi^0$ integral is over one
period. In principle, by modifying the integrand, one could try to
convert this again to an integral over the entire imaginary axis,
and then the expression of $Z$ derived from this modified integrand
will also contain as sum over $k^0$. In practice, such modifications
do not affect the $1/N$ expansion of (\ref{osvlike}), since this
depends only on the neighborhood of the saddle point.

\subsection{$K3$ Divisors in $K3 \times T^2$}

The rigid examples considered thus far are rather special. In
particular we only considered the rank $r=1$ case. To see if perhaps
we reproduce more of the topological string amplitude in some large
$r$ limit, we apply the same idea to our basic example, $X=K3 \times
T^2$, with $r=p^1$ coincident D4-branes on $S=K3$.

The degeneracies are now given by
\begin{equation}
 \Omega_{\infty}(p,q_0,q_1,\vec q;u) = \delta_{q_1,0} \, p_{24}(N), \qquad N = 1 - q_0 p^1 + \frac{1}{2}\vec
 q^2.
\end{equation}
where $\vec q \in II^{19,3}$. The factor $\delta_{q_1,0}$ arises
because in the $R \to \infty$ limit, there are no bound states of
D-branes with 6 mutually Dirichlet-Neumann directions. This also
fits with the fact that there is no attractor point when $q_1 \neq
0$. At finite $R$ and sufficiently large $\vec q$ or $B$-field, BPS
states with nonzero $q_1$ may appear
\cite{Brunner:1999jq,Douglas:2000ah,Douglas:2000gi,Witten:2000mf}.
The supergravity solutions corresponding to those states will be
multicentered, decaying at some point when $R \to \infty$
\cite{Denef:2000nb,Denef:2002ru,Bates:2003vx}. Since we are
considering the strict limit $R \to \infty$ here, we do not need to
consider these.

The computation of the partition function is similar to the previous
subsection. There is one new element: in solving the level matching
condition for $q_0$, we must ensure integrality of
$q_0=-\frac{1}{p^1}(N-1-\vec q^2/2)$. This is easily achieved by
inserting a projector:
\begin{equation}\label{zoent}
 Z_0 = \sum_{N,\vec{q}} {1\over p^1} \sum_{k^0=0}^{p^1-1} e^{-2\pi i {k^0\over
p^1} (N-1-\vec q^2/2)} \, p_{24}(N) \, e^{\pi \frac{\phi^0}{p^1}(N-1
- \vec q^2/2) - \pi \vec\phi \cdot \vec q}.
\end{equation}
As before, this sum is divergent, but can again be
regularized.\footnote{Because the D2 charge lattice now has 3
positive norm squared directions, spanned by $(\omega,{\rm Re}\,
\Omega,{\rm Im}\, \Omega)$, where $\omega$ is the K\"ahler form and
$\Omega$ the holomorphic 2-form on K3, the regularization will
involve $\Omega$ as well as $\omega$. This is special to cases with
$\CN=4$ supersymmetry.} We will not do this in detail, but use its
existence as justification for the formal manipulations in the
following. Carrying out the sum over $N$, we get
\begin{equation}\label{zient}
 Z_0 ={1\over p^1} \sum_{k^0=0}^{p^1-1} \eta^{-24}(e^{\pi (\phi^0-2 i k^0)/p^1})
 \sum_{\vec q}
e^{ -\frac{\pi}{p^1}(\phi^0-2 i k^0) \vec q^2 - \pi \vec \phi \cdot
\vec q}.
\end{equation}
Finally, performing a modular transformation on the Dedekind
function and a Poisson resummation over $\vec q$, we obtain
the main formula of this subsection:
\begin{equation}\label{thingens}
Z_0 = {1\over 2 i (p^1)^2} \sum_{k^0=0}^{p^1-1} (\phi^0 - 2 i k^0)
\sum_{\vec k \in II^{3,19} } \exp \biggl(\frac{\pi}{2} \frac{p^1
(\vec \phi - 2 i \vec k)^2} {\phi^0-2 i k^0}
 -\log \eta^{24} (e^{\frac{4 \pi p^1}{\phi^0-2 i k^0}}) \biggr)\ .
\end{equation}
This is very similar to what we found in section \ref{sec:rigdiv},
with the addition of a finite sum over shifts of $\phi^0$. We can
also write this in integral form:
\begin{equation} \label{exactforK3}
 \Omega(p,q) = {1\over 2 i (p^1)^2} \int_{-i p^1}^{i p^1} d\phi^0 \int_{-i\infty}^{i \infty}
 d\vec{\phi} \,\, \phi^0 \exp \biggl( \frac{\pi}{2} \frac{p^1
 \vec \phi^2} {\phi^0} -\log \eta^{24} (e^{\frac{4 \pi
 p^1}{\phi^0}}) + q_0 \phi^0 + \vec q \cdot \vec\phi \biggr),
\end{equation}
which should be compared to the conjectured
\begin{equation} \label{conjectforK3}
 \Omega(p,q) \qeq  \int d\phi^0 d\phi^1 d\vec{\phi} \,
 \exp \biggl( \frac{\pi}{2} \frac{p^1 \vec \phi^2} {\phi^0}
 -\log \eta^{24} (e^{\frac{2 \pi (p^1+i\phi^1)}{\phi^0}})
 -\log \eta^{24} (e^{\frac{2 \pi (p^1-i\phi^1)}{\phi^0}})
 + q_0 \phi^0 + \vec q \cdot \vec\phi \biggr).
\end{equation}
This is similar to the exact expression (\ref{exactforK3}), but
clearly not quite the same. Working formally, we can fourier expand
the $1/\eta^{24}$ functions in (\ref{conjectforK3}) and integrate
$\phi^1$ over $(0, \phi^0)$. This gives
\begin{equation}
 \Omega(p,q) \qeq \int d\phi^0 d\vec{\phi} \,\,
 \phi^0
 \exp \biggl( \frac{\pi}{2} \frac{p^1 \vec \phi^2} {\phi^0}
 + \log \sum_n \bigl( p_{24}(n) \bigr)^2 e^{\frac{4 \pi p^1}{\phi^0}(n-1)}
 + q_0 \phi^0 + \vec q \cdot \vec\phi \biggr).
\end{equation}
Unfortunately, this differs from (\ref{exactforK3}) in that
$p_{24}(N)$ appears squared here, but not so in the expansion of the
$1/\eta^{24}$ in (\ref{exactforK3}).

In conclusion, at least for the case of $K3$, we see that
considering arbitrary rank still does not fully reproduce the
topological amplitude.

\subsection{General Case and Monodromy Invariance}

Despite the arbitrary rank, the K3 case is still somewhat
degenerate, insofar as it does not correspond to a regular,
``large'' black hole. Unfortunately, exact counting of microstates
of general D4-D2-D0 systems is considerably harder than the cases we
considered so far. However, some information about the form of the
partition function can be obtained purely from monodromy invariance,
where the monodromy under consideration is around large radius,
i.e.\ integral shifts of the B-field.

To make this precise, let us first review the general relation of
electric and magnetic charges to microscopic quantities. The
magnetic charge of $r$ coincident D4-branes wrapped around a divisor
$S=m^A J_A$ is $p^A=r m^A$. The D0 and D2 electric charges $q_0$ and
$q_A$ corresponding to a rank $r$ coherent sheaf with Chern classes
$c_i$ are given by
\begin{eqnarray}
 q_A &=& \int_S \iota^* J_A \wedge (c_1 + \frac{1}{2} \, r \, c_s) \label{qAform}\\
 q_0 &=& -\biggl( \frac{\Delta}{2r} - \int_S \frac{1}{2 r} (c_1+ \frac{1}{2} \, r \, c_s)^2
 - \frac{1}{24} \, r \, \chi(S) \biggr) \label{inducedq0} \\
 {\rm with~} \Delta &:=& \int_S 2 r c_2 + (1 - r) c_1^2.
\end{eqnarray}
Here $c_s$ is again the Chern class of a spin$^c$ structure on $S$,
as discussed earlier, $\chi(S)=S^3 + c_2(X) \cdot S$ is the Euler
characteristic of $S$, $\iota^*$ is the pullback map to $S$, and
$\Delta$ is the Bogomolov discriminant \cite{Bogomolov,Friedman}.
For semistable sheaves $\Delta \geq 0$. When $\Delta$ is
sufficiently large, the dimension of the sheaf moduli space is
$d=\Delta - (r^2-1) \chi({\cal O}_S)$. As before, the electric
charges as defined above in general may have nonintegral shifts.
More precisely
\begin{equation} \label{Qquantization}
 q_A \in \frac{r \, c_{s,A}}{2} + \IZ, \qquad q_0 \in -\frac{r \, c_2(X) \cdot
 S}{24} + \IZ.
\end{equation}

%If $S$ is very ample, the pullback map $H^2(X) \to H^2(S)$ is
%injective because of the Lefshetz hyperplane theorem. If the
%pullback map is not injective (e.g.\ for $S=K3$ in $X=T^2 \times
%K3$), we choose a basis $\{ J_A \}=\{ J_I \} \cup \{ J_{I'} \}$ of
%$H^2(X)$ such that the kernel of the pullback map is spanned by $\{
%J_{I'} \}$. For the system under consideration, we then have
%$q_{I'}=0$, and the nonzero D2 electric charges are labelled by
%$q_I$, $I=1,\ldots,n$. The number $n$ is thus the rank of the
%pullback of $H^2(X)$ to $S$. The induced intersection form is
%$C_{IJ}:= \int_S J_I \wedge J_K = m^A C_{AIJ}$, where $C_{ABC}:=J_A
%\cdot J_B \cdot J_C$. The form $C_{IJ}$ is nondegenerate. We can
%rewrite (\ref{inducedq0}) as
%\begin{equation} \label{inducedq0bis}
%  q_0 = -\biggl( \frac{\Delta}{2r} - \frac{1}{2 r} C^{IJ} q_I q_J
% - \frac{1}{24} \, r \, \chi(S) \biggr),
%\end{equation}
%where $C^{IJ} := (C^{-1})^{IJ}$. Note that because $\chi(S) = S^3 +
%c_2(X)\cdot S$ is cubic in $S$, a single smooth D4-brane in the
%class $r S$ does \emph{not} have the same D0-charge as $r$ D4 branes
%in the class $S$, unless $S^3=0$.

One universal feature of the D-brane moduli space $\CM$ in the limit
$R \to \infty$ is that it is invariant under monodromy of the
charges around large radius. These monodromies can be thought of as
induced by shifts $B \to B + n^A J_A$, $n^A \in \IZ$. At the level
of sheaves, this corresponds to tensoring with a line bundle, which
maps
 $$
 c_1 \to c_1 + r n^A \iota^* J_A
 $$
and leaves $\Delta$ invariant. The $\omega$-stability condition for
sheaves is that every subsheaf of rank $r'$ and first Chern class
$c_1'$ must satisfy $c_1' \cdot \omega / r' < c_1 \cdot \omega / r$
(with $\omega$ the K\"ahler form), so monodromy does not affect this
condition and the BPS spectrum is preserved.\footnote{This is only
true for physical BPS states when $R=\infty$. At finite $R$,
$\Pi$-stability is the proper physical criterion rather than
$\omega$-stability \cite{Douglas:2000ah,Douglas:2000gi}. For any
arbitrarily large but fixed $R$, $\Pi$-stability becomes
qualitatively different from $\omega$-stability for sufficiently
large charges $q_A$.} The monodromy action on the charges is
\begin{equation}
 q_0 \to q_0 + q_A n^A + \frac{r}{2} C_{AB} n^A n^B, \qquad q_A \to q_A + r C_{AB}
 n^B,
\end{equation}
where $C_{AB}:=C_{ABC} m^C$. The shift by $c_s$ in (\ref{qAform}) is
precisely such that the change in $q_0$ is guaranteed to be
integral. Invariance of the degeneracies $\Omega_{\infty}(p,q;u)$
under this transformation implies that they will only depend on the
monodromy invariant $\Delta$ and a label $s$ in a finite set giving
the value of the D2 charge modulo monodromies. The number of
monodromy inequivalent classes grows with $P$. Nevertheless, at any
finite value of $P$, monodromy invariance constrains the
$\phi^A$-dependence of the partition sum to be given by a finite sum
of theta functions.

To see this, let us consider the unregularized partition function
(\ref{zosv}) and work formally (this can again be regularized and
justified as before). We write $q_A = s_A + r C_{AB} n^B$, where
$s_A$ parametrizes D2-charges modulo monodromies and $n^B \in \IZ$.
Assuming $C_{AB}$ is nondegenerate,\footnote{This is guaranteed if
$S$ is very ample. In other cases, such as the $K3$ example studied
above, it may happen that the $q_A$ induced on $S$ take values in a
linear subspace of the full charge space (because $\iota^*:H^2(X)
\to H^2(S)$ fails to be injective), so the quadratic form $C_{AB}$
will be degenerate. In such cases, we can restrict to that linear
subspace, generically the restricted quadratic form will be
invertible, and essentially all of what follows goes through. If the
restricted quadratic form is still degenerate, there will be an
infinite number of monodromy inequivalent classes, and the
discussion needs to be changed somewhat.} the $s_A$ take values in a
\emph{finite} set ${\cal Q}$ of order $|\det r C_{AB}|$.
Correspondingly, we decompose the partition sum as
\begin{eqnarray}
 Z_0 &=& \sum_{q_0,s_A,n^A} \Omega_{\infty}(p,q_0,s_A+r C_{AB}n^B;u)\,
 e^{-\pi (\phi^0 q_0 + \phi^A s_A  + \phi^A r C_{AB} n^B ) }
\end{eqnarray}
Using monodromy invariance and shifting $q_0$ then gives
\begin{eqnarray}
 Z_0&=& \sum_{q_0,s_A,n^A} \Omega_{\infty}(p,q_0-s_A n^A - \frac{r}{2} C_{AB}
 n^A n^B,s_A;u)\,
 e^{-\pi [\phi^0 q_0 + \phi^A s_A  + \phi^A r C_{AB} n^B ] } \\
 &=& \sum_{q_0',s_A,n^A} \Omega_{\infty}(p,q_0',s_A;u)\,
 e^{-\pi [\phi^0 q_0' + \phi^0(s_A n^A + \frac{r}{2} C_{AB} n^A n^B) + \phi^A s_A  + \phi^A r C_{AB} n^B]
 }\\
 &=& \sum_{s \in {\cal Q}} {\cal Z}_s(\phi^0) \, \Theta_s(\phi^0,\phi^A).
\end{eqnarray}
In the last line we defined
\begin{eqnarray}
 {\cal Z}_s(\phi^0) &:=& \sum_{q_0} \Omega_{\infty}(p,q_0,s_A;u) \, e^{-\pi
 \phi^0 q_0} \label{calZdef} \\
 \Theta_s(\phi^0,\phi^A) &:=& \sum_{n^I}
 e^{-\pi [\frac{\phi^0 r}{2} C_{AB} n^A n^B  + n^A( s_A \phi^0 + r C_{AB} \phi^B) + \phi^A
 s_A]}.
 \label{thetasdef}
\end{eqnarray}
Thus we see that the $\phi^A$ dependence of the partition sum is
given by a \emph{finite} sum of theta functions $\Theta_s$. After a
modular transformation, this could be brought in a form analogous to
(\ref{thingens}), but in any case, the $\phi^A$ dependence will
still be given by a finite sum of theta functions. Brought in
integral form, analogous to (\ref{exactforK3}), this will give a
finite sum of \emph{Gaussian} functions in the $\phi^A$.

This should be compared to the $\phi^A$ dependence of the
topological string amplitude squared, which is given by an intricate
series of instanton corrections determined by a typically infinite
set of Gromov-Witten invariants, or by an infinite product
determined by the likewise infinite set of BPS invariants. It is of
course very unlikely that this will in general match a finite set of
Gaussian functions.

However, when $P = r S \to \infty$, the number of these Gaussian
terms goes to infinity. Therefore, this result does not contradict
the weaker form of the conjecture, i.e.\ asymptotically for $P \to
\infty$.

Clearly, more results on exact BPS degeneracies of D-brane systems
corresponding to large black holes would be very useful to make
further progress using the approach of this section.

\section{Conclusion}

In this work, we have studied the detailed degeneracies of small black holes,
using their dual description as perturbative heterotic BPS states. The
comparison with the macroscopic Bekenstein-Hawking-Wald entropy including
the leading $R^2$ corrections, and assuming a mixed statistical ensemble,
shows a remarkable agreement to all orders in an asymptotic expansion
in inverse charges, in a large set of models with $\CN=2$ and $\CN=4$
supersymmetry. At the same time, we found apparent discrepancies in
special models, where however the macroscopic computation is not under
good control since the moduli are attracted to the boundary of the
K\"ahler cone. It would be very interesting to generalize our analysis
to the case of ``large'' black holes, with non-vanishing entropy at
tree-level, where these effects do not occur. This would require
improving our understanding of the effective conformal field theory
which describes the micro-states. It would also
be interesting to understand the relation with other approaches
which postulate a statistical ensemble \cite{Sen:2004dp,Sen:2005pu,
Sen:2005ch,Cardoso:2004xf}, or more drastically trade the singular
black hole geometry with a sum over smooth geometries \cite{Mathur:2005zp}.

\vskip1cm

{\noindent \it Acknowledgments :} It is a pleasure to thank E.
Diaconescu, B. Florea, J. Kappeli, E. Kiritsis, C. Kounnas, Y. Oz,
A. Sen, A. Strominger, C. Vafa, E. Verlinde and B. de Wit for
valuable discussions. BP is grateful to the the Theory Group 
at Rutgers University for hospitality during part of this work.
The work of FD and GM is supported in part by
DOE grant DE-FG02-96ER40949. The work of BP is supported in part by
the European Research Network MRTN-CT-2004-512194.

\newpage

\appendix
\vskip 1cm

\centerline{\bf \large Appendices}

\section{The Rademacher Expansion}\label{rademacher}

Here we state briefly the Rademacher expansion. For more details
and information see \cite{Dijkgraaf:2000fq}.

Suppose  we have a ``vector-valued nearly holomorphic modular
form,''  i.e.,  a collection of functions $f_\mu(\tau)$ which form
a finite-dimensional unitary representation of the modular group
of weight $w<0$. Under the standard generators we have
\bea\label{rep}
f_\mu(\tau+1) & = &e^{2\pi i \Delta_\mu} f_\mu(\tau)\\
f_\mu(-1/\tau) & = &(-i \tau)^w S_{\mu\nu} f_\nu(\tau)
\eea
We assume the $f_\mu(\tau)$ have
 no singularities for $\tau$ in the upper half plane,
except at the cusps $\IQ \cup i \infty$. We may assume they have
an absolutely convergent Fourier expansion
\be
\label{collec}
f_\mu(\tau) = q^{\Delta_\mu} \sum_{m \geq 0} F_\mu(m) q^m \qquad
\mu = 1, \dots, r
\ee
with $F_\mu(0)\not=0$ and that the
$\Delta_\mu$ are real.  We wish to give a formula for the Fourier
coefficients $F_\mu(m)$.

Define:
\be
\label{integrl} \hat I_\nu(z)
=  -i (2\pi)^{\nu}  \int_{\epsilon-i\infty}^{\epsilon+i\infty} t^{-\nu-1}
 e^{(t + z^2/(4t))} dt = 2\pi ({z \over  4\pi} )^{-\nu} I_\nu(z)
\ee for $Re(\nu)>0, \epsilon>0$,
where $I_\nu(z)$ is the standard modified Bessel
function of the first kind.

Then we have:
 \bea\label{radi} F_\nu(n) &= & \sum_{c=1}^\infty\sum_{\mu=1}^{r}
  c^{w-2} K\ell(n,\nu,m,\mu;c)
\sum_{m+ \Delta_\mu < 0} F_\mu(m) \\
&&\vert m+\Delta_\mu
\vert^{1-w}  \hat I_{1-w}
 \biggl[ {4\pi\over c} \sqrt{\vert m+\Delta_\mu\vert(n + \Delta_\nu)}
\biggr] .
\eea
The coefficients $ K\ell(n,\nu,m,\mu;c)$ are
generalized Kloosterman sums, defined as
\be
\label{kloos}
{\rm Kl}(n,\nu;m,\mu;c):=
\sum_{0<d<c; d\wedge c=1}
e^{2\pi i \frac{d}{c} (n+\Delta_\nu)} M(\gamma_{c,d})^{-1}_{\nu\mu}
e^{2\pi i \frac{a}{c} (m+\Delta_\mu)}
\ee
where
\be
\gamma_{c,d} = \begin{pmatrix} a & (ad-1)/c \\ c & d \end{pmatrix}
\ee
is an element of $Sl(2,\IZ)$ and $M(\gamma)$ its matrix representation.
For $c=1$ in particular, we have:
\be\label{radiii}
K\ell(n,\nu,m,\mu;c=1)   = S^{-1}_{\nu\mu}
\ee
The series \eqref{radi} is
convergent. Moreover the asymptotics of $I_\nu$ for large $Re(z)$
is given by
\be\label{inuas}
 I_\nu(z) \sim \frac{e^z}{\sqrt{2\pi z}}\left[ 1- \frac{(\mu
    -1)}{8z} + \frac{(\mu
    -1)(\mu -3^2)}{2!(8z)^2} - \frac{(\mu
    -1)(\mu -3^2)(\mu -5^2)}{3!(8z)^3}+ \ldots \right],
\ee
where $\mu = 4 \nu^2$.

\section{Modular Cornucopia \label{cornu}}
In this section, we collect definitions and useful identities of
modular forms. The Jacobi theta function is defined by\footnote{This differs from the definition in  \cite{Kiritsis:1997gu}
by a factor of 2 in the characteristics.}
\be
\th[^a_b](v|\t)=\sum_{n\in \IZ}q^{{1\over 2}\left(n-a\right)^2}
e^{2\pi i\left(v-b\right)\left(n-a\right)}
\,,\label{t1}\ee
where $a,b$ are real and $q=e^{2\pi i\t}$. It satisfies the modular properties
\bea
\th[^a_b](v|\t+1)&=&e^{-i\pi a(a-1)}\th[^a_{a+b-\frac12}](v|\t) \\
\th[^a_b]\left(\frac{v}{\t}|-\frac{1}{\t}\right)&=&
e^{2i\pi ab + i \pi \frac{v^2}{\t}}
\th[^a_{b}](v|\t)
\eea
The Jacobi-Erderlyi theta functions are the values at half periods,
\be
\th_1(z|\tau)=\th[^\half_\half](z|\t),\quad
\th_2(z|\tau)=\th[^\half_0](z|\t),\quad
\th_3(z|\tau)=\th[^0_0](z|\t),\quad
\th_4(z|\tau)=\th[^0_\half](z|\t)
\ee
In particular,
\be
\label{modt1}
\theta_1(v/\tau,-1/\tau) = i \sqrt{-i \tau}  e^{i\pi  v^2/\tau}
\theta_1(v,\tau)
\ee
The Dedekind $\eta$ function is defined as
\be
\eta(\t)=q^{1\over 24}\prod_{n=1}^{\infty}(1-q^n)
\,.\label{t10}
\ee
It satisfies the modular property
\be
\label{modde}
\eta\left( - \frac{1}{\t} \right) = \sqrt{-i\t} \eta(\tau)
\ee
It is related to the Jacobi-Erderlyi theta functions by the identities
\bea
{\partial\over \partial v}\th_1(v)|_{v=0} &= & 2\pi{}~\eta^3(\t)
\label{t11} \\
\th_2(0|\t)\th_3(0|\t)\th_4(0|\t)&=&2\eta^3
\label{t13}
\eea
The Riemann identity allows to carry out sums over spin structures,
\be
{1\over
2}\sum_{a,b=0}^1~(-1)^{a+b+ab}~\prod_{i=1}^4
\theta[^{\frac{a}{2}}_{\frac{b}{2}}](v_i)
=-\prod_{i=1}^4
{}~\theta_1(v_i')
\,,\label{tt14}\ee
where
\be
v_1'={1\over 2}(-v_1+v_2+v_3+v_4)\;\;\;,\;\;\;v_2'={1\over
2}(v_1-v_2+v_3+v_4) \,,
\ee
\be
v_3'={1\over 2}(v_1+v_2-v_3+v_4)\;\;\;,\;\;\;v_4'={1\over
2}(v_1+v_2+v_3-v_4)\,.
\ee
A generalized form holds provided $\sum_i h_i=\sum_i g_i=0$:
\be
{1\over 2}\sum_{a,b=0}^1(-1)^{a+b+ab}\prod_{i=1}^4
\theta[^{\frac{a+h_i}{2}}_{\frac{b+g_i}{2}}](v_i)=-\prod_{i=1}^4
\theta[^{\frac{1-h_i}{2}}_{\frac{1-g_i}{2}}](v_i')
\,.\label{t15}\ee
The Jacobi and Dedekind function satisfy the following ``doubling identities'':
\begin{subequations}
\label{doubling}
\be
\theta_2(\tau)= {2 [\eta(2\tau)]^2 \over \eta(\tau)}
\ ,\quad
\theta_3(\tau)= e^{\frac{i\pi}{12}} {[\eta({\tau+1\over 2})]^2
\over \eta(\tau)} \ ,\quad
\theta_4(\tau)= {[\eta({\tau\over 2})]^2 \over \eta(\tau)}
\ee
\begin{equation}
\th_2(2\tau)=\frac{1}{\sqrt{2}}\sqrt{\th_3^2(\tau)-\th_4^2(\tau)}\ ,\quad
\th_3(2\tau)=\frac{1}{\sqrt{2}}\sqrt{\th_3^2(\tau)+\th_4^2(\tau)}
\end{equation}
\begin{equation}
\th_4(2\tau)=\sqrt{\th_3(\tau)\th_4(\tau)}
\ ,\quad
\eta(2\tau)=2^{-2/3}\th_2^{2/3}(\tau)(\th_3(\tau)\th_4(\tau))^{1/6}
\end{equation}
\begin{equation}
\th_2(\tau/2)=\sqrt{2\th_2(\tau)\th_3(\tau)}\ ,\quad
\th_3(\tau/2)=\sqrt{\th_3^2(\tau)+\th_2^2(\tau)}
\end{equation}
\begin{equation}
\th_4(\tau/2)=\sqrt{\th_3^2(\tau)-\th_2^2(\tau)}\ ,\quad
\eta(\tau/2)=2^{-1/6}\th_4^{2/3}(\tau)(\th_2(\tau)\th_3(\tau))^{1/6}
\end{equation}
\begin{equation}
\th_2\left(\frac{\tau+1}{2}\right)=e^{\frac{i\pi}{ 8}}
\sqrt{2\th_2(\tau)\th_4(\tau)}\ ,\quad
\th_3\left(\frac{\tau+1}{2}\right)=\sqrt{\th_4^2(\tau)+i\th_2^2(\tau)}
\end{equation}
\begin{equation}
\th_4\left(\frac{\tau+1}{2}\right)=\sqrt{\th_4^2(\tau)-i\th_2^2(\tau)}\ ,\quad
\eta\left( \frac{\tau+1}{2}\right)=2^{-1/6}~e^{\frac{i\pi}{24}}~\th_3^{2/3}(\tau)(\th_2(\tau)\th_4(\tau))^{1/6}
\end{equation}
\begin{equation}
\eta(2\tau)~
\eta(\tau/2)~
\eta\left( (\tau+1)/2 \right)
=e^{-i \pi/24} \eta^3(\tau) \ .
\end{equation}
\end{subequations}
Another convenient set of modular forms are the Eisenstein series,
\be
E_2={12\over i\pi}\partial_{\tau}\log\eta=1-24\sum_{n=1}^{\infty}
{nq^n\over 1-q^n}
\,,\label{E2}\ee
\be
E_4={1\over 2}\left(\vartheta_2^8+\vartheta_3^8+\vartheta_4^8\right)
=1+240\sum_{n=1}^{\infty}{n^3q^n\over 1-q^n}
\,,\label{E3}\ee
\be
E_6={1\over 2}\left(\vartheta_2^4+\vartheta_3^4\right)
\left(\vartheta_3^4+\vartheta_4^4\right)
\left(\vartheta_4^4-\vartheta_2^4\right)=
1-504\sum_{n=1}^{\infty}{n^5q^n\over 1-q^n}
\,.\label{E4}
\ee
$E_4$ and $E_6$ have modular weight 4 and 6, and
generate the ring of modular forms under $Sl(2,\IZ)$. $E_2$ is not
a proper modular form as it transforms
inhomogeneously under the modular group.

It is also useful to define the following function
\be
\xi(v)=\prod_{n=1}^{\infty}{(1-q^n)^2\over (1-q^ne^{2\pi iv})
(1-q^ne^{-2\pi iv})}={\sin\pi v\over \pi}{\vartheta_1'\over
\vartheta_1(v)}
\,.\label{E19}\ee
which often appears in generating functions of helicity supertraces.
Its first $v$-derivatives at $v=0$ are
\be
\xi(0)=1\ ,\quad \xi'(0)=0\ ,\quad
\xi^{(2)}(0)=-{\pi^2\over 3}(1-E_2)\ .
\label{E21}
\ee

\section{Counting $J=0$ DH States in the (4,24) Model \label{appj}}

Although there do not exist regular BPS spherically symmetric
spinning black hole
solutions of the tree-level supergravity, heterotic DH states in general
may carry angular momentum $J$. It is conceivable that these states
  correspond to multi-centered black holes, or require the inclusion
of higher derivative corrections.
In this section, we examine the degeneracies of DH states
in Het/$T^6$ with a prescribed value of the angular momentum $J$,
and show that the restriction to DH states with $J=0$
leads to different subleading corrections for the entropy as
compared to the case where all values of the angular momentum
are summed over. This suggests that the statistical ensemble
implicit in the Bekenstein-Hawking-Wald entropy allows
for arbitrary fluctuations of the angular momentum,
at vanishing potential $\Omega$ conjugate to $J$.

Let us start by recalling that the angular momentum of DH states
arises from bosonic and fermionic oscillators in the two non-compact
coordinates transverse to the light-cone. Right-moving oscillators
map one state to another in the same supersymmetry multiplet (unless
they break the BPS property), so the angular momentum of the highest
weight state of a given multiplet arises from left-movers only.
Introducing a parameter $v$ conjugate to the left-moving helicity $J_3^L$ of
the highest weight, the partition function of DH states is given by
\be
\label{b4qv}
\Omega_4(v,q) = \Tr[ (J_3^R)^4  e^{i\pi v J_3^L} q^{L_0} {\bar q}^{\bar L_0} ]
= \frac{3\sin\pi v} {\eta^{21}(\tau) \theta_1(v;\tau) }
\ee
Using $\theta_1'(0)=2\pi\eta^3$, this reproduces
\eqref{b4622} when $v=0$.
The right-hand side of this equation may be
viewed \eqref{b4qv} as the character of the trivial
representation of affine $Sl(2)_k$, and decomposed into
contributions of fixed $U(1)$ charge
using a generalization of the Kac-Peterson formula,
\be
\chi_{Sl(2)_k}^0 = \frac{2q^{1/8} \sin \pi v}{\theta_1(v,\tau)}
= \sum_{m=-\infty}^{\infty} e^{2\pi i m v} q^{-\frac{m^2}{k}}
\hat c_{m}^{j=0} (\tau)
\ee
The $Sl(2,\IR)$ level $k$ string functions $\hat c_{m}^{j=0}$
have been computed in \cite{Bakas:1991fs,Distler:1991wr} and read
\be
\label{sfsl2}
\hat c_m^{j=0} = \frac{q^{|m|+\frac{m^2}{k}}}{q^{-1/8} \eta^3}
\left( 1 + (1+q^{|m|}) \sum_{n=1}^{\infty} (-1)^n q^{\frac12
[n^2 + (2|m|+1) n - 2 |m|]} \right)
\ee
(Notice that the level $k$ does not affect the spectrum, except for an
overall shift.)
This allows us to extract the partition function of states of given
left-moving helicity $m=h_L>0$,
\be
Z_{hel}(m,q) = \frac{3}{\eta^{21}} q^{-\frac18 - \frac{m^2}{k}
\hat c_m^{j=0} }
=\frac{3}{2\eta^{24}}
\left( q^m + (1+q^{m}) \sum_{n=1}^{\infty} (-1)^n q^{\frac12
[n^2 + (2m+1) n ]} \right)
\ee
Since each multiplet of spin $J$
contributes $2J+1$ states with $m$ ranging from $-J$ to $J$, one can
obtain the partition function of given angular momentum $J$ by
\be
\label{kjss}
Z_{spin}(J,q) = Z_{hel}(J,q) - Z_{hel}(J+1,q)
\ee
Using \eqref{sfsl2}, this may be rewritten as
\be
Z_{spin}(J,q)=\frac{3}{2\eta^{24}}
\left( 1 + q^J + (2+q^{J}+q^{-(J+1)})
\sum_{n=1}^{\infty} (-1)^n q^{\frac12
[n^2 + (2J+1) n ]} \right)
\ee
In particular, for $J=0$, we find $
Z_{spin}(0,q) = \frac{1}{\eta^{24}} \cdot S_0(q)$
where
\be
S_0(q) = 1 -3 q + q^2 + 3 q^3 - q^5 - 3 q^6 +  \dots
= 2 + (1+3q) \sum_{n=1}^{\infty} (-1)^{n} q^{ \frac12 n(n+1) -1}
\ee
Working out \eqref{kjss} at low levels, we obtain (up to an
overall factor of 3/2)
\be
\begin{array}{llllllllllll}
J=0 &: q^{-1}& + &21& +& 253 q& +& 2255 q^2 &+& 16446 q^3 &+& ...\\
J=1 &:    &&  1 &+& 22  q &+& 276  q^2 &+& 2552 q^3 &+& ...\\
J=2 &:   &&&&     q &+&  22  q^2 &+& 277  q^3 &+& ...\\
J=3 &:  &&&&&&           q^2 &+&  22  q^3 &+& ...
\end{array}
\ee
reproducing the total partition function,
\be
Z(q) = \sum_{J=0}^{\infty} (2J+1) Z_{spin}(J,q) = \frac32 \left(
\frac{1}{q}+ 24 + 324 q + 3200 q^2 + 25650 q^3 + ...\right)
\ee
(notice that the degeneracy of each Regge trajectory stabilize to a constant
as the excitation level becomes large, 1,22,277,2576,19574,... )

Let us now extract the asymptotics of the degeneracies $\Omega(J;N)$.
Although the string functions have modular weight $-1/2$, their
behavior under modular transformations is ill-understood,
so that the Rademacher formula does not apply directly.
Relatedly, the partition function \eqref{b4qv} is not a weak Jacobi
form. Nevertheless, we may try and obtain
the leading asymptotics by saddle point
methods\footnote{Degeneracies of strings with prescribed angular momentum
were studied in \cite{Russo:1994ev}, for a different scaling of the
charges.}. Using \eqref{b4qv} and  \eqref{kjss}, we have
\be
\Omega_{spin}(N,J) = 4i \int_{iL-\frac12}^{iL+\frac12} d\tau \int_0^1 dv
\ e^{-2\pi i (N-1)\tau + 2i\pi (J+\frac12) v}
\frac{\sin^2\pi v}{\eta^{21}(\tau)\ \theta_1(v,\tau)}
\ee
In this expression, the range of the $\tau$ integration is chosen
such that it corresponds to a small circle around the origin in the
$q=e^{2\pi i\tau}$ variable. Using the modular properties \eqref{modt1} and
\eqref{modde} and approximating $ \eta(-1/\tau) \sim \tilde q^{1/24},
\theta_1 (v/\tau,-1/\tau) \sim 2 q^{1/8} \sin(\pi v/\tau)$
with $\tilde q=e^{-2\pi i/\tau}$,
we obtain
\be
\Omega_{spin}(N,J) \sim -2i \int_{iL-\frac12}^{iL+\frac12} d\tau (-i\tau)^{11}
\int_0^1 dv\
e^{-2\pi i (N-1)\tau + \frac{2\pi i}{\tau} +i\pi \frac{v^2}{\tau} +
2i \pi(J+\frac12) v}
\frac{\sin^2(\pi v)}{\sin(\pi v/\tau)}
\ee
Rescaling the variables as
\be
\tau= \frac{x}{\sqrt{N-1}}\ ,\quad v=-\frac{J+1/2}{\sqrt{N-1}} + x y
\ee
the integral becomes
\be
\begin{split}
\Omega_{spin}(N,J) \sim -2i \int dx dy  &
\left(\frac{-ix}{\sqrt{N-1}}\right)^{12}
\frac{\sin^2\left[\pi x ( y - \pi \frac{J+\frac12}{\sqrt{N-1}})\right]}
{\cos(\pi y \sqrt{N-1})} \\
& e^{2\pi i \sqrt{N-1}
\left( - x + \frac{1}{x} + x
\frac{(J+\frac12)^2}{N-1} + \frac12 xy^2 \right)}
\end{split}
\ee
Unfortunately, saddle point methods do not seem to apply straightforwardly,
due to the large oscillations in the denominator. For $J=0$, we find
numerically that
\be
\Omega_{spin}(N,J=0) \sim N^{-33/4}
e^{4\pi \sqrt N}
\ee
which is suppressed by $O(N^{-3/2})$ compared to the
all-$J$ result \eqref{boltzentropy}.
In particular, the success of the OSV conjecture appears to depend on
choosing an ensemble where the angular momentum is free to
fluctuate at zero conjugate potential $\Omega=0$.

\section{Other $Het(4, n_V)$  and $Het(2,n_V)$ Models \label{oth4}}
In this appendix, we discuss other heterotic orbifold models with
$\CN=4$ or $\CN=2$ supersymmetry and reduced rank.
We start with a different construction of the
$(4,16)$ model discussed in Section 3.3,
now based on the $SO(32)$ heterotic string in ten dimensions.
This construction can be easily generalized to produce
models with rank 12, 10 and 9.

\subsubsection*{Another $Het(4,16)$ model}

As explained in \cite{Kiritsis:2000zi},
the heterotic string at a point of enhanced symmetry $SO(16)\times
SO(16)$ may be obtained by orbifolding the  $SO(32)$ heterotic string
compactified on $S_1$ by a $\IZ_2$ action $g_1$, which
shifts the $U(1)$ charges of $8$ out of the 16 left-moving
bosons by half a unit, as well as acts by a translation by half a
period along the circle $S_1$. The partition function is most
easily written by decomposing the level 1 characters of $SO(32)$ under
$SO(16)\times SO(16)$, using the general formula
\bea
\label{o2noo}
O_{2n} &=& O_{n}O_{n}+V_{n}V_{n} \\
V_{2n} &=& O_{n}V_{n}+V_{n}O_{n} \\
S_{2n} &=& S_{n}S_{n}+C_{n}C_{n} \\
C_{2n} &=& S_{n}C_{n}+C_{n}S_{n}
\eea
relating the level 1 characters of $SO(2n)$ in the
O,V,S,C conjugacy classes to the level 1 characters of $SO(n)$.
Either of them are expressed in terms of free fermion partition functions,
\be
\begin{pmatrix} O_{n} \\ V_{n} \end{pmatrix}
= \frac12 \left(\theta_3^{n/2} \pm \theta_4^{n/2} \right)\ ,\quad
\begin{pmatrix} S_{n} \\ C_{n} \end{pmatrix}
= \frac12 \left(\theta_2^{n/2} \pm (-i \theta_1)^{n/2} \right)
\ee
In this fashion, the partition function for the Narain lattice
$\Gamma_{1,17}$ at the $SO(16)\times SO(16)$ point can be written as
\be
\label{z117}
\frac14 Z_{1,1}\ar{0}{0} \left( \sum_{a,b=0,1} \theta^{16}\ar{a}{b} \right)
+ \frac12 Z_{1,1}\ar{0}{\half} \theta_3^8\theta_4^8
+ \frac12 Z_{1,1}\ar{\half}{0} \theta_2^8\theta_3^8
+ \frac12 Z_{1,1}\ar{\half}{\half} \theta_2^8\theta_4^8
\ee
or, decomposing into the various sectors,
\be
\begin{split}
Z_+^0
\left( O_{16}O_{16}+ S_{16} S_{16} \right)
+
Z_-^0
\left( V_{16}V_{16}+ C_{16} C_{16} \right) \\
+
Z_+^\half
\left( O_{16}S_{16}+ S_{16} O_{16} \right)
+
Z_-^\half
\left( V_{16}C_{16}+ C_{16} V_{16} \right)
\end{split}
\ee
where
\be
Z_{\pm}^{h/2}
= \frac12\left( Z_{1,1}\ar{h/2}{0} \pm  Z_{1,1}\ar{h/2}{\half} \right)
\ee
denotes the projected lattice sum in the $h$-th twisted sector.
Compactifying this model further on $S_1'\times T^4$ to four dimensions,
we may now take a further $\IZ_2$ freely acting orbifold which exchanges the
two $SO(16)$ factors and acts as a translation by half a period
on $S_1'$: the untwisted, unprojected sector contributes
\be
\begin{split}
\label{untunp}
&\frac{Z_{6,6}\ar{00}{00} +  Z_{6,6}\ar{00}{\half 0} }{4}
\left( O_{16}O_{16}+ S_{16} S_{16} \right)
+
\frac{Z_{6,6}\ar{00}{00} -  Z_{6,6}\ar{00}{\half 0} }{4}
\left( V_{16}V_{16}+ C_{16} C_{16} \right) \\
&+ \frac{Z_{6,6}\ar{\half 0}{00} +  Z_{6,6}\ar{\half 0}{\half 0} }{4}
\left( O_{16}S_{16}+ S_{16} O_{16} \right)
\frac{Z_{6,6}\ar{\half 0}{00} -  Z_{6,6}\ar{\half 0}{\half 0} }{4}
\left( V_{16}C_{16}+ C_{16} V_{16} \right)
\end{split}
\ee
while the untwisted, projected sector reads
\be
\frac{Z_{6,6}\ar{00}{0 \half} +  Z_{6,6}\ar{00}{\half \half} }{4}
\left[ O_{16} (2\tau) + S_{16} (2\tau)\right]
+\frac{Z_{6,6}\ar{00}{0 \half} -  Z_{6,6}\ar{00}{\half \half} }{4}
\left[ V_{16} (2\tau) + C_{16} (2\tau) \right]
\ee
The twisted, unprojected sector can be obtained by modular S transformation,
\be
\frac{Z_{6,6}\ar{0 \half}{00} +  Z_{6,6}\ar{\half \half}{00} }{4}
\left[ O_{16} \left(\frac{\tau}{2}\right)
     + S_{16} \left(\frac{\tau}{2}\right) \right]
+\frac{Z_{6,6}\ar{0 \half}{00} -  Z_{6,6}\ar{\half \half}{00} }{4}
\left[ V_{16} \left(\frac{\tau}{2}\right)
     + C_{16} \left(\frac{\tau}{2}\right) \right]
\ee
and finally, the twisted, projected sector is obtained by a further
T transformation,
\be
\begin{split}
\frac{Z_{6,6}\ar{0 \half}{0 \half} +  Z_{6,6}\ar{\half \half}{\half \half} }{4}
\left[ O_{16} \left(\frac{\tau+1}{2}\right)
     + S_{16} \left(\frac{\tau+1}{2}\right) \right] \\
+\frac{Z_{6,6}\ar{0 \half}{0 \half} -  Z_{6,6}\ar{\half \half}{\half \half} }{4}
\left[ V_{16} \left(\frac{\tau+1}{2}\right)
     + C_{16} \left(\frac{\tau+1}{2}\right) \right]
\end{split}
\ee
In order to obtain the degeneracies of states with given electric
charges under the diagonal $SO(16)$, we need to change basis
and rewrite the product of level 1 characters in \eqref{untunp} into a
sum of products of $D_8=SO(16)$ level 2 theta functions with
characteristics. One may check that the finite group
$D_8/2D_8$ decomposes into 7
orbits, with respective length 1, 1, 1, 1, 56, 140, 56
corresponding to (i)
the orbit of the origin
(ii) the orbits of one half the highest weights of the
(level 1) V,S,C representations
(iii) the orbits of the highest weights of the
(level 2) $\Lambda_2$,$\Lambda_4$,$\Lambda_6$ representations,
of dimension 120, 1820 and 8008. In cases (i) and (ii),
the theta function with characteristics is simply
obtained by doubling the argument of the level 1 case, i.e.
\be
\theta_{D_8[2];O}(\tau) = O_{16}(2\tau)\ ,\quad
\theta_{D_8[2];V}(\tau) = V_{16}(2\tau)\ ,\mbox{etc}
\ee
while, in case (iii), an explicit computation shows that
\be
\theta_{D_8[2];120}(\tau)= \frac12 \theta_2^2 \theta_3^6 (2\tau) \ ,\quad
\theta_{D_8[2];1820}(\tau)= \frac12 \theta_2^4 \theta_3^4 (2\tau) \ ,\quad
\theta_{D_8[2];8008}(\tau)= \frac12 \theta_2^6 \theta_3^2 (2\tau)
\ee
Generalizing the identity \eqref{thth}, we may now use these theta
series to decompose the product of two level-1 theta series into
a sum of products of level-2 theta series:
\bea
O_{16}^2 &=& \theta^2_{D_8[2];O}+\theta^2_{D_8[2];O}+\theta^2_{D_8[2];O}
+\theta^2_{D_8[2];O}\nn\\
&&+ 56\ \theta^2_{D_8[2];120}+ 135\ \theta^2_{D_8[2];1820}
+ 56\ \theta^2_{D_8[2];8008}\\
V_{16}^2 &=& 2\ \theta_{D_8[2];O}\theta_{D_8[2];V} + 2\
\theta_{D_8[2];S}\theta_{D_8[2];C} \nn\\
&&+ 56\ \theta^2_{D_8[2];120}+ 135\ \theta^2_{D_8[2];1820}
+ 56\ \theta^2_{D_8[2];8008}\\
S_{16}^2 &=& 2\ \theta_{D_8[2];O}\theta_{D_8[2];S} + 2\
\theta_{D_8[2];V}\theta_{D_8[2];C} \nn \\
&&+ 112\ \theta_{D_8[2];120}\theta_{D_8[2];8008}
+ 135\ \theta^2_{D_8[2];1820} \\
C_{16}^2 &=& 2\ \theta_{D_8[2];O}\theta_{D_8[2];C} + 2\
\theta_{D_8[2];V}\theta_{D_8[2];S} \nn \\
&&+ 112\ \theta_{D_8[2];120}\theta_{D_8[2];8008}
+ 135\ \theta^2_{D_8[2];1820}
\eea
As in \eqref{ththe}, we view each term on the right hand side as the
product of the partition function for the lattice of physical electric charges
$P_1+P_2$, times the partition function of the lattice of
unphysical electric charges $P_1-P_2$. It is the latter
which, together with the partition function of the oscillators, determines
the degeneracies of DH states.

In all cases, the level-2 theta series with characteristics
are modular forms of
weight 4. Taking into account the action on the left-moving bosonic
oscillators, we find
that the degeneracies in the untwisted sector are enumerated by
\be
\frac14 \left( \frac{\theta_{D_8[2], \lambda} }{\eta^{24}}
\pm \delta_{0,\wp} \frac{2^4 \theta_2^4}{\eta^{12}} \right)
\ee
where $\lambda$ is any element in the finite group $D_8/2D_8$,
while those in the twisted sectors are counted by
\be
\frac12 \biggl(  {1\over \eta^{12} \vartheta_4^4}
\pm
 {1\over \eta^{12} \vartheta_3^4 } \biggr)
\ee
In particular, the asymptotics are governed by the same
formulae \eqref{deg414}. As in any $\CN=4$ heterotic models,
the absolute degeneracies are equal to (2/3 times) the helicity
supertraces $\Omega_4$.

\subsubsection*{$Het(4,12)$ model}
A similar construction as in \eqref{z117} allows to construct the
point of enhanced symmetry $SO(8)^4$ of the $SO(32)$ heterotic string:
one simply needs to orbifold
the heterotic string compactified on $S_1'\times S_2''$
by $\IZ_2\times \IZ_2$, where the two generators $g_1$ and $g_2$
both act by shifting  the $U(1)$ charges of a different set
of 8 left-moving bosons (4 of which being common to $g_1$ and $g_2$),
and by a translation by half a period in either of the two circles.
The partition function of the $\Gamma_{2,18}$ Narain lattice at
the $SO(8)^4$ point is therefore
\be
\begin{split}
Z_{D_4^4} =&\frac{1}{4\eta^{16}} \left[
\frac12 Z_{2,2}\ar{00}{00} \sum_{a,b=0,1} \theta^{16}\ar{a}{b} \right.\\
&\left. + \sum_{dd} \left( Z_{2,2}\ar{00}{dd} \theta_3^8\theta_4^8
+ Z_{2,2}\ar{dd}{00} \theta_2^8\theta_3^8
+ Z_{2,2}\ar{dd}{dd} \theta_2^8\theta_4^8 \right) \right]
\end{split}
\ee
where the sum runs over the 2-digit binary numbers
$dd=00,0\half ,\half 0,\half \half$ \cite{Kiritsis:2000zi}.
Using \eqref{o2noo}, this may be decomposed into characters of $SO(8)^4$,
\bea
\label{zd842}
Z_{D_4^4} &=& Z_{++}^{00}
\left( O_{8}^4 + V_8^4+S_8^4 + C_8^4 \right)
+
2 \left[ Z_{+-}^{00}  +Z_{-+}^{00}  +Z_{--}^{00}   \right]
\left( O_{8}^2 V_8^2 + S_8^2 C_8^2 \right) \nn\\
&+&
\left[ Z_{++}^{0\half } + Z_{++}^{\half 0} + Z_{++}^{\half \half }
+ Z_{--}^{\half \half } + Z_{+-}^{\half 0} + Z_{-+}^{0\half }  \right]
\left( O_{8}^2 + V_8^2 \right) \left( S_8^2 + C_8^2 \right) \nn \\
&+&
4 \left[ Z_{--}^{0\half } + Z_{--}^{0\half } + Z_{-+}^{\half 0}
+ Z_{-+}^{\half \half } + Z_{+-}^{0\half } + Z_{+-}^{\half \half }  \right]
O_{8} V_8 S_8 C_8
\eea
where
\be
Z_{\eps_1 \eps_2}^{h_1 h_2}
= \frac1{4\eta^{16}}
 \left( Z_{2,2}\ar{\frac{h_1}{2}\frac{h_2}{2}}{0\ 0} +
\eps_1 Z_{2,2}\ar{\frac{h_1}{2}\frac{h_2}{2}}{\half\  0} +
\eps_2 Z_{2,2}\ar{\frac{h_1}{2}\frac{h_2}{2}}{0\ \half }
+  \eps_1 \eps_2 Z_{2,2}\ar{\frac{h_1}{2}\frac{h_2}{2}}{\half\  \half } \right)
\ee
denotes the projected lattice sum in the $(h_1,h_2)$ twisted sector
of the $\IZ_2\times \IZ_2$ orbifold. The resulting theory can
be orbifolded by an element $g_3:=e$ of order 4 permuting the four
$SO(8)$ factors cyclically, together with a translation
of order 4 along one of the circles in the torus $T^4$.
The partition function in the untwisted sector, with an insertion
of an odd power of the generator is thus given by
\be
\begin{split}
Z_{D_4[4]}\ar{0}{\frac{g}{4}} =
\frac1{16 \eta^4(4\tau)}
\left(
Z_{3,3}\ar{000}{00\frac{g}{4}} +
Z_{3,3}\ar{000}{\half 0\frac{g}{4}} +
Z_{3,3}\ar{000}{0\half \frac{g}{4}} +
Z_{3,3}\ar{000}{\half \half \frac{g}{4}} \right) \\
\times \left[ O_{8}(4\tau) + V_8(4\tau) +S_8(4\tau)  + C_8(4\tau) \right]
\end{split}
\ee
with $g=1,3$, while for an insertion of $e^2$,
\be
\begin{split}
Z_{D_4[4]}\ar{0}{\half} =
\frac1{8 \eta^8(2\tau)} \left\{ \frac12
\left(
Z_{3,3}\ar{000}{00\half } +
Z_{3,3}\ar{000}{\half 0\half } +
Z_{3,3}\ar{000}{0\half \half } +
Z_{3,3}\ar{000}{\half \half \half } \right) \right. \\
\times \left[ O_{8}^2(2\tau) + V_8^2(2\tau)
+S_8^2(2\tau)  + C_8^2(2\tau) \right]\\
+
\left(
3 Z_{3,3}\ar{000}{00\half } -
Z_{3,3}\ar{000}{\half 0\half } -
Z_{3,3}\ar{000}{0\half \half } -
Z_{3,3}\ar{000}{\half \half \half } \right) \\
\times \left[ O_{8} (2\tau) V_8 (2\tau)
+ S_8(2\tau)   C_8(2\tau) \right] \\
+
\left(
Z_{3,3}\ar{0\half 0}{00\half } +
Z_{3,3}\ar{0\half 0}{0\half \half } +
Z_{3,3}\ar{\half 00}{00\half } +
Z_{3,3}\ar{\half 00}{\half 0\half } +
Z_{3,3}\ar{\half \half 0}{00\half } +
Z_{3,3}\ar{\half \half 0}{\half \half \half } \right) \\
\left. \times \left[ O_{8} (2\tau) + V_8 (2\tau) \right]
\left[ S_8(2\tau)  + C_8(2\tau) \right] \right\}
\end{split}
\ee
and, in the absence of any insertion,
$Z_{D_4[4]}\ar{0}{0} = \frac14 Z_{D_4^4}$.
The twisted sectors can be obtained as usual by modular
transformations, leading to
\be
\begin{split}
Z_{D_4[4]}\ar{\frac{h}{4}}{\frac{g}{4}} =
\frac{e^{-2\pi i g/3}}{16}
\left(
Z_{3,3}\ar{00\frac{h}{4}}
          {00\frac{g}{4}} +
Z_{3,3}\ar{\frac{h}{2}0\frac{h}{4}}
          {\frac{g}{2}0\frac{g}{4}} +
Z_{3,3}\ar{0\frac{h}{2}\frac{h}{4}}
          {0\frac{g}{2}\frac{g}{4}} +
Z_{3,3}\ar{\frac{h}{2}\frac{h}{2}\frac{h}{4}}
          {\frac{g}{2}\frac{g}{2}\frac{g}{4}} \right) \\
\times \left[ \frac{O_{8} + V_8 + S_8 + C_8}{\eta^4} \right]
\left(\frac{\tau+g}{4}\right)
\end{split}
\ee
for $h=1,3$, $g=0,1,2,3$,
\be
\begin{split}
Z_{D_4[4]}\ar{\half}{\frac{g}{4}} =
\frac{e^{-2\pi i g/3}}{8 \eta^8\left(\frac{\tau+g/2}{2}\right)}
\left\{ \frac12
\left(
Z_{3,3}\ar{00\half }{00\frac{g}{4}} +
Z_{3,3}\ar{\half 0\half }{\frac{g}{2}0\frac{g}{4}} +
Z_{3,3}\ar{0\half \half }{0\frac{g}{2}\frac{g}{4}} +
Z_{3,3}\ar{\half \half \half }{\frac{g}{2}\frac{g}{2}\frac{g}{4}}
\right)
\right. \\
\times \left[ O_{8}^2 + V_8^2
+S_8^2  + C_8^2 \right]\left( \frac{\tau+g/2}{2} \right)\\
+\left(
3 Z_{3,3}\ar{00\half}{00\frac{g}{4}} -
Z_{3,3}\ar{\half 0\half}{\frac{g}{2}0\frac{g}{4}} -
Z_{3,3}\ar{0\half \half}{0\frac{g}{2}\frac{g}{4}} -
Z_{3,3}\ar{\half \half \half}{\frac{g}{2}\frac{g}{2}\frac{g}{4}} \right)
\times \left[ O_{8} V_8
+ S_8  C_8 \right] \left( \frac{\tau+g/2}{2} \right) \\
+
\left(
Z_{3,3}\ar{00\half}{0 \frac{g+1}{2} \frac{g}{4}} +
Z_{3,3}\ar{0\half \half}{0 \frac{g+1}{2} 0} +
Z_{3,3}\ar{00\half}{\frac{g+1}{2} 0 \frac{g}{4}} +
Z_{3,3}\ar{\half 0\half}{\frac{g+1}{2} 0 0} \right. \\
\left. \left.  +
Z_{3,3}\ar{00\half}{\frac{g+1}{2} \frac{g+1}{2} \frac{g}{4}} +
Z_{3,3}\ar{\half \half \half}{\frac{g+1}{2} \frac{g+1}{2} 0} \right)
 \times \left[ O_{8}  + V_8 \right]
\left[ S_8  + C_8 \right] \left( \frac{\tau+g/2}{2} \right) \right\}
\end{split}
\ee
for $g=0,2$ and
\be
\begin{split}
Z_{D_4[4]}\ar{\half}{\frac{g}{4}} =
\frac{e^{-2\pi i g/3}}{8 \eta^4\left(\frac{\tau+(g-1)/2}{2}\right)} \left\{ \frac12
\left( Z_{3,3}\ar{00\half}{00\frac{g}{4}} +  Z_{3,3}\ar{\half 0\half}{\frac{g}{2}0\frac{g}{4}} +
       Z_{3,3}\ar{0\half \half}{0\frac{g}{2}\frac{g}{4}} + Z_{3,3}\ar{\half \half \half}{\frac{g}{2}\frac{g}{2}\frac{g}{4}} \right) \right. \\
\times \left[ O_{8} + V_8
+S_8  + C_8 \right]\left( \frac{\tau+(g-1)/2}{2} \right)
\end{split}
\ee
for $g=1,3$.

In order to extract the degeneracies of states of given electric
charge under the invariant (level 4) $SO(8)$, one
may work in two steps. Let us first assume that the four charge vectors
$P_1,\dots P_4$ are in the root lattice of $D_4$, as in the first term
of \eqref{zd842}. We decompose
\bea
\label{p1p2p3p4}
P_1+P_3 = 2 \Sigma + \wp &,\quad& P_2+P_4 = 2 \Sigma' + \wp'\\
P_1-P_3 = 2 \Delta - \wp &,\quad& P_2-P_4 = 2 \Delta' - \wp'
\eea
where $\wp,\wp'$ take value in the finite group $D_4/2D_4$.
Next we further decompose
\bea
\Sigma + \Sigma' = 2 \Sigma'' + \wp'' \\
\Sigma - \Sigma' = 2 \Delta'' - \wp''
\eea
where $\wp'' \in D_4/2E_4$. Since the physical charge is
\be
Q= P_1+P_2+P_3+P_4=4 \Sigma''+ 2 \wp'' + \wp + \wp'\ ,
\ee
we obtain the degeneracies of DH states with
a given charge $Q$ by summing over $\Delta,\Delta',\Delta'', \wp,\wp',\wp''$
at fixed values of $\Sigma''$ and $\P_0=\wp''/2 + (\wp + \wp')/4$ in the
discrete group $D_4/4D_4$. For the other terms in \eqref{zd842}, the same
decomposition holds, upon shifting $\wp$, $\wp'$, or $\wp''$ by
$2\lambda$ where $\lambda$ is in the weight lattice of $D_4$.
Decomposing the square of the charge vector as
\be
\begin{split}
\sum_{i=1}^4 P_i^2
=&
2 \left( \Delta -\frac12 \wp \right)^2
+2 \left( \Delta' -\frac12 \wp' \right)^2  \\
&+4 \left( \Delta''- \frac{\wp''}{2}+ \frac{\wp-\wp'}{4} \right)^2
+4 \left( \Sigma''+ \frac{\wp''}{2}+ \frac{\wp+\wp'}{4} \right)^2
\end{split}
\ee
we see that the partition function of the Narain lattice $\Gamma_{2,18}$ at
the $SO(8)^4$ point may be written as a sum of products of
two level 2 theta series
\be
\label{thpd42}
\Theta_{D_4[2], \wp}(\tau) := \sum_{\Delta\in D_4(1)}
=e^{2\pi i \tau( \Delta - \half  \wp)^2 }
\ee
times  two level 4 theta series,
\be
\label{thpd44}
\Theta_{D_4[4], \wp}(\tau) := \sum_{\Delta\in D_4(1)}
=e^{4\pi i \tau( \Delta - \frac14  \wp)^2 }
\ee
corresponding
to sums over the lattice vectors $\Delta, \Delta', \Delta'', \Sigma''$
respectively. Under this decomposition, the last factor can be viewed
as the partition function for the physical charges.

In order to compute the required theta series,
note that $D_4/2D_4$ decomposes into five orbits,
of respective length 1,1,1,1,12:
(i) the zero orbit
(ii) the orbit of one half the highest weights of the $V,S,C$ irreps
and (iii) the orbit of the highest weight of the adjoint
representation $A$. The corresponding level 2 theta series are given by
\be
\label{thd42e}
\theta_{D_4[2];O}(\tau) =O_8(2\tau)\ ,\quad
\theta_{D_4[2];V}(\tau) =V_8(2\tau)\ ,\quad
\ee
\be
\theta_{D_4[2];S}(\tau) =S_8(2\tau)\ ,
\theta_{D_4[2];C}(\tau) =C_8(2\tau)\ ,\quad
\theta_{D_4[2];A}(\tau) =\frac12 \theta_2^2\theta_3^2 (2\tau)
\ee
These can then be used to decompose the product of two identical
level 1 theta series according to
\bea
\label{o8o8}
O_{8}^2 &=& \theta^2_{D_4[2];O}+\theta^2_{D_4[2];V}+\theta^4_{D_8[2];S}
+\theta^4_{D_8[2];C}
+12\ \theta^2_{D_8[2];A} \\
\label{v8v8}
V_{8}^2 &=& 2\ \theta_{D_4[2];O} \theta_{D_4[2];V}
+2\ \theta_{D_4[2];S} \theta_{D_4[2];C}
-12\ \theta^2_{D_8[2];A}
\eea
as well as the relations which follow from \eqref{v8v8} by triality.

The level 4 theta series \eqref{thpd44}
can now be obtained by repeating this
procedure twice. They fall into $5\times 5$ orbits of the Weyl group,
corresponding to the two-stage decomposition $D_4/4D_4 =
(D_4/2 D_4) \times 2D_4/4D_4$. The theta series corresponding to
$\wp=4\lambda$ where $\lambda$ is the highest weight of the
$O,S,C,V,A$ representations are simply obtained from
\eqref{thd42e} by doubling the argument $\tau \to 2\tau$.

Using the duplication identities in Appendix \ref{cornu}, one may
rewrite the partition functions of the oscillators in the untwisted
sector as
\be
\frac{1}{\eta^8(\tau) \eta^4(4\tau)} = \frac{2^4}
{ (\theta_3^2-\theta_4^2) \theta_2 \eta^{9}} \ ,\quad
\frac{1}{\eta^8(\tau) \eta^8(2\tau)} = \frac{2^4}{\theta_2^4 \eta^{12} }
\ee
We thus find that the degeneracies in the untwisted sector are enumerated by
\bea
\frac1{16} \sum_{\substack{ \wp,\wp',\wp'' \in D_4/2D_4,\lambda\\
\wp+\wp'+2\wp''+\lambda=\wp_0}}
&&\left(
\frac{\theta_{D_4[2], \wp+2\lambda}\theta_{D_8[2], \wp'+2\lambda}
\theta_{D_4[4], -\wp+\wp'+2\wp''} }{\eta^{24}}
\right. \\ &&\left.
\pm \delta_{0,\wp}\ \delta_{0,\wp'} \delta_{0,\lambda}
2^4 \frac{\theta_{D_4[2], \wp''}(2\tau)}{\theta_2^4 \eta^{12}}
\pm
\delta_{0,\wp}\ \delta_{0,\wp'}\ \delta_{0,\wp''} \delta_{\lambda,0}\
\frac{2^4}{\theta_2 (\theta_3^2-\theta_4^2)\eta^{9}}
\right) \nn
\eea
The three terms behaves as
\be
\BesselI{7}{4}{Q^2/2}\ , \BesselI{7}{4}{\frac12 Q^2/2}
\ ,\BesselI{7}{4}{\frac38 Q^2/2}
\ee
respectively, so that the degeneracies are dominated by the
untwisted, unprojected contribution.

In the sector twisted by the order $2$ element $e^2$,
the momenta automatically have $\Delta=\wp=0$ and $\Delta'=\wp'=0$
but one still needs to sum over the unphysical charges $\Delta''$
using the level 2 identities \eqref{o8o8} with $\tau\to \tau/2$.
Using the duplication identities
\be
\frac{1}{\eta^8(\tau) \eta^4(\tau/2)} = \frac{2^2}{\theta_4^2
  \eta^{10} }
\ ,\quad
\frac{1}{\eta^8(\tau) \eta^8(\tau/2)} = \frac{2^4}{\theta_4^4 \eta^{12} }
\ee
we find that the degeneracies are given by
\be
\label{dt17}
\frac{1}{2}
\left(  {1\over \eta^{12} \vartheta_4^4}
\pm {1\over \eta^{12} \vartheta_3^4 } \right)
\theta_{D_8[2], \wp''}(\tau/2)
+
\delta_{0,\wp''}
\left(  {1\over \eta^{10} \vartheta_4^2}
\pm {1\over \eta^{10} \vartheta_3^2 } \right)
\ee

Finally, in the sectors twisted by the order 4 element $e$ or $e^3$,
one may rewrite the partition functions for the twisted oscillators as
\bea
\frac{1}{\eta^8(\tau) \eta^4(4\tau)} &=& \frac{2^4}
{ (\theta_3^2-\theta_4^2) \theta_2 \eta^{9}} \nn \\
\frac{1}{\eta^8(\tau) \eta^4\left(\frac{\tau}{4}\right)} =
\frac{1}{(\theta_3^2-\theta_2^2)\theta_4\eta^{9}} &,&
\frac{1}{\eta^8(\tau) \eta^4\left(\frac{\tau+1}{4}\right)} =
\frac{e^{i\pi/12}}{(\theta_4^2-i\theta_2^2)\theta_3\eta^{9}} \nn\\
\frac{1}{\eta^8(\tau) \eta^4\left(\frac{\tau+2}{4}\right)} =
\frac{e^{i\pi/6}}{(\theta_3^2+\theta_2^2)\theta_4\eta^{9}} &,&
\frac{1}{\eta^8(\tau) \eta^4\left(\frac{\tau+3}{4}\right)} =
\frac{e^{i\pi/4}}{(\theta_4^2+i\theta_2^2)\theta_4\eta^{9}}
\eea
We thus find that the degeneracies are enumerated by
\be
\frac14 \left( \frac{1}{(\theta_3^2-\theta_2^2)\theta_4\eta^{9}}
+\eps_1 \frac{1}{(\theta_4^2-i\theta_2^2)\theta_3\eta^{9}}
+\eps_2 \frac{1}{(\theta_3^2+\theta_2^2)\theta_4\eta^{9}}
+\eps_3 \frac{1}{(\theta_4^2+i\theta_2^2)\theta_3\eta^{9}}
\right)
\ee
where, depending on the moding of the momenta along the three circles,
$(\eps_1,\eps_2,\eps_3)$ is any vector in $(1,1,1), (-i,-1,i),
(-1,1,-1),(i,-1,-i),$ (the corresponding ground state dimensions
are $\Delta=3/8, 1/8, -1/8, \Delta= -3/8$, respectively.)

In all sectors, applying the Rademacher formula we find that
the degeneracies grow uniformly as
\bea
\label{deg48}
\Omega_{abs}(Q) = \frac23 \Omega_4(Q)  \sim
\BesselI{7}{4}{Q^2/2} + \cdots
\eea
The exponentially suppressed corrections however depend sensitively
on the details of the charges.

\subsubsection*{A Het(2,8) model}
Let us now consider an $\CN=2$ variant of the $(4,12)$ model.
We start from the $SO(32)$ heterotic string on $T^2$ at the point of
enhanced symmetry $SO(8)^4$, further compactify on a square $T^4$,
\be
\label{g622d44}
\Gamma_{6,22} = D_4(-1) \oplus D_4(-1) \oplus D_4(-1) \oplus D_4(-1)
\oplus II^{2,2} \oplus II^{4,4}
\ee
and perform a $\IZ^4$ orbifold acting on the momenta as
\be
g \vert P_1,P_2,P_3,P_4,P_5,P_6\rangle
= e^{2\pi i \delta \cdot P_5} \vert P_2,P_3,P_4,P_1,P_5, R(g) P_6 \rangle
\ee
where $R(g)$ acts by a $\IZ_4$ rotation in a two-plane inside $T^4$,
breaking the supersymmetry to $\CN=2$.
The degeneracies can be obtained easily from the $(4,12)$ model
by dropping the untwisted, unprojected sector and
multiplying by $\eta^4$ times the partition function of four $\IZ_4$-twisted
left-moving bosons (the contribution of the right-moving bosons is absorbed
into the helicity supertrace). The orbifold blocks for four
$\IZ_4$-twisted chiral bosons can be obtained by the following
simple trick: Consider the orbifold of $4\times 4=16$ chiral bosons by
cyclic permutations of the four blocks of four.
The partition function with one insertion
of the $\IZ_4$ generator is $1/\eta^4(4\tau)$. On the other hand,
diagonalizing the oscillators, it should be the product of
four untwisted, four $\IZ_2$-twisted boson and eight $\IZ_4$-twisted
chiral bosons:
\be
\frac{1}{\eta^4(4\tau)} = \frac{1}{\eta^4(\tau)} \times
\frac{2^2 \eta^2 (\tau)}{\theta_2^2(\tau)} \times
\left(Z_{4}\ar{0}{\frac14} \right)^2
\ee
hence
\be
Z_4\ar{0}{\frac14} =Z_4\ar{0}{\frac34} = \frac{\eta^2(2\tau)}{\eta^2(4\tau)}
= 4 \sqrt{ \frac{\eta \theta_2}{\theta_3^2 - \theta_4^2} }
= \frac{2\eta}{\th\ar{\frac12}{\frac14}(0 | \tau)}
\ee
The other orbifold blocks can be obtained by modular transformations,
\be
Z_4\ar{\frac14}{\frac{g}4} = Z_4\ar{\frac34}{\frac{g}4}
=  2 \frac{\eta^2((\tau+g)/2)}{\eta^2((\tau+g)/4)}\ ,\quad
\ee
\be
Z_4\ar{0}{0}=\frac{Z_{4,4}}{\eta^4}\ ,
Z_4\ar{0}{\frac12} = \frac{\eta^4(\tau)}{\eta^4(2\tau)}\ ,\quad
Z_4\ar{\frac12}{0} = 4\frac{\eta^4(\tau)}{\eta^4(\tau/2)}\ ,\quad
\ee
\be
Z_4\ar{\frac12}{\frac14} =2\frac{\eta^2(\tau)}{\eta^2(\tau/2)}\ ,\quad
Z_4\ar{\frac12}{\frac12} =4 \frac{\eta^4(\tau+1)}{\eta^4((\tau+1)/2)}\ ,\quad
Z_4\ar{\frac12}{\frac34} =2\frac{\eta^2(\tau+1)}{\eta^2((\tau+1)/2)}
\ee
Using the same notation as in the (4,12) model,
we thus find that the second helicity supertraces $\Omega_2$
in the untwisted sector are generated by
\be
\frac1{16}
\delta_{0,\wp}\ \delta_{0,\wp'}\
2^6 \frac{\theta_{D_4[2], \wp''}(2\tau) }{\theta_2^6 \eta^{6}}
\pm
\delta_{0,\wp}\ \delta_{0,\wp'}\ \delta_{\wp'',0}\
\frac{2^6}{\sqrt{\theta_2 (\theta_3^2-\theta_4^2)^3\eta^{9}}}
\ee
Importantly, the untwisted unprojected term does not contribute, due to its
extended $\CN=4$ supersymmetry.
The second term grows as
\be
\BesselI{5}{4}{\frac{3}{16} Q^2/2}
\ee
and is suppressed with respect to the first.

Finally, in the sector
twisted by the order $2$ element, we find that the second helicity
supertraces are generated by
\be
\label{dt16} \frac{1}{2} \left(  {1\over
\eta^{6} \vartheta_4^6} \pm {1\over \eta^{6} \vartheta_3^6 }
\right) \theta_{D_8[2], \wp''}(\tau/2) + \delta_{0,\wp''} \left(
{1\over \eta^{4} \vartheta_4^4} \pm {1\over \eta^{4} \vartheta_3^4
} \right)
\ee
The degeneracies from the second term grow as \be
\BesselI{5}{4}{\frac23 Q^2/2} \ee

Finally, in the sectors twisted by the order 4 element $e$ or $e^3$,
we find that the second helicity supertraces
are enumerated by
\be
\frac{1}{\sqrt{(\theta_3^2-\theta_2^2)^3\theta_4\eta^{9}}}
+\eps_1 \frac{1}{\sqrt{(\theta_4^2-i\theta_2^2)^3\theta_3\eta^{9}}}
+\eps_2 \frac{1}{\sqrt{(\theta_3^2+\theta_2^2)^3\theta_4\eta^{9}}}
+\eps_3 \frac{1}{\sqrt{(\theta_4^2+i\theta_2^2)^3\theta_3\eta^{9}}}
\ee
where $(\eps_1,\eps_2,\eps_3)$ is any vector in $(1,1,1),
(-i,-1,i), (-1,1,-1),(i,-1,-i),$. The corresponding ground state energies
are $\Delta=3/8, 1/8, -1/8,
\Delta= -3/8$ respectively. In these four cases, the second
helicity supertraces
grow as \be \BesselI{5}{4}{ Q^2/2} \ee

\section{Some Properties of the Mac-Mahon Function \label{mcmah}}

In this section, we derive some properties of the
Mac-Mahon function
\be
f(\lambda) := \sum_{n=1}^{\infty} n \log ( 1 - q^n)
\ee
with $q=e^{i n \lambda}$. This is an entire
function of $\la$ in the upper half plane.
Taylor-expanding the logarithm and carrying out the sum over $n$,
it may be rewritten as
\be
\label{deffls}
f(\la) =  \sum_{d=1}^\infty {1\over d} {1\over (2\sin {d\la \over 2})^{2}}
\ee
We would like to derive the asymptotic expansion for $\la\to 0$.

Let us recall the standard argument. From the standard expansion
\be\label{bern}
{x \over e^x-1} = 1-{x\over 2} + \sum_{n\geq 1} {B_{2n}\over (2n)!} x^{2n}
\ee
in terms of the Bernoulli numbers $B_n$, we get
%
%\be\label{bernii}
%{1\over (e^{x/2} - e^{-x/2})^2} = {1\over x^2} - \sum_{n\geq 1} {2n-1\over (2n)!} B_{2n} x^{2n-2}
%\ee
%
%so
%
\be\label{bernii}
{1\over (2\sin(x/2))^2} = {1\over x^2} + \sum_{n\geq 1} {2n-1\over (2n)!} B_{2n}(-1)^{n-1} x^{2n-2}
\ee
Note that $B_{2n}= (-1)^{n-1}\vert B_{2n}\vert$.

If we substitute \eqref{bernii} into \eqref{deffls}  and
exchange the sum on $n$ and $d$ we find the series
\be\label{formasymp}
  \lambda^{-2} \zeta(3) + \sum_{n=1}^\infty {(2n-1) \vert B_{2n} \vert  \over (2n)!} \lambda^{2n-2}(\sum_{d\geq 1}  d^{2n-3})
\ee
Note that the sums on $d$ are infinite. While one may try
and define them for $n\neq 1$ by zeta function regularization,
the $n=1$ term is still infinite. If we simply discard the $n=1$
term and use this regularization we get
\be\label{formasympi}
  \lambda^{-2} \zeta(3) - \sum_{n=0}^\infty \lambda^{2n+2} {\vert B_{2n+4} \vert \over (2n+4)!}
 {(2n+3)  \over (2n+2)  }    B_{2n+2}
\ee
Using the relations between Bernoulli numbers and Rieman zeta functions,
\be
\zeta(3-2g) = - \frac{B_{2g-2}}{2g-2}\ ,\quad
\zeta(2g) = (-1)^{g+1} \frac{B_{2g} 2^{2g-1} \pi^{2g}}{(2g)!}
\ee
valid for $g\geq 2$, $g\geq 1$, respectively, one recovers
the standard result in the topological string literature.

However the manipulation used above is not valid. One way to see it is
that an entire function such as $f(\la)$ cannot
possibly have an infinite term $\la^0 \zeta(1)$ in its
asymptotics. Nevertheless, the amazing agreement between the
coefficients of the terms  $\la^{\geq 2}$ with the
integrals on moduli space \cite{faber} and with the
predictions of heterotic/typeII duality \cite{Marino:1998pg}
suggest the higher terms are indeed correct.
This will prove to be the case.

One valid way to derive the asymptotics is to proceed as follows.
We use the series
\be\label{sers}
{1\over \sin^2(\pi x)} = {1\over \pi^2} \sum_{n\in \IZ} {1\over (x+n)^2}
\ee
Substituting into \eqref{deffls}, the double sum on $n,d$
is absolutely convergent. We can therefore exchange
the sum on $n,d$. Defining $z:= \la/(2\pi)$ we have
\be\label{iij}
f(\la) = {\zeta(3)\over \la^2} + {1\over 4\pi^2} \sum_{n\not=0} \sum_{d=1}^\infty {1\over d(dz+n)^2}
\ee
Define
\be
g(z) =  {1\over 4\pi^2} \sum_{n\not=0} \sum_{d=1}^\infty {1\over d(dz+n)^2}
\ee
In order to study the $z\to 0$ asymptotics, we should apply
the Poisson summation formula to the sum on $d$.

Care is however needed
due to the incomplete summation on $d$.
While the Poisson summation formula as usually stated applies
to continuous functions, we wish to apply it
to the function
\be\label{psof}
f(x) := \begin{cases}  {1\over x(xz+n)^2} & x \geq 1 \\
0 & x<1 \end{cases}
\ee
This falls off nicely at infinity, but has a discontinuity at $x=1$.

Suppose, generally, that  $f(x)$ has a discontinuity at $x=1$.
The standard procedure to prove the Poisson formula is to
construct the periodic function $F(x) = \sum_{n\in \IZ} f(x+n)$,
expand it in Fourier series,
$F(x) = \sum_{m\in \IZ} \hat F_m  e^{2\pi i m x}$,
and evaluate at $x=0$. For piecewise continuous functions,
the Fourier series only converges
to the average $\half(F(0+) + F(0-))$ at points of discontinuity.
If $f(x)=0$ for $x<1$ then we get
\be\label{psfdesc}
\half f(1) + \sum_{d=2}^\infty f(d) = \sum_{\ell\in \IZ} \int_1^\infty  e^{2\pi i \ell x} f(x) dx
\ee
Taking this into account we have the Poisson summation formula
\be\label{poisson}
\sum_{d=1}^\infty {1\over d(dz+n)^2} = {1\over 2(z+n)^2} + \sum_{\ell\in \IZ} \int_1^\infty e^{2\pi i \ell x} {1\over x(xz+n)^2} dx
\ee
Now we write $g(z) = g_0(z) + g_1(z)$ where
\bea
\label{dfgzer}
g_0(z)&:=& {1\over 4\pi^2} \sum_{n\not=0}\biggl( {1\over 2(z+n)^2} + \int_1^\infty   {1\over x(xz+n)^2} dx \biggr)\\
\label{dfgone}
g_1(z)&:=& {1\over 4\pi^2} \sum_{n\not=0}\sum_{\ell\not=0 } \int_1^\infty e^{2\pi i \ell x} {1\over x(xz+n)^2} dx
\eea
To compute the integrals we write
\be\label{parfracs}
{1\over x(xz+n)^2}= - {z\over n (xz+n)^2} + {1\over x n^2} - {z\over (xz + n)n^2} =
{d\over dx} \biggl[ {1\over n^2} \log{x\over xz+n}  + {1\over n (xz+n)} \biggr]
\ee
Let us analyze first $g_0(z)$. The integral on $x$ is elementary and we get:
\be\label{egnzne}
g_0(z)= {1\over 4\pi^2} \sum_{n\not=0}\biggl( {1\over 2(z+n)^2} + {1\over n^2}\bigl[\log(1/z) - \log({1\over z+n})\bigr] - {1\over n(z+n)} \biggr)
\ee
Expanding the various terms and recalling that $z= \la/(2\pi)$ we find
\be\label{gzerexp}
g_0(z) = {1\over 12} \log{2\pi \over \la} + {i \pi \over 24} + {1\over 4 \pi^2} \sum_{n\geq 1} {\log n^2\over n^2}
+ \sum_{n=0}^\infty \lambda^{2n+2} {\vert B_{2n+4} \vert \over (2n+4)!} \biggl( n+\half  - {1\over 2n+2}\biggr)
\ee
or, equivalently,
\be\label{gzerexp2}
g_0(z) = {1\over 12} \log{2\pi i \over \la} -  {1\over 2 \pi^2}
\zeta'(2)
+ \sum_{n=0}^\infty \lambda^{2n+2} {\vert B_{2n+4} \vert \over (2n+4)!} \biggl( n+\half  - {1\over 2n+2}\biggr)
\ee

Now we turn to \eqref{parfracs}. We can write $g_1$ as a sum of three terms:
\bea\label{sumgone}
g_1(z) & = & h_1+ h_2 + h_3 \\
h_1(z)&:= & {1\over 4\pi^2} \sum_{n\not=0}\sum_{\ell\not=0 } {1\over n^2} \int_1^\infty e^{2\pi i \ell x} {1\over x}  dx \\
h_2(z)&:= & - {1\over 4\pi^2} \sum_{n\not=0}\sum_{\ell\not=0 } {z\over n^2} \int_1^\infty e^{2\pi i \ell x} {1\over xz+n}  dx \\
h_3(z)&:= & -  {1\over 4\pi^2} \sum_{n\not=0}\sum_{\ell\not=0 } {z\over n } \int_1^\infty e^{2\pi i \ell x} {1\over (xz+n)^2}  dx
\eea
The first term, $h_1$ is just a constant in $z$, but is only convergent
when we group together the $\ell$ and $-\ell$ terms in the sum.
The integral can be computed in terms of the cosine integral function
${\rm Ci}(x)$ defined in \cite{abrastegun} 5.2.27:
\be\label{hone}
h_1 = {1\over 4\pi^2}{\pi^2\over 3} \sum_{\ell=1}^\infty \int_1^\infty 2 \cos(2\pi \ell x) {dx \over x}
= -{1\over 6} \sum_{\ell=1}^\infty {\rm Ci}(2\pi \ell)
\ee
Since ${\rm Ci}(2\pi x) \sim 1/(2\pi x)^2$ for
large integer $x$, the sum over $\ell$ converges. Indeed,
$h_1=\frac{1}{12}\gamma_E$ where $\gamma_E$ is the Euler-Mascharoni
constant.

For the second term we use the identity 5.1.28 in \cite{abrastegun}:
\be\label{expints}
\int_1^\infty e^{2\pi i \ell x} {1\over xz+n}  dx = {1\over z} e^{-2\pi i \ell(n/z)} E_1(-2\pi i \ell(1+n/z))
\ee
Note that $z$ has a nonzero imaginary part so the argument
of the exponential integral, and the denominator
in the integral is never zero even if $n$ is negative  ($E_1$
is a variant of the exponential integral).
Then we use the asymptotic expansion AS 5.1.51 to get
\be\label{expints2}
\int_1^\infty e^{2\pi i \ell x} {1\over xz+n}  dx \sim  {1\over z} \sum_{s=0}^\infty {(-1)^s s! \over (-2\pi i \ell(1+n/z))^{s+1} }
\ee
For $z$ pure imaginary, $z\to 0$, say,
this is in the valid range for the expansion. Now we sum over $\ell$ and get:
\be\label{expintsii}
\sum_{\ell\not=0} \int_1^\infty e^{2\pi i \ell x} {1\over xz+n}  dx \sim
\sum_{k=0}^\infty {(-1)^k \vert B_{2k+2} \vert \over 2k+2} {z^{2k+1} \over (z+n)^{2k+2} }
\ee
Taking a derivative with respect to $n$ gives
\be\label{expintsiii}
\sum_{\ell\not=0} \int_1^\infty e^{2\pi i \ell x} {1\over (xz+n)^2}  dx \sim
\sum_{k=0}^\infty (-1)^k \vert B_{2k+2} \vert  {z^{2k+1} \over (z+n)^{2k+3} }
\ee
Next we expand the denominators in a power series in $z/n$ and include the
sum over $n$. In this way we get:
\bea
\label{expintsii3}
h_2 &\sim&  - \sum_{n=0}^\infty \lambda^{2n+2} {\vert B_{2n+4} \vert \over (2n+4)!} {1\over 2n+2} \sum_{k=0}^n {2n+2 \choose 2k+2} B_{2k+2} \\
\label{expintsii4}
h_3 &\sim&  - \sum_{n=0}^\infty \lambda^{2n+2} {\vert B_{2n+4} \vert \over (2n+4)!}   \sum_{k=0}^n {2n+2 \choose 2k+2} B_{2k+2}
\eea
Now, the Bernoulli polynomial $B_n(x) = \sum_{k=0}^n {n \choose k} B_k x^{n-k} $at $x=1$ is  $B_n(1) = (-1)^n B_n $ so we may simplify
\be\label{expintsii5}
h_2 +h_3  \sim  - \sum_{n=0}^\infty \lambda^{2n+2} {\vert B_{2n+4} \vert \over (2n+4)!}
\biggl( {2n+3 \over 2n+2} B_{2n+2} + {2n+3\over 2n+2} n \biggr)
\ee

Putting it all together, the asymptotics for $f(\la)$ for $\la\to 0$
in the upper half-plane are
\be\label{realasymp}
f(\la)  \sim   \lambda^{-2} \zeta(3)
- \sum_{n=0}^\infty \lambda^{2n+2} {\vert B_{2n+4} \vert \over (2n+4)!}
 {(2n+3)  \over (2n+2)  }    B_{2n+2}
+ {1\over 12} \log{2\pi i \over \la} - {1\over 2 \pi^2} \zeta'(2)
+ \frac{1}{12} \gamma_E
\ee
This differs from the standard expression by the last three terms.
While the constant is not so important, the logarithmic term is
indeed important.

\medskip

We close this section by an observation which hints at possibly
interesting modular properties of the Mac-Mahon function. By
analogy with the Dedekind $\eta$ function, let us compute \be
E_3(\tau) := - q \frac{d}{dq} f(\lambda) = \sum_{n=1}^{\infty}
\frac{n^2 q^{n}}{1-q^n} \ee where $q=e^{2\pi i \tau}=e^{i n
\lambda}$ (the reason for this notation will become clear
shortly). Expanding the denominators, we obtain \be E_3(\tau) =
\sum_{N=1}^{\infty} \left( \sum_{n|N} n^2 \right) q^N =
\sum_{n=1}^{\infty} \sum_{m=1}^{\infty} n^2 q^{mn} =
\sum_{m=1}^{\infty} \frac{q^m ( 1+q^m)}{(1-q^m)^3} \ee Now we use
the identity \be \sum_{n\in \IZ} \frac{1}{(n+ z)^3} = 4i\pi^3
\frac{p (1+p)}{(1-p)^3} \ee where $p=e^{2\pi i z}$. This allows to
rewrite \be E_3(\tau) = \frac{1}{4i\pi^3} \sum_{m=1}^{\infty}
\sum_{n=-\infty}^{\infty}  \frac{1}{(n+ m \tau)^3} \ee While this
expression is similar to the usual modular invariant Eisenstein
series $E_{2n}$, it is important to note that, due to the
restriction $m>0$, $E_3$ is {\it not} modular invariant. Instead,
its orbit under $Sl(2,\IZ)$ is an infinite family of functions \be
E^{(p,q)}_3(\tau) = \frac{1}{4i\pi^3} \sum_{(m,n) \in \IZ, p m + n
q > 0} \frac{1}{(n+ m \tau)^3} \ee In particular,
$E_3(\tau)=E(\tau)=E_3^{(1,0)}(\tau)$ is mapped under $\tau\to
-1/\tau$ to $E_3^{(0,1)}(\tau)$ which does not admit a
$q$-expansion. Indeed, $f(\lambda)$ at $\lambda\to 0$ is not
exponentially suppressed but rather consists of an infinite power
series, as discussed above.

\medskip

%\bibliographystyle{JHEP-2}
%\bibliographystyle{utphys}
%\bibliographystyle{plain}
%\bibliographystyle{apsrev}

%\bibliography{ref}

\providecommand{\href}[2]{#2}\begingroup\raggedright\endgroup

\end{document}